\documentclass[aps,prd,amsmath,amssymb,superscriptaddress,eqsecnum,notitlepage,showpacs]{revtex4-1}
\usepackage{color}
\usepackage[pdftex]{graphicx}

\newcommand{\leaveout}[1]{}
\newcommand{\fixme}[1]{}
\newcommand{\fixmeAO}[1]{}

\newcommand{\Dg}{\Delta g_{\ell}}
\newcommand{\g}{g_{\ell}}
\newcommand{\gt}{g_{\ell}}

\newcommand{\gm}{g_{\ell-}}
\newcommand{\gmt}{g_{\ell-}}
\newcommand{\gp}{g_{\ell+}}
\newcommand{\hnt}{\tilde h_{n,\ell}}
\newcommand{\Dgt}{\Delta \tilde \g}
\newcommand{\DGl}{\Delta G_{\ell}}
\newcommand{\DGw}[3]{\DGl(#1,#2;#3)}
\newcommand{\Dh}{\Delta h_{\ell}}
\newcommand{\htd}{h_{\ell}}
\newcommand{\Dht}{\Delta \tilde h_{\ell}}
\newcommand{\Dhn}{\Delta h_{n,\ell}}
\newcommand{\Dhnt}{\Delta \hnt}
\newcommand{\f}{f_{\ell}}

\newcommand{\an}[1]{a_{#1}}
\newcommand{\anp}[1]{a_{{#1} +}}
\newcommand{\anm}[1]{a_{{#1} -}}
\newcommand{\Tnm}[1]{T_{#1}^{(-)}}
\newcommand{\Tnp}[1]{T_{#1}^{(+)}}
\newcommand{\Tno}[1]{T_{#1}^{(0)}}
\newcommand{\ant}[1]{\tilde a_{#1}}
\newcommand{\antm}[1]{\tilde a_{#1 -}}
\newcommand{\antp}[1]{\tilde a_{#1 +}}
\newcommand{\anb}[1]{\hat a_{#1}}
\newcommand{\nb}{\bar\nu}
\newcommand{\fb}{\hat f_{\ell}}
\newcommand{\h}{h_{\ell}}
\newcommand{\hn}{h_{n,\ell}}
\newcommand{\W}[1]{W\left(#1\right)}
\newcommand{\Wh}{\hat W}

\newcommand{\ob}{\bar{\omega}}
\newcommand{\rb}{\bar{r}}
\newcommand{\gpt}{g_{\ell+}}
\newcommand{\J}{J_{\ell}}

\newcommand{\Jn}{J_{n,\ell}}
\newcommand{\Jb}{\hat J_{\ell}}
\newcommand{\Jnb}[1]{\hat J_{#1,\ell}}
\newcommand{\nAS}{\nu_{AS}}
\newcommand{\Wimpind}{\zeta}
\newcommand{\imax}{I}
\newcommand{\thB}{\zeta_0}
\newcommand{\sumB}{j}

\newcommand{\Aout}{A^{out}_{\ell,\omega}}
\newcommand{\Ain}{A^{in}_{\ell,\omega}}

\begin{document}


\author{Marc Casals}
\email{marc.casals@ucd.ie}
\affiliation{School of Mathematical Sciences and Complex \& Adaptive Systems
Laboratory, University College Dublin, Belfield, Dublin 4, Ireland}

\author{Adrian Ottewill}
\email{adrian.ottewill@ucd.ie}
\affiliation{School of Mathematical Sciences and Complex \& Adaptive Systems
Laboratory, University College Dublin, Belfield, Dublin 4, Ireland}

\title{
Analytic Investigation of the Branch Cut of the Green Function in Schwarzschild Space-time
}

\begin{abstract}
The retarded Green function for linear field perturbations in Schwarzschild black hole space-time possesses a branch cut in the complex-frequency plane.
This branch cut has remained largely unexplored: only asymptotic analyses 
either for small-frequency
(yielding the known tail decay at late times of an initial perturbation of the black hole) or for large-frequency (quasinormal modes close to the branch cut
 in this regime
have been linked to quantum properties of black holes) have been carried out
in the literature.
The 
 regime along the cut
 inaccessible to these asymptotic analyses
has so far remained
essentially
 unreachable.
We present a new method for the analytic calculation of the branch cut \textit{directly on} the cut
for general-spin fields in Schwarzschild space-time.
This method is valid for \textit{any} values of the frequency on the cut and so it provides analytic access to the whole branch cut for the first time.
We calculate the modes along the cut and investigate their properties and connection with quasinormal modes.
We also investigate the contribution from these branch cut modes to the self-force acting on a point particle 
on a 
Schwarzschild background space-time.
\end{abstract}

\date{\today}
\maketitle





\section{Introduction}

The study of spin-field perturbations of black holes is important for many reasons.
Classically they are important, for example, for investigating the classical stability of black holes, for the detection of field waves emitted by black holes
and for the calculation of the
self-force on a point particle moving in a black hole background space-time (which serves to model a black hole inspiral in the extreme mass ratio).
Black hole perturbations are also important for understanding the quantum properties of black holes.

A crucial object for the study of Schwarzschild black hole perturbations is the retarded Green function of the wave equations they obey.
These equations may be separated by performing a Fourier  transform in time and a multipole decomposition in the angular separation of the
spacetime points.
Thus,  the calculation of  black hole perturbations is reduced to that of the Fourier modes in the complex-frequency ($\omega$) plane followed by 
a sum/integral of the modes.
Leaver~\cite{Leaver:1986,PhysRevD.38.725} \fixme{how to put the paper and the Erratum together in same Ref.?}
deformed the Fourier integral along the real-frequency axis into the complex-frequency plane, thus picking up the singularities of
the Fourier modes of the Green function.
These modes possess two types of singularities as functions of complex frequency:
an infinite number of simple poles  (the so-called quasinormal mode frequencies) and a branch cut (BC) which lies on the negative-imaginary axis (NIA).
Leaver showed that the two main contributions to the Green function then come from a series of modes (quasinormal modes, QNMs) at the poles 
and an integral of modes around the branch cut, which we shall refer to as BC modes.
While QNMs have been extensively studied (see, e.g.,~\cite{Berti:2009kk} for a review), very little is so far known about the BC
modes.

To date, only the leading asymptotic behaviour of the BC modes for small frequencies along the
NIA
has been studied at length in the literature.
This small-frequency regime in the BC
 is known to yield a leading
power-law
 tail decay at late times of an initial black hole perturbation (see, e.g., the pioneering work by Price~\cite{Price:1971fb,Price:1972pw},
details of the tail at large radius in~\cite{Leaver:1986} and details 
at arbitrary radius as well as a
higher-order logarithmic behaviour
in~\cite{PhysRevLett.109.111101,Casals:Ottewill:2011smallBC}).
The BC modes for large frequencies along the cut have only been studied by Maassen van den Brink~\cite{MaassenvandenBrink:2003as}
and by the authors~\cite{Casals:2011aa}. 
In~\cite{Casals:2011aa} it was shown that the
BC modes at large frequencies lead to a divergence
in the BC contribution to both
the Green function at `very early' times as well as
to
the black hole response to a noncompact Gaussian distribution as initial data
(it is expected these divergences in the BC contributions are cancelled by similar divergences in the QNM contributions).
The fact that highly-damped QNMs approach the BC enabled~\cite{PhysRevLett.109.111101,Casals:Ottewill:2011smallBC,MaassenvandenBrink:2003as} to apply the large-frequency
asymptotic analyses of the BC to the calculation of highly-damped QNMs. 
These modes have been associated to quantum properties of black holes (e.g., see~\cite{Maggiore:2007nq} in relation to black hole area quantization
and~\cite{Keshet:2007be} in relation to Hawking radiation in the case of rotating black holes).

To the best of our knowledge, the only investigations of the BC modes
for frequencies which are neither asymptotically large nor small
(we will refer to this regime as the `mid'-frequency regime) are the following ones,
 which were carried out in the gravitational case only.
 The BC in the Green function modes is due to a corresponding BC of a particular solution, $\g (r,\omega)$, of the radial equation, Eq.(\ref{eq:radial ODE}) below.
In~\cite{Leung:2003ix,Leung:2003eq} the authors obtained the radial solution $\g (r,\omega)$
for frequencies  near, but off, 
the NIA via a numerical integration of the radial equation.
They 
thus calculated
the radial solution on both sides of -- but away from --  the NIA,
 evaluated the difference and then extrapolated it onto the NIA, 
thus obtaining the BC `strength'.
This is a rather tricky numerical evaluation,  since the difference in values of $\g$ between the two sides of the NIA becomes
exponentially-small  as the frequency approaches the NIA.
The only other investigation of BC modes in the `mid'-frequency regime on the NIA was carried out by Maassen van den Brink who,
in a different and impressive work~\cite{MaassenvandenBrink:2000ru}, performed an asymptotic analysis of the
BC modes about the
so-called algebraically-special frequency  $\omega_{AS}$
~\cite{1973JMP....14.1453W,citeulike:11195363,MaassenvandenBrink:2000ru}.

The algebraically special frequency
lying within the `mid'-frequency regime on the NIA 
occurs only
for the case of field perturbations of spin $s=2$ (axial gravitational) 
and correspondingly the rest of this paragraph applies to this case only.
The BC modes have a distinct `dipole-like' behaviour near $\omega_{AS}$, unlike at other frequencies~\cite{Leung:2003ix,Leung:2003eq}.
The algebraically special frequency, though not a QNM itself for axial gravitational perturbations 
(it is a QNM for polar gravitational perturbations)~\cite{MaassenvandenBrink:2000ru} is intimately linked
to QNMs: the dipole-like behaviour of the BC modes may be explained in terms of poles in the `unphysical' complex-frequency
 Riemann sheet~\cite{Leung:2003ix,Leung:2003eq}. 
Furthermore, a QNM frequency very close to (or exactly equal to) 
$\omega_{AS}$
marks the start of the highly-damped region of QNMs 
(e.g.,~\cite{Berti:2009kk}).
As the rotation of the black hole is increased from zero (i.e., the Schwarzschild case studied in this paper), 
multiplets of QNMs emerge from -- exactly at or very near to, depending on the azimuthal
angular number --
 the algebraically special frequency $\omega_{AS}$~\cite{Leung:2003ix,Leung:2003eq,PhysRevD.68.124018},
at least in the case of the lowest multipole angular momentum number $\ell=s=2$. 

No analytic method exists so far for calculating the BC of the Green function in the `mid'-frequency regime (except, as mentioned above, near $\omega_{AS}$ for $s=2$).
However, the above works (see Ref.18 in~\cite{Leung:2003ix} and Sec.VI~\cite{MaassenvandenBrink:2000ru}) suggest the tantalizing possibility of calculating
the BC by expressing 
the BC `strength' 
via
a convergent series of irregular confluent hypergeometric functions evaluated directly on the NIA, so
that no extrapolation onto the NIA would be required.
In this paper we take up this suggestion. Thus, we provide a new method for calculating analytically the BC modes for general integral spin
 directly on the NIA for arbitrary values of the frequency. 
We prove that 
our
new series for 
the BC modes
 is convergent for \textit{any} values of the
 frequency along the NIA, thus providing analytic access for the first time
to the whole `mid'-frequency regime.
We note that our method is also valid in the  small- and large- frequency regimes, but it is not useful there 
since convergence becomes slower as the frequency becomes small while, for large-frequencies, the BC modes grow and oscillate for fixed radii.
Asymptotic analyses are therefore necessary in practise in these regimes.
 
We calculate the BC modes using our new method and we investigate their properties and connection with QNMs.
We also re-analyze \fixme{?} the so-called Jaff\'e series (which is a series representation of the radial solution which is purely ingoing into
the event horizon and possesses no BC) and, in particular, the behaviour of the Jaff\'e coefficients.
Finally, we apply our calculation of the BC modes to investigate their contribution to the self-force  (see, e.g.,~\cite{Poisson:2011nh})  acting on a point particle moving on a
Schwarzschild background space-time.
In~\cite{PhysRevLett.109.111101} we `sketched out' the main idea for our new method for the calculation of the BC modes for arbitrary frequency, in this paper
we `flesh out' the details.
We note that the method we present here provided the results for the plots of quantities in the `mid'-frequency regime in~\cite{Casals:2011aa}, where it was shown
that these `mid'-frequency results overlap with the large-frequency asymptotics presented there.
In~\cite{Casals:Ottewill:2011smallBC} we will present a thorough small-frequency analysis of the BC modes and we will show 
that these `mid'-frequency results also overlap with that analysis in the small-frequency regime.

In Sec.\ref{sec:BC} we introduce the main perturbation equations and expressions for the Green function modes.
In Sec.\ref{sec: series} we present the various series representations which we use for the calculation of the BC modes; 
in particular, Eq.(\ref{eq:Leaver-Liu series for Deltagt}) is the new series that we derive and use for the calculation of the pivotal 
quantity, the
BC `strength'.
In Sec.\ref{sec:Jaffe} we analyse the so-called Jaff\'e coefficients $\an{n}$ (in particular, we correct the large-$n$ asymptotics
of these coefficients given in the literature), which are fundamental in the calculation of all the series representations we use.
In Secs.\ref{sec:calc f}--\ref{sec:BCMs} we calculate the various quantities required for the BC modes and these modes themselves.
In Sec.\ref{sec:SF} we investigate the contribution of the BC modes to the self-force.
Finally, in Appendix~\ref{sec:App} we give some properties of the irregular confluent hypergeometric function, which we require for the calculation
of the BC modes in the main body of the paper.

In this paper we take units $c = G =1$.
We will frequently use a bar over a quantity to indicate that it has been made dimensionless via the introduction of 
an appropriate factor of the radius of the event horizon, $r_h=2M$, where $M$ is the mass of the black hole.


\section{Branch Cut} \label{sec:BC}

After a Fourier-mode decomposition in time $t$ and a multipole-$\ell$ decomposition in the angular distance $\gamma$, 
the retarded Green funcction for linear field perturbations in Schwarzschild space-time is expressed as
\begin{align} \label{eq:Green}
&
G_{ret}(x,x')=
\frac{1}{r\cdot r'}
\sum_{\ell=0}^{\infty}(2\ell+1)P_{\ell}(\cos\gamma)G^{ret}_{\ell}(r,r'; t),\quad
G^{ret}_{\ell}(r,r'; t)\equiv
\frac{1}{2\pi}
\int_{-\infty+ic}^{\infty+ic} d\omega\ G_{\ell}(r,r';\omega)e^{-i\omega  t},
\end{align}
\fixme{Eq.10Dolan\& Ottewill'11: we're missing a minus sign? Missing $1/(rr')$ in Eq.1~\cite{Casals:2011aa}. Is Eq.\ref{eq:Green} 
only valid in the scalar case?}
where $c>0$ and
the Fourier modes of the Green function are given by
\begin{align}
G_{\ell}(r,r';\omega)=\frac{\f(r_<,\omega)\g(r_>,\omega)}{\W{\omega}},
\end{align}
where  $r_>\equiv \max(r,r'),\ r_<\equiv \min(r,r')$ and $r$ is the Schwarzschild radial coordinate.
The function 
\begin{align}  \label{eq:Wronskian}
\W{\omega}\equiv 
W\left[\g(r,\omega),\f(r,\omega)\right]=
\gt\f'-\f\gt',
\end{align}
where a prime indicates a derivative with respect to $r_*$,
is the Wronskian of two solutions $\f(r,\omega)$ and $\g(r,\omega)$ of
the following second order radial ODE:
\begin{align} \label{eq:radial ODE}
&\left[\frac{d^2}{dr_*^2}+\omega^2-V(r)\right]u_{\ell}(r,\omega)=0
\\
&V(r)\equiv \left(1-\frac{r_h}{r}\right)\left[\frac{\ell(\ell+1)}{r^2}+\frac{r_h(1-s^2)}{r^3}\right]
\nonumber
\end{align}
where $r_*\equiv r+r_h\ln\left(
\frac{r}{r_h}
-1\right)$ is the radial `tortoise coordinate', $r_h\equiv 2M$ is the radius of the event horizon and $M$ is the mass of the Schwarzschild black hole.
For $\omega\in \mathbb{R}$\fixme{should it be $\ob\in \mathbb{R}$?}, the solutions $\f$ and $\g$
 obey the `physical' boundary conditions
of, respectively, purely-ingoing waves into the black hole 
\begin{align} \label{eq:f,near hor}
&
\f(r,\omega) \sim e^{-i\omega r_*}, & \rb_*\to -\infty,
\\&
\f(r,\omega) \sim \Aout e^{+i\omega r_*}+\Ain e^{-i\omega r_*}, & \rb_*\to +\infty
 \label{eq:f,large-r}
\end{align}
and purely-outgoing waves out to radial infinity,
\begin{equation} \label{eq:g}
\gt(r,\omega)\sim  e^{+i\omega r_*}, \quad
\ \rb_*\to \infty
\end{equation}
The complex-valued coefficients $\Ain$ and $\Aout$ are, respectively, incidence and reflection coefficients and
it is straight forward to check that $W=-2i\omega \Ain$.
The boundary conditions (\ref{eq:f,near hor}) and (\ref{eq:g})
also define, respectively, the radial solutions $\f$ and $\gt$ unambiguously for $\text{Im}(\omega)\ge 0$ when $r_*\in\mathbb{R}$.
In $\text{Im}(\omega) < 0$, with $r_*\in\mathbb{R}$ \fixme{should these $\omega$ and $r_*$ actually be $\ob$ and $\rb$?}, the solution $\g$ must 
be defined by analytic continuation.

\begin{figure}[h!]
\begin{center}
   \includegraphics[width=8cm]{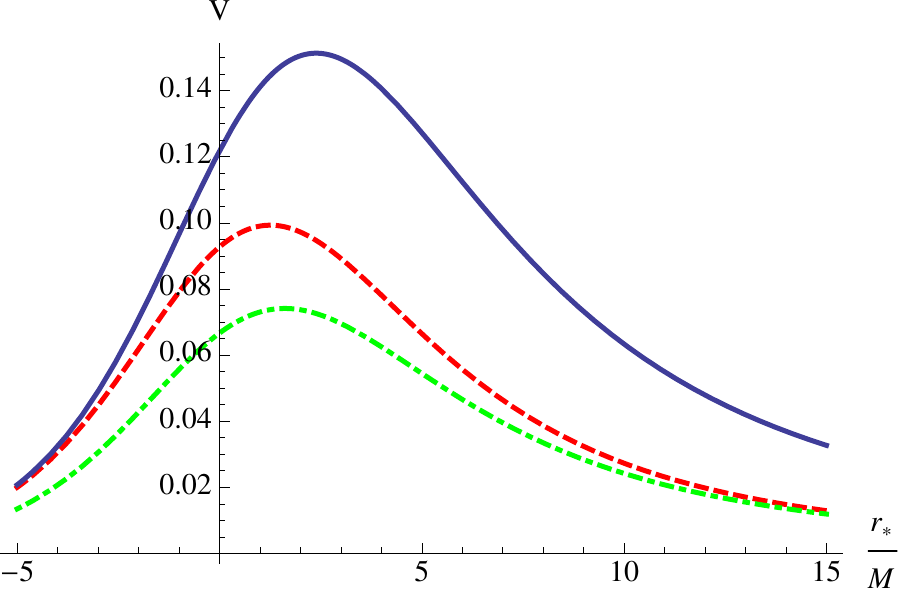}
    \end{center}
\caption{
Radial potential Eq.(\ref{eq:radial ODE}) as a function of $r_*/M$.
Continuous blue curve: $s=2$, $\ell=2$; dot-dashed green curve: $s=1$, $\ell=1$; dashed red curve: $s=0$, $\ell=1$.}
\label{fig:potential}
\end{figure}

The parameter $s=0,1,2$ in the potential in Eq.(\ref{eq:radial ODE})
 is the helicity  of the field perturbation, to which, with an abuse of language, we will refer to as `spin':
$s=2$ corresponds to axial -- also called `odd' -- gravitational perturbations (in which case Eq.(\ref{eq:radial ODE}) becomes the Regge-Wheeler equation~\cite{Regge:1957td}), 
$s=1$ to electromagnetic perturbations~\cite{Wheeler:1955zz} 
and $s=0$ to scalar perturbations~\cite{Price:1971fb,Price:1972pw}.
Polar -- or `even' -- gravitational perturbations obey the Zerilli equation~\cite{PhysRevLett.24.737,PhysRevD.2.2141}
and solutions to 
this
 equation can be obtained from the solutions, and their radial derivatives, to the Regge-Wheeler equation~\cite{Chandrasekhar}.
At the algebraically special frequency $\omega_{AS}$~\cite{1973JMP....14.1453W,citeulike:11195363,MaassenvandenBrink:2000ru}, 
however, this relationship 
between solutions to the Zerilli equation and solutions to  the Regge-Wheeler equation
becomes singular.
In Fig.\ref{fig:potential} we plot the potential $V(r)$ for some token values of spin $s$ and multipole number $\ell$.

It can be shown~\cite{Leaver:1986a} that the radial solution $\f$ has no branch cut in the complex-frequency plane whereas $\g$ has a branch cut 
down the negative imaginary axis NIA (see, e.g., Eq.(\ref{eq:Leaver-Liu}) below).
This BC in $\g$ can be explained~\cite{Ching:1994bd,Ching:1995tj} in terms of the radial potential (minus the centrifugal barrier) falling off
slower than exponentially at radial infinity.
On the other hand, the exponential decay with $r_*$ of the potential near the horizon leads to a series of poles in $\f$ down in the NIA (see Sec.\ref{sec:calc f} below). 
We note that the Wronskian $W$ `inherits' the BC from $\g$ and the poles in the NIA from $\f$.
We define 
$\Delta A(\nu)\equiv  A_+(-i\nu)-A_-(-i\nu)$
for any function $A=A(\omega)$ possessing a BC along the NIA,
where $A_{\pm}(-i\nu)\equiv \lim_{ \epsilon\to 0^+}A(\pm \epsilon-i\nu)$, with $\nu>0$.
We will equally refer to both quantities $\omega$ and $\nu\equiv i\omega$ as `frequencies'; we note that $\nu>0$ along the NIA.

We note the symmetries
\begin{equation} \label{eq:symms f,g}
\gt(r,\omega)=\gt^*(r,-\omega^*),\quad  \f(r,\omega)=\f^*(r,-\omega^*),
\quad
W(\omega)=W^*(-\omega^*)
\quad \text{if} \quad r_*\in \mathbb{R}
\end{equation}
which follow from the radial ODE (\ref{eq:radial ODE}) and the boundary conditions
(\ref{eq:f,near hor}) and (\ref{eq:g}).
These symmetries lead to 
$\gm=\gp^*$
and
$W_+=W_-^*$
if
$r_*\in \mathbb{R}$,
so that the branch cuts of $\g(r,\omega)$ and $W$ along the NIA are only in their imaginary parts, their real parts having no branch cut.
In particular, then, the absolute value of the Wronskian, $|W|$, has no BC.

The Fourier integral along the real frequency line in Eq.(\ref{eq:Green}) can be deformed into the complex-frequency plane~\cite{Leaver:1986,PhysRevD.38.725}.
The two main contributions to $G^{ret}_{\ell}$ are, then, a series over the residues at the poles of the Fourier modes $G_{\ell}(r,r';\omega)$
(the QNM frequencies, which are located at the zeros of the Wronskian $W$) and an integral around the BC.
The branch cut contribution $G^{BC}$ to the retarded Green function is given by
\begin{equation}
\label{eq:GBC}
G^{BC}(r,r';\gamma;t)=
\frac{1}{r\cdot r'}
\sum_{\ell=0}^{\infty}(2\ell+1)P_{\ell}(\cos\gamma)G^{BC}_{\ell}(r,r';t),\quad
G_{\ell}^{BC}(r,r';t)\equiv
-\frac{i}{2\pi}
\int_{0}^{-\infty} d\nu\ 
\DGw{r}{r'}{\nu}
e^{-\nu t},
\end{equation}
\fixme{Eq.10Dolan\& Ottewill'11: we're missing a minus sign?}
where the 
BC modes
$\DGl$ can be expressed as~\cite{Casals:2011aa,Leung:2003ix}
\begin{align} \label{eq:DeltaG in terms of Deltag}
\DGw{r}{r'}{\nu}=
-2i\nu \f(r,-i\nu)\f(r',-i\nu)
\frac{
q(\nu)
}{
|W|^2
},
\quad r_*\in \mathbb{R}
\end{align}

We denote the function $q(\nu)$ as the branch cut `strength' as it is defined via the equation
\begin{equation} \label{eq:dg=qg}
\Dgt(r,\nu)
=iq(\nu)\gt(r,+i\nu).
\end{equation}
where, here, $\Dgt(r,\nu)\equiv \gpt(r,-i\nu)-\gmt(r,-i\nu)$ (the extra tilde in the notation is justified in the next section).
\fixme{Swop notations between $\Dgt$ and $\Dg$?}
From the symmetries (\ref{eq:symms f,g}) and the fact that $\gt(r,+i\nu)$ (that is, $\g$ evaluated on the positive-imaginary axis) is real-valued it
follows that $q(\nu)$ is also a real-valued quantity.
We note that `$-iq$' here corresponds to the quantity `$K$' in Eq.31~\cite{Leaver:1986}.

As mentioned in the Introduction,
we will use a bar over a quantity to indicate that it has been made dimensionless via the introduction of 
an appropriate factor $r_h$, e.g.,  $\rb \equiv r/r_h$, $\ob\equiv \omega r_h$, $\bar{\nu}\equiv \nu r_h$, etc.





\section{Series representations for the radial solutions} \label{sec: series}

If one tried to find the radial solution $\f$ or $\g$ in the region $\text{Im}(\omega)<0$ by na\"ively solving numerically the 
radial Eq.(\ref{eq:radial ODE}) and imposing the `boundary conditions' (\ref{eq:f,near hor}) and (\ref{eq:g}), respectively, one would run into
computational problems. The reason is that these `boundary conditions' are exponentially dominant over the other, linearly independent
solution at the radial endpoint where the condition is imposed (that is, at $\rb_*\to -\infty$ for $\f$ and at $\rb_*\to +\infty$ for $\g$). 
Therefore, if one tried to numerically integrate the radial ODE starting with the `boundary condition' at one endpoint towards the other
endpoint, any accidental inclusion -- no matter how small -- of the other, wrong solution would grow 
exponentially and so would the numerical error\fixme{explain this better}.
There are various methods around this problem. For example, one could solve the radial equation in the region $\text{Im}(\omega)\ge 0$,
where the boundary conditions are well-posed, and then analytically continue onto the region $\text{Im}(\omega)<0$.
Also, Leaver's Eqs.32--36~\cite{Leaver:1986} provides a framework for calculating the BC contribution to the retarded Green function.
However, this method is rather difficult to implement (except in the asymptotic small-$\nb$ regime) due to the presence of Leaver's
`phase parameter', which is required because of the use of a particular series representation for $\g$ in terms of Coulomb wave functions.
In this paper we choose to use certain series representations \fixme{Aren't these effectively analytic continuations from $\text{Im}(\omega)\ge 0$?}
 for $\f$ and $\g$ which do not involve Leaver's `phase parameter' and which we show are convergent in the desired region on the frequency plane.

Leaver~\cite{Leaver:1986a} provides various series representations for the radial solutions $\f$ and $\g$.
All calculations of the BC modes in this paper are carried using
a specific choice of series representation for each one of the two solutions, which we give in Secs.\ref{sec:series for f} and \ref{sec:series for g}.
However, while the new series representation for $\Dgt$ (and therefore for the BC `strength' $q(\nu)$) which we present in Sec.\ref{sec:series for Delta tilde g}
 is fundamentally based on our choice of series representation for $\g$,
our calculation of the BC modes does not depend in an important way on the specific choice of series for calculating $\f$:
one could  just as well use any different method valid in the `mid'-frequency regime for calculating $\f$.
We present the various series that we use in the following subsections and we
 investigate their convergence properties in the following sections.


\subsection{Series for $\f$}\label{sec:series for f}

In order to calculate the radial function $\f$, we will use the well-known Jaff\'e series~\cite{Leaver:1986a}
\begin{align} \label{eq:Jaffe}
&
\f(r,\omega)=\left(
\bar r
-1\right)^{-i\bar\omega} 
\bar r
^{2i\bar\omega}e^{i\omega r}
\J(r,\omega),
\\&
\J(r,\omega)\equiv \sum_{n=0}^{\infty}
\Jn(\omega),\quad
\Jn\equiv
\an{n}(\omega)\left(
1-\frac{1}{\bar r}
\right)^n
\nonumber
\end{align}
We will refer to the complex-valued coefficients $\an{n}(\omega)$ as the Jaff\'e series coefficients, even though they also appear as coefficients
in the series representation that we will use for $\g$, Eq.(\ref{eq:Leaver-Liu}) below.
The Jaff\'e series coefficients are functions of the series index $n$, the frequency $\omega$ and, although not indicated explicitly, the multipole number $\ell$ and
the spin value $s$.
The Jaff\'e series coefficients satisfy a 3-term recurrence relation which we give and analyze in the following section.
The initial value $\an{0}$ remains undetermined by the recurrence relation; the specific value
$\an{0}=e^{-2i\bar\omega}$ yields the desired normalization (\ref{eq:f,near hor}) for $\f$ and, therefore, this will always be our choice of value for
 $\an{0}$ when using the Jaff\'e series for $\f$.


\subsection{Series for $\g$}\label{sec:series for g}

Our choice of series representation for $\g$ is also given in~\cite{Leaver:1986a}:
\begin{align} \label{eq:Leaver-Liu}
&
\g(r,\omega)=
\bar r
^{1+s}\left(
\bar r
-1\right)^{-i\bar\omega}e^{i\omega r}\h(r,\omega),\quad
\h(r,\omega)\equiv \sum_{n=0}^{\infty}\hn(\omega),\qquad
\hn\equiv 
\ant{n}
(\omega)
\Tnm{n}
\\ &
\Tnm{n}
\equiv (-2i\bar\omega+1)_n\ U(s+1-2i\bar\omega+n,2s+1,-2i\omega r), \nonumber
\end{align}
where $\ant{n}(\omega)$ 
satisfy the same recurrence relations 
as the Jaff\'e series coefficients $\an{n}$ in Eq.(\ref{eq:Jaffe}) but it is $\ant{0}\neq \an{0}$
-- that is, $\ant{n}$ and $\an{n}$ only differ by an overall normalization factor which we give below.
The series (\ref{eq:Leaver-Liu}) has been broadly ignored in the literature, possibly due to the fact that the irregular confluent hypergeometric $U$-functions  are rather
hard to manage.
We will refer to Eq.(\ref{eq:Leaver-Liu}) as the `Leaver-$U$ series'.

It is clear from the Leaver-$U$ series  Eq.(\ref{eq:Leaver-Liu}) and the properties of the irregular confluent hypergeometric function~\cite{bk:onlineAS}
 that the radial solution $\g(r,\omega)$ has a branch cut running along the line $\omega r:0\to -\infty\cdot i$.
If $r>0$, then $\g(r,\omega)$ has a branch cut along the NIA, $\omega:0\to -\infty\cdot i$.

The principal branch of $U(a,b,z)$ is given by $\arg (z)\in(-\pi,+\pi]$.
Therefore, we can evaluate 
directly  {\it on} the NIA 
the confluent hypergeometric $U$-function appearing in Eq.(\ref{eq:Leaver-Liu})
 and calculate the corresponding
$\Tnm{n}$ via Eq.(\ref{eq:Leaver-Liu}). 
That is, $\Tnm{n}$ may be evaluated {\it on} the NIA and its value will correspond to the principal branch value,
i.e., to the limiting value as the frequency $\omega$ approaches the NIA from the third quadrant in the complex-frequency plane.
The corresponding value of $\gt$ will then give $\gpt$ provided that the series Eq.(\ref{eq:Leaver-Liu}) converges.
It will be understood, when we
do not
say it explicitly, that any quantities possessing a BC 
along the NIA which are evaluated on the NIA via the use of Eq.(\ref{eq:Leaver-Liu})
will correspond to their limiting value approaching the NIA from the third quadrant.

In order to check what boundary condition the Leaver-$U$ series (\ref{eq:Leaver-Liu}) satisfies for $r\to\infty$, we 
use Eq.13.5.2~\cite{bk:AS} and we obtain
\begin{equation} \label{eq:g r*nu->inf mine}
\g(r,\omega)\sim \bar r^{1+s}\left(
\bar r
-1\right)^
{-i\bar\omega}(-2i\omega r)^{-s-1+2i\bar\omega}
 \ant{0} 
 e^{+i\omega r} 
\quad \text{for}\ |\omega r|\to \infty\ \text{and}
\ |\pi/2-\arg(\omega)-\arg(r)|<3\pi/2
\end{equation}
Therefore, when $r>0$, the Leaver-$U$ series yields the asymptotics
\begin{equation} \label{eq:g r->inf mine}
\g(r,\omega)\sim
(-2i\bar\omega)^{-s-1+2i\bar\omega}
 \ant{0} 
 e^{+i\omega r_*} 
\quad \text{for}\ r\to \infty\ \text{and}
\ |\pi/2-\arg(\omega)|<3\pi/2,\quad r>0
\end{equation}
We note that Eq.(\ref{eq:g r->inf mine}) does not agree with Eq.75~\cite{Leaver:1986a}; we believe that Eq.75~\cite{Leaver:1986a} is missing the first factor
on the right hand side of Eq.(\ref{eq:g r->inf mine}).

With the specific normalization choice of 
$\ant{0}= (-2i\bar\omega)^{+s+1-2i\bar\omega}$
the function $\g$ calculated using the Leaver-$U$ series satisfies the desired normalization Eq.(\ref{eq:g}); therefore, 
this will always be our choice (different from the choice $\an{0}=e^{-2i\bar\omega}$ above for the Jaff\'e series for $\f$)  when using the Leaver-$U$ series for $\g$.
We note that $\ant{n}$ themselves have a branch cut along the NIA
(this was already noted in Ref.18 of~\cite{Leung:2003ix}), as
we have
\begin{equation}\label{eq:Delta a}
\Delta \ant{n}=
\left[e^{-4\pi\ob}-1\right]
\antm{n}=
\left[e^{4\pi\ob}+1\right]\antp{n}
\end{equation}
where 
$\tilde a_{n\pm}\equiv \lim_{\epsilon\to 0^+}\ant{n}(\omega=\pm\epsilon-i\nu)$
and where
we have assumed $s\in \mathbb{Z}$.


\subsection{Series for $\Dgt$}\label{sec:series for Delta tilde g}



From Eqs.(\ref{eq:Leaver-Liu}) and (\ref{eq:Delta z^aU}) it follows that
\begin{align} \label{eq:Leaver-Liu series for Deltagt}
&
\Dgt(r,\nu)
\equiv \gpt(r,-i\nu)-\gmt(r,-i\nu)
=\bar r^{1+s}\left(\bar r-1\right)^{-\bar\nu}e^{\nu r}
\Dht(r,\nu),
\quad
\Dht(r,\nu)
= \sum_{n=0}^{\infty}
\Dhnt
\\ & \Dhnt\equiv
\frac{2\pi ie^{-2\nu r}e^{\pi i (s+1-2\bar\nu)}}{\Gamma(1-2\bar \nu)}
\antm{n}\cdot
\Tno{n},
\quad
\Tno{n}\equiv \frac{(-1)^n\Gamma(1+n-2\bar\nu)U(s-n+2\bar\nu,2s+1,2\nu r)}{\Gamma(1+s+n-2\bar\nu)\Gamma(1-s+n-2\bar\nu)}
\notag
\end{align}
\fixme{take out of the definition of $\Dhnt$ the factor which does not depend on $n$?}
This is a series for calculating $\Dgt$ by evaluating quantities directly on the NIA.
The principal branch is to be taken for
the confluent hypergeometric $U$-function in Eq.(\ref{eq:Leaver-Liu series for Deltagt}).


\subsection{Series for $\Dg$}

The series in this subsection, which we denote by $\Dg$, would correspond to $\Dgt$ if the coefficients 
$\ant{n}$
 did not have a branch cut; specifically, we may view $\Dg$ as the discontinuity of $\g$ across the NIA  if we replace $\ant{n}$ by $\an{n}$ in Eq.(\ref{eq:Leaver-Liu}).
Since that is not actually the case, we will not be using the series for $\Dg$ anywhere. However, we include it here
for completeness, as the factors $\Tnp{n}$ in the terms of this series 
satisfy the same recurrence relation (Eq.(\ref{eq:recurr rln Tn}) below) as the factors $\Tnm{n}$ and $\Tno{n}$ introduced above for $\g$ and 
$\Dgt$ respectively.
The solution $\Tnp{n}$ to the recurrence
relation Eq.(\ref{eq:recurr rln Tn}) is linearly independent from the solutions $\Tnm{n}$ and $\Tno{n}$.
If we replace $\ant{n}$ by $\an{n}$ in Eq.(\ref{eq:Leaver-Liu}), we can calculate the  discontinuity across the NIA of the resulting quantity as:
\begin{align} \label{eq:Delta h}
&
\Dg(r,\nu)
\equiv \gpt(r,-i\nu)\frac{\anp{n}}{\antp{n}}-\gmt(r,-i\nu)\frac{\anm{n}}{\antm{n}}
=
\bar r
^{1+s}\left(
\bar r
-1\right)^{-\bar\nu}e^{+\nu r}
\Dh(r,\nu),
\quad
\Dh(r,\nu)
= \sum_{n=0}^{\infty}
\Dhn
\\ & \Dhn\equiv
\frac{(-1)^{2s}2\pi i\an{n}}
{\Gamma (1+2s)\Gamma(1-2\bar\nu)}\Tnp{n},\quad
\Tnp{n}\equiv
\frac{\Gamma(1-2\bar\nu+n)M(1-2\bar\nu+n+s,2s+1,-2\nu r)}{\Gamma(1-2\bar\nu+n-s)}
\nonumber
\end{align}
\fixme{take out of the definition of $\Dhn$ the factor which does not depend on $n$?}
where we have used Eq.13.1.6 in~\cite{bk:AS}.
We note the appearance of the regular confluent hypergeometric (Kummer) function $M$ in 
(\ref{eq:Delta h}) for $\Dg$, as opposed to the irregular confluent hypergeometric function $U$ in Eq.(\ref{eq:Leaver-Liu series for Deltagt})
for $\Dgt$.


\subsection{Series for the radial derivatives}

An expression for calculating the $r_*$-derivative of the radial solution $\f$ follows straightforwardly from Eq.(\ref{eq:Jaffe}):
\begin{align} \label{eq:df/dr*}
&
\frac{d\f}{dr_*}=
\left(\bar r-1\right)^{1-\nb}
\bar r^{2\nb -1}e^{\nu r}\left[\frac{d\J}{dr}+\nu  \frac{(\bar r^2-2)}{\bar r(\bar r-1)}\J\right]
\\ &
\frac{d\J}{dr}=\frac{1}{r_h\bar r^2}\sum_{n=0}^{\infty}(n+1)\an{n+1}\left(1-\frac{1}{\bar r}\right)^n
\nonumber
\end{align}

In order to obtain an expression for the $r_*$-derivative of $\gt$ we use Eqs.4.22--4.24~\cite{Liu-1997}:
\begin{align} \label{eq:d tilde g/dr*}
&
\frac{d\gt}{dr_*}=\left(1-\frac{1}{\bar r}\right)\left[\left(\frac{1+s}{r}-\frac{\nb}{r-r_h}+\nu\right)\gt+\frac{\gt}{\htd}\frac{d\htd}{dr}\right]
\\ &
\frac{d\htd}{dr}=\sum_{n=0}^{\infty}\ant{n}\frac{d\Tnm{n}}{dr},\quad 
\frac{d\Tnm{n}}{dr}=\frac{s+1-2\nb+n}{r}\left[\frac{-2\nb+n+1-s}{-2\nb+n+1}\Tnm{n+1}-\Tnm{n}\right]
\nonumber
\end{align}


\section{Jaff\'e series coefficients} \label{sec:Jaffe}

Both the series coefficients $\an{n}(\omega)$ appearing in the Jaff\'e series Eq.(\ref{eq:Jaffe}) for $\f$ 
and the series coefficients $\ant{n}(\omega)$ appearing in the Leaver-$U$ series Eq.(\ref{eq:Leaver-Liu}) for $\g$
 satisfy the following 3-term recurrence relation,
\begin{equation} \label{eq:recurrence rln a_n}
\alpha_n \an{n+1}+\beta_n \an{n}+\gamma_n \an{n-1}=0,\quad n=1,2,\dots
\end{equation}
with $a_n=0$ for $n<0$ and where
\begin{align}
\alpha_n&\equiv (n+1)(n-2\bar{\nu}+1)  \notag \\ 
\beta_n&\equiv -[2n^2+(2-8\bar{\nu})n+8\bar{\nu}^2-4\bar{\nu}+\ell(\ell+1)+1-s^2]\\ 
\gamma_n&\equiv n^2-4\bar{\nu} n+4\bar{\nu}^2-s^2  \notag
\end{align}
We note that although in this section we use the notation $\an{n}$ to indicate a solution of Eq.(\ref{eq:recurrence rln a_n})
 the results in this section apply equally to the coefficients $\ant{n}$ since these results are independent of the specific choice of the
 $n=0$ coefficient.


\subsection{Singularities of $\an{n}$} \label{sec:sings of an}

From Eq.(\ref{eq:recurrence rln a_n}) it follows that, in principle, the coefficients $\an{n}$ will have a simple pole where $\alpha_{n-1}=0$, i.e., at $n-2\bar\nu=0$.
Therefore, if $\bar\nu=k/2$ for some $k\in \mathbb{N}$ then $a_n$ will have a simple pole $\forall n\ge k$ (see, e.g., App.B~\cite{Leung:2003ix}).
However, such a pole will not occur if at the same time it happens that $\beta_{k-1} \an{k-1}+\gamma_{k-1}\an{k-2}=0$.
This occurs \fixme{check/justify} for $s=2$ at the algebraically-special frequency $\ob_{AS}=-i\nb_{AS}$, where
$\nb_{AS}\equiv (\ell-1)\ell (\ell+1)(\ell+2)/6$~\cite{MaassenvandenBrink:2000ru}.
Therefore, the coefficients $\an{n}$ do not have a pole at $\nb=\nb_{AS}$ for $s=2$ while they do have a simple pole there for $s=0,1$.

 Suppose that $\left\{b_n\right\}$ and $\left\{c_n\right\}$ are two sets of solutions to a recurrence relation, then, 
 if $\lim_{n\to \infty} b_n/c_n=0$ it is said that $b_n$ are minimal and $c_n$ are dominant. 
 If the solution one seeks is dominant, then one can find the desired solution by solving the recurrence relation using standard forward recursion.
 However, if one wants to obtain a minimal solution, using forward recursion would be unstable and one must resort to finding the desired solution
 using, e.g., Miller's algorithm of backward recursion (see, e.g.,~\cite{Gautschi1961}).
In order to investigate whether the solutions 
to the recurrence relation Eq.(\ref{eq:recurrence rln a_n}) are minimal, dominant or neither, we require the large-$n$ behaviour of the coefficients $\an{n}$.
We also require the 
large-$n$ behaviour of $\an{n}$
in order to study the convergence properties of 
any series involving these coefficients.

\subsection{Large-$n$ asymptotics}

In order to obtain the large-$n$ asymptotics of the coefficients $a_n$ 
we follow
App.B~\cite{Wimp}.
We thus express the asymptotic behaviour as the so-called {\it Birkhoff series}
\begin{equation}
\an{n} = e^{\mu_0n\ln n+\mu_1 n}n^{\thB}e^{\sum_{\sumB=0}^{\imax}\Wimpind_{\sumB+1}n^{\beta-\sumB/\rho}+O\left(n^{\beta-(\imax+1)/\rho}\right)}
\end{equation}
for a certain chosen value of $\imax\in\mathbb{N}$, where $\beta\in [0,1)$, $\Wimpind_1\neq 0$, $\rho\in \mathbb{Z}$, $\rho\ge 1$
and 
$\mu_{\sumB}, \Wimpind_{\sumB} \in\mathbb{C}$ 
 for all $\sumB$.
Substituting this expression into the recurrence relation (\ref{eq:recurrence rln a_n}) we obtain
\begin{align}
\label{eq:a_n large-n Wimp}
&
\an{n} = n^{-\bar\nu-3/4}e^{\pm 2\sqrt{2\bar\nu n}i+\sum_{\sumB=1}^{\imax}\Wimpind_{\sumB+1}n^{1/2-\sumB/2}+O\left(n^{1/2-(\imax+1)/2}\right)}
\\ &
\Wimpind_3=\frac{i (-9 - 48 \ell - 48 \ell^2 - 48 \bar\nu +  64 \bar\nu^2)}{48 \sqrt{2} \sqrt{\bar\nu}}
\nonumber
\\& 
\Wimpind_4= \frac{3 + 16 \ell + 16 \ell^2 - 48 \bar\nu + 64 s^2 \bar\nu + 128 \bar\nu^2 +  128 \bar\nu^3}{128 \bar\nu}
\nonumber
\\ &
\Wimpind_5=
\nonumber
\\ &
\frac{i}{30720 \sqrt{2} \nb^{3/2}}
\left[315-1280 \ell^4-2560 \ell^3+160 \ell^2+1440 \ell+\left(7680 \ell^2+7680 \ell+13728\right) \nb+
\right.\nonumber \\& \left.
\left(-30720 \ell^2-30720 \ell-3200\right) \nb ^2 -10240 \nb ^3\right]
   \nonumber
\\ &
\Wimpind_6=
\nonumber
\\ &
 \frac{1}{24576 \nb   ^2}  
\left[-81+768 \ell^4+1536 \ell^3+480 \ell^2-288 \ell+\left(-1536 \ell^2-1536 \ell-288\right) \nb+\left(6144 \ell^2+6144 \ell+1152\right) \nb ^2 +
\right.\nonumber \\& \left.
\left(24576 s^2-16384\right)\nb ^3 +24576 \nb ^4+16384 \nb ^5\right]
   \nonumber   
\end{align}
The coefficients $\Wimpind_{\sumB}$ are real for $\sumB$ even and they are purely imaginary for $\sumB$ odd \fixme{is this just true for $\bar\nu\in\mathbb{R}$?}.
We note that the spin dependence does not appear until the term $\Wimpind_6$.
The coefficient $\Wimpind_2$ corresponds to an undetermined overall normalization and
the `$\pm$' sign corresponds to the two linearly independent solutions of the recurrence relation.
Since the recurrence relation (\ref{eq:recurrence rln a_n}) is unchanged under $n\to ne^{2\pi i}$, one solution can be obtained
from the other under this change; this is essentially 
equivalent to changing the sign of $\Wimpind_{\sumB+1}$ for $\sumB$ even in (\ref{eq:a_n large-n Wimp}).
 On the NIA, where $\bar\nu>0$, the two solutions behave similarly
 (that is, no solution is  dominant over the other) and an appropriate linear combination of them should be taken.
Off the NIA, if $\omega$ is not a QNM frequency then the $a_n$ are dominant~\cite{Leaver:1986}
 and they are generated by forward recursion; 
whereas if $\omega$ is a QNM frequency then the $a_n$ are minimal and they can be generated by Miller's algorithm of backward recursion.
Indeed, requiring for the solutions $a_n$ to be minimal has become a widely used, successful method for finding QNM frequencies of 
black holes~\cite{Leaver:1985}.

Finally, we note that the leading order of Eq.(\ref{eq:a_n large-n Wimp}) differs from Eq.46~\cite{Leaver:1986a}
 in having a power of $n$ equal to `$-\bar\nu-3/4$' instead of `$-2\bar\nu-3/4$'; we have checked numerically
 for specific values of the parameters (both for $\omega$ on and off the NIA) that Eq.(\ref{eq:a_n large-n Wimp}) gives the correct asymptotic behaviour.

\subsection{Plots}

\begin{figure}[h!]
\begin{center}
\includegraphics[width=8cm]{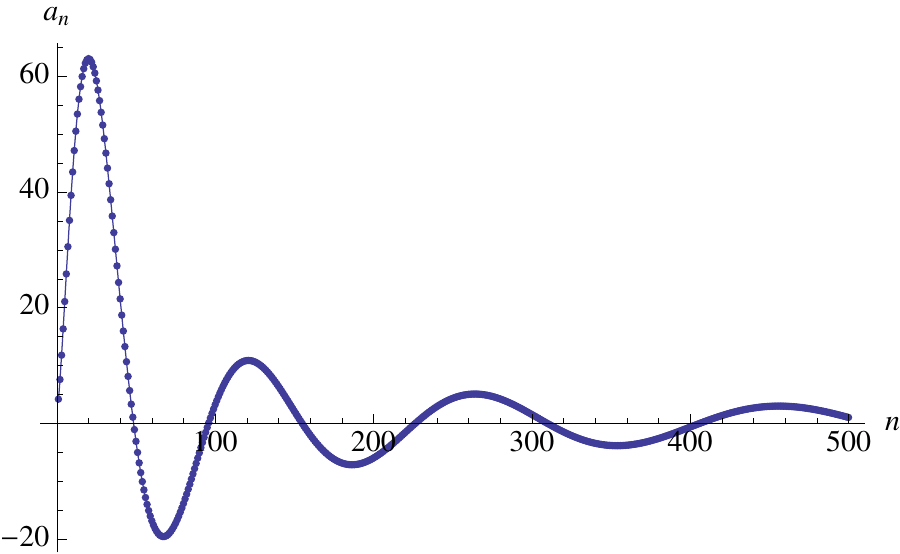} 
\includegraphics[width=8cm]{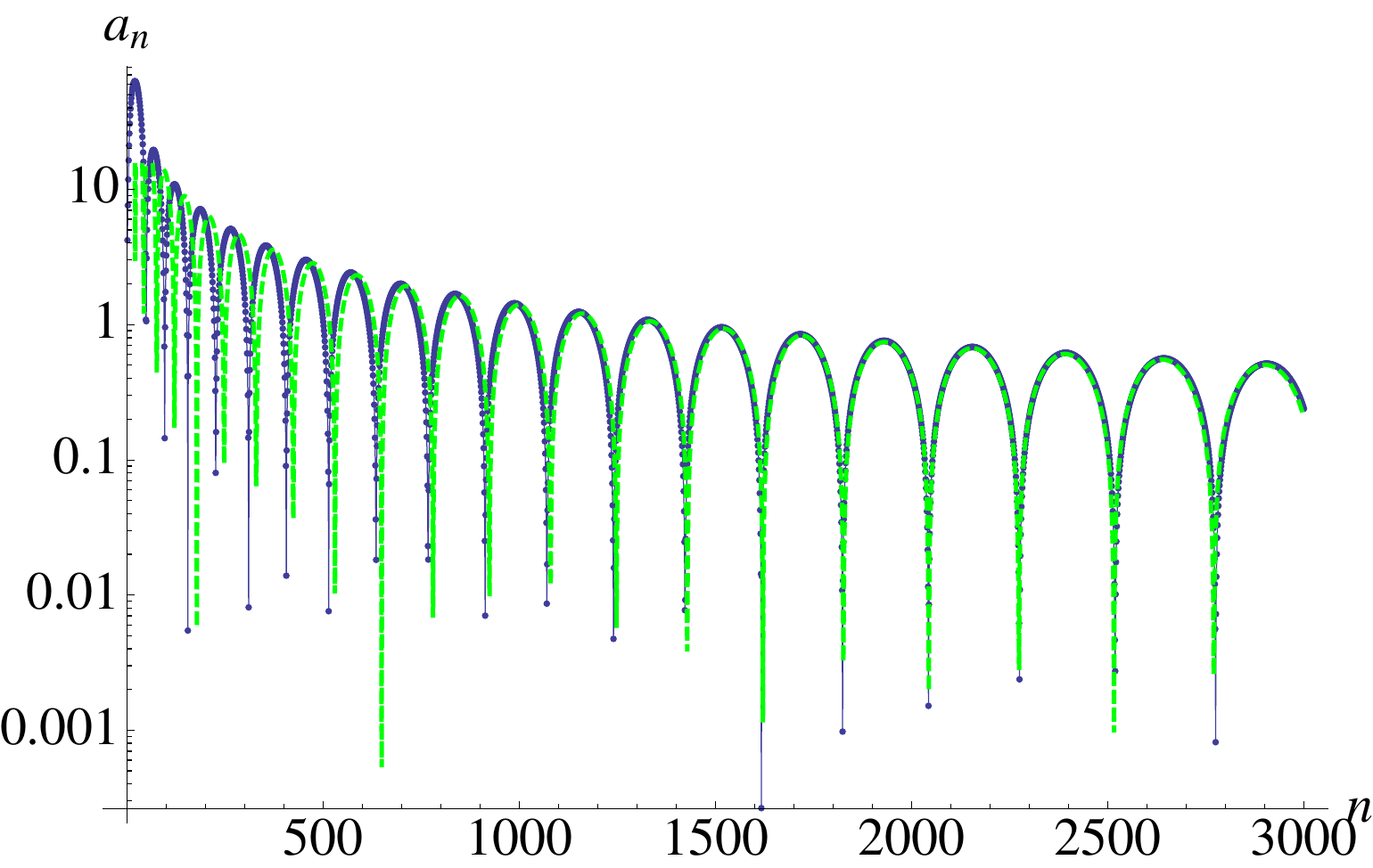} 
\includegraphics[width=8cm]{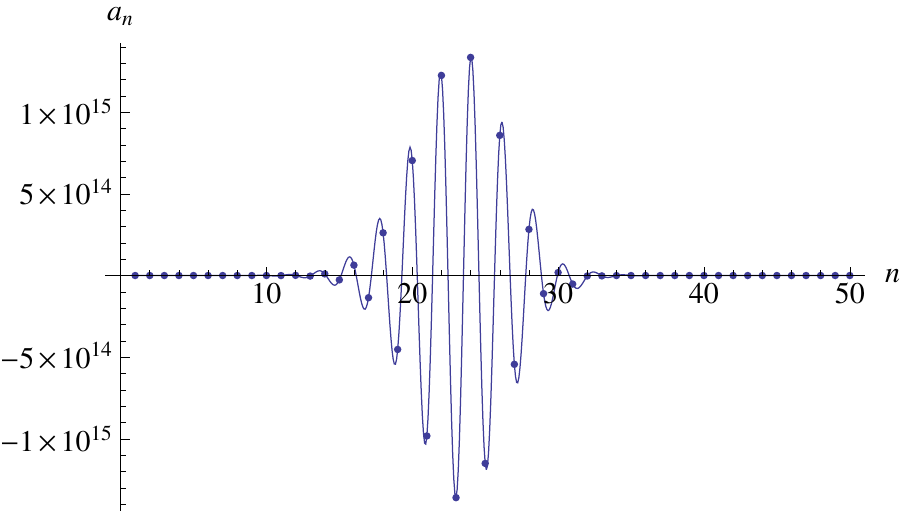} 
\includegraphics[width=8cm]{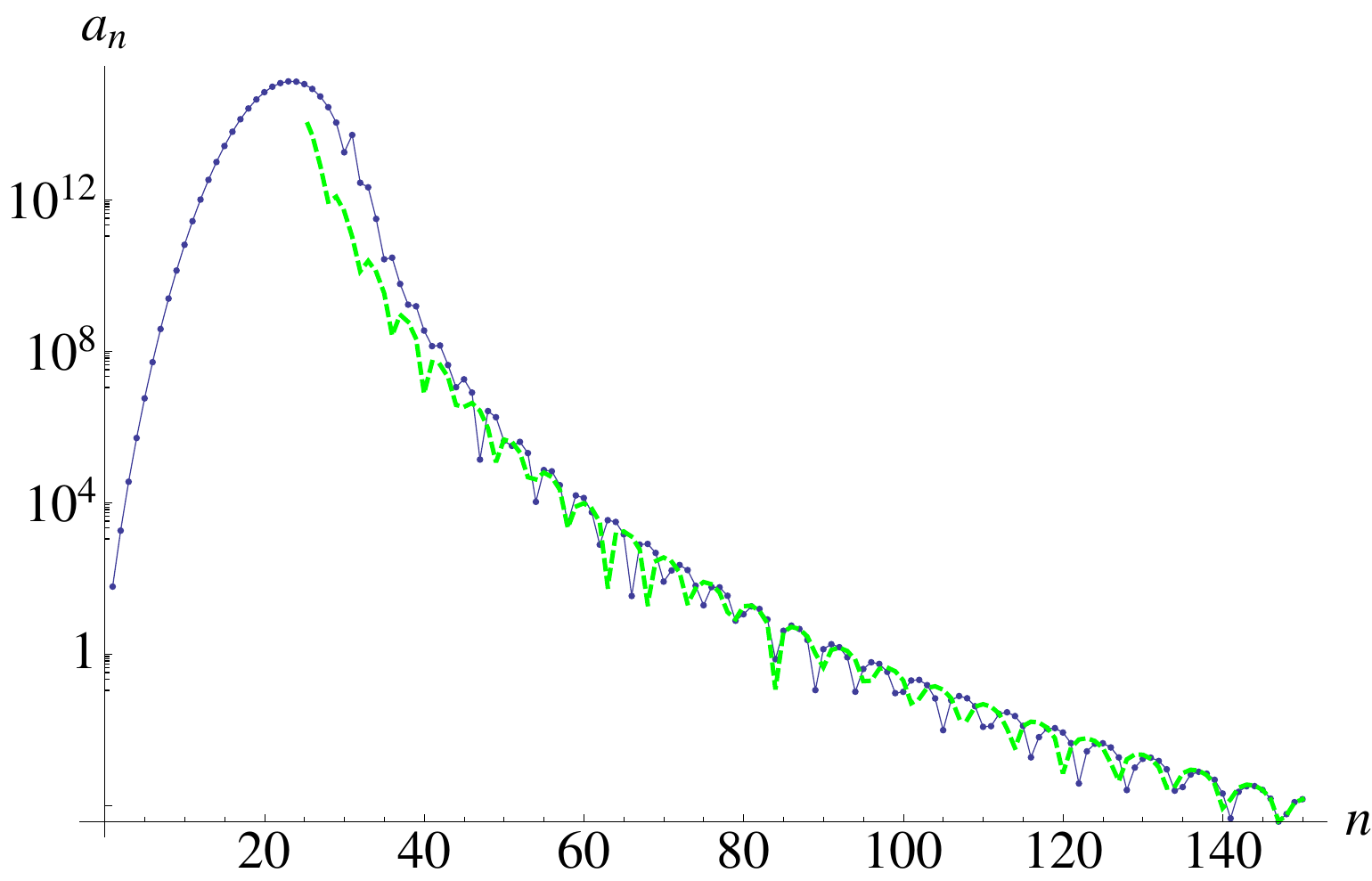} 
\end{center}
\caption{
Jaff\'e coefficient $\an{n}$ as a function of $n$ for $s=2$, $\ell=2$. Figs.(a) and (b) are for $\nb=0.2$ and figs.(c) and (d)  for $\nb=15.4$.
Figs. (b) and (d) are log-plot versions of (a) and (c) respectively.
Blue dots: exact solution $\an{n}$ to the recurrence relation (\ref{eq:recurrence rln a_n}) with $\an{0}=1$; the continuous blue curve is an interpolation of the blue dots.
Dashed green curve: large-$n$ asymptotics Eq.(\ref{eq:a_n large-n Wimp}) where we have taken a linear combination of the two linearly independent asymptotic solutions
such that the linear combination matches the exact value of $a_n$ at both  $n=1000$ and $n=2000$ for $\nb=0.2$, and at $n=145$ and $n=150$ for $\nb=15.4$.
We note the `pulse' centered around $n=25$ in the case  $\nb=15.4$ (the equivalent `pulse' in the case $\nb=0.2$ is centered around its first peak at $n\approx 20$); 
the coefficient $\an{n}$ reaches its maximum magnitude at the `pulse' and then the magnitude decays rapidly with $n$.
}
\label{fig:a_n asympts w=-0.1001i}
\end{figure}

\begin{figure}[h!]
\begin{center}
\includegraphics[width=16cm]{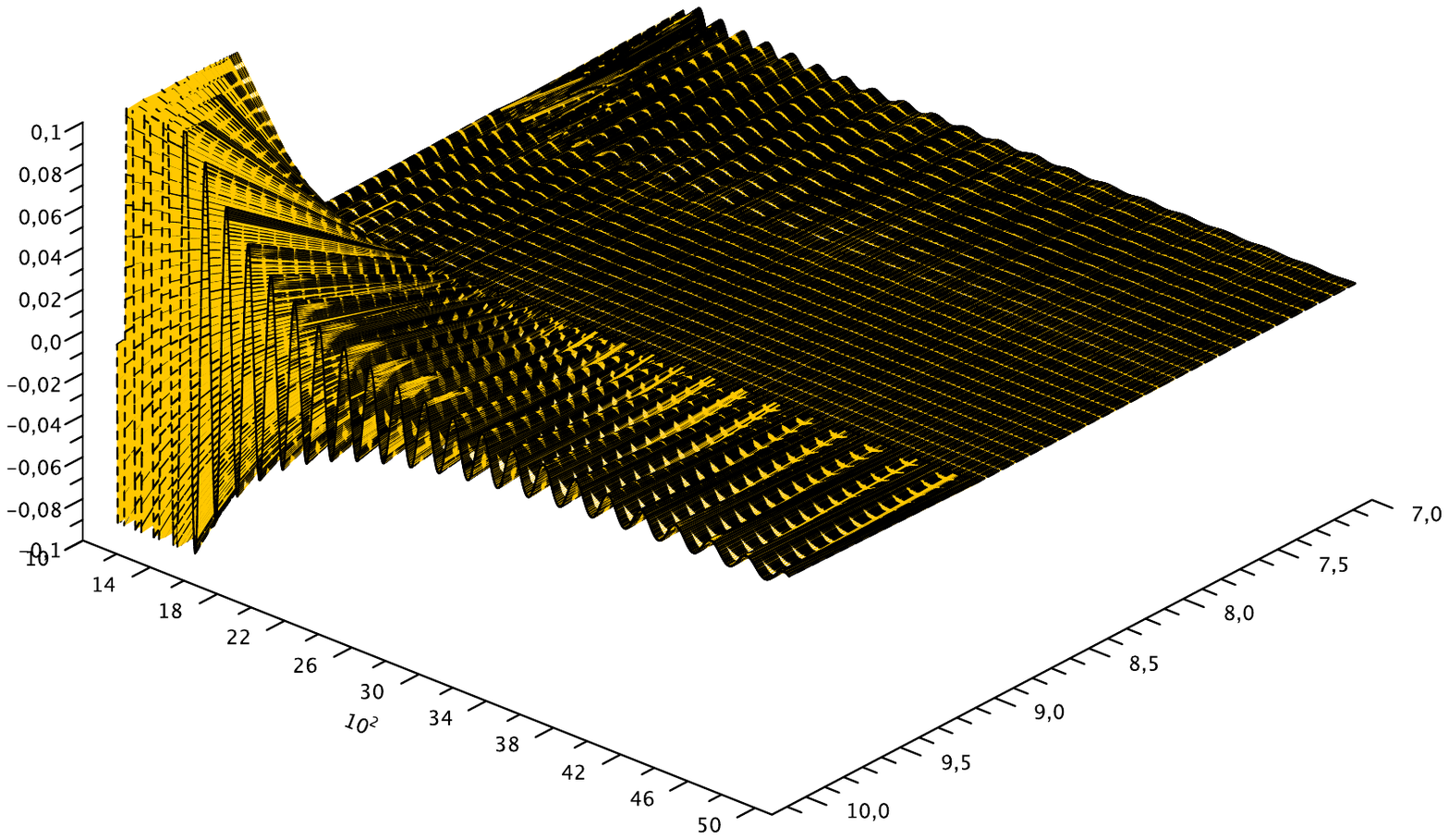} 
\end{center}
\caption{Exact solution $a_n$ to the recurrence relations with $s=0$, $\ell=2$ as a function of both $M\nu:7.1001\to 10.1001$ and $n=1000\to 5000$.
\fixme{This plot should be checked/redone}\fixme{What value of $\an{0}$ was used? Try log-plot? Take out leading order for large-$n$? better for $\ell=2$?}
}
\label{fig:a_n asympts w=-7.1001i->-10.1001i}
\end{figure}

In Fig.\ref{fig:a_n asympts w=-0.1001i}
we show that the large-$n$ asymptotics given in Eq.(\ref{eq:a_n large-n Wimp})
match
the exact solution to the
recurrence relation (\ref{eq:recurrence rln a_n}).
We note the appearance of a `pulse',
after which the values of $\an{n}$ decay rapidly.
Fig.\ref{fig:a_n asympts w=-7.1001i->-10.1001i} is a 3D-plot of $a_n$ as a function of both $n$ and $M\nu$.
We have only included plots for $s=2$ as representative of the behaviour of the coefficients $a_n$, as the behaviour is similar for other spins.
The behaviour is also similar at the algebraically-special frequency $\omega_{AS}$.
At the poles described in Sec.\ref{sec:sings of an} the behaviour
of `$\sin(2\pi\nb)\an{n}$' is also similar, except that the first $n<2\nb$ terms are exactly zero. 

\section{Calculation of $\f(r,\omega)$} \label{sec:calc f}

We calculate the radial solution $\f$ using the Jaff\'e series Eq.(\ref{eq:Jaffe}).
As shown by Leaver in Sec.IV.A~\cite{Leaver:1986a}, the Jaff\'e series is absolutely convergent $\forall \omega\in\mathbb{C}$ and for 
any $r\in [r_h,\infty)$, since then: $\lim_{n\to\infty}\left|a_{n+1}(1-1/\bar r)^{n+1}/(a_{n}(1-1/\bar r)^{n})\right|=|1-1/\bar r|<1$.
By the same argument the Jaff\'e series is uniformly convergent on $r\in [r_h,r_{max}]$ for any finite $r_{max}$ but will generally not be so
at radial infinity, provided the coefficients are not singular (see below). However, as shown by Leaver, the Jaff\'e series is uniformly convergent -- including radial infinity --
 if $\sum_n a_{n}$ is finite, which is guaranteed if the sequence
$\{a_{n}\}$ is minimal 
and this occurs
at the QNM frequencies.
At these frequencies it is $\Ain=0$ and $\Aout=e^{-2i\ob}\sum_n\an{n}$.

As shown in Sec.~\ref{sec:sings of an}, $\an{n}$ have simple poles $\forall n\ge k$ when $\nb=k/2$ for some $k\in\mathbb{N}$.
 The exception is the case $\nu=\nu_{AS}\equiv i\omega_{AS}$ for $s=2$, which is not a pole. These poles carry over to $\f$ so that this radial solution
 has  simple poles at $\nb=k/2$ (these poles of $\f$ were shown in~\cite{Jensen:1985in,Jensen:1985in:err1} 
using a different method, namely, a Born series), except at $\nb_{AS}$ when $s=2$.
However, the BC modes 
$\Delta G_{\ell}$ are independent of the normalization of $\f$, and so it is useful to define
\begin{equation} \label{eq:anb}
\anb{0}\equiv -\an{0}\sin(2\pi i\ob)
\end{equation}
with $\an{0}=e^{-2i\bar\omega}$.
We denote the corresponding quantities $\an{n}$, $\Jn$, $\J$, $\f$ and $W$ obtained using this normalization by $\anb{n}$, $\Jnb{n}$, $\Jb$, $\fb$ and $\hat W$ respectively.
We note that at the pole $\nb=k/2$, the first nonzero value of $\anb{n}$ will be for $n=k$.
Therefore, at $\nb=k/2$ it is $\fb\sim e^{+i\omega r_*}$ as $r_*\to -\infty$ 
and so
 $\fb(r,\omega)\propto \fb(r,-\omega)$~\cite{MaassenvandenBrink:2000ru}.
In the particular case of the algebraically special frequency $\nb_{AS}$, exact solutions to the radial equation have been found~\cite{citeulike:11195363}.\fixme{are these exact solutions for $\f$, $\g$ or both?}

Therefore, as a function of $\omega\in \mathbb{C}$,  the radial solution $\f$ only has singularities at the simple poles $\ob=-ik/2$, 
$k\in\mathbb{N}$, on the NIA  (except at $\nb_{AS}$ for $s=2$)
while $\fb$ is analytic in the whole frequency plane.

In Fig.\ref{fig:ln(f),l=2,s=0} we illustrate, for frequencies on the positive-imaginary axis (PIA) of the complex-frequency plane, 
the convergence properties of the Jaff\'e series and we plot $\f$
(there is no need to calculate $\fb$ on the PIA since $\f$ has no poles there) as a function of $|\nb|$.
In Fig.\ref{fig:ln(f),l=2,s=0 NIA} we do similarly but for $\fb$ on the NIA instead of $\f$ on the PIA.
In this case we do not plot the partial term $\Jnb{n}$ since the behaviour is essentially the same as that of $\an{n}$ in Fig.\ref{fig:a_n asympts w=-0.1001i}(d).
The radial derivative of $\fb$ as a function of the frequency has a similar behaviour to that of $\fb$.
In Fig.\ref{fig:ln(f),l=2,s=0 wrt r} we plot, on the NIA, $\ln|\fb|$ and $\ln|d\fb/dr_*|$ as  functions of the radius:
for some values of $\nb$
the solution $\fb$ has a zero and for other values of $\nb$ it does not. 
We note that in Figs.\ref{fig:ln(f),l=2,s=0}--\ref{fig:ln(f),l=2,s=0 wrt r} we only include plots for $s=2$ as the behaviour for other spins is very similar.
In~\cite{Casals:Ottewill:2011smallBC} we show that the Jaff\'e series for $\fb$ agrees well with a small-$\nb$ series expansion.

\begin{figure}[h!]
 \includegraphics[width=8cm]{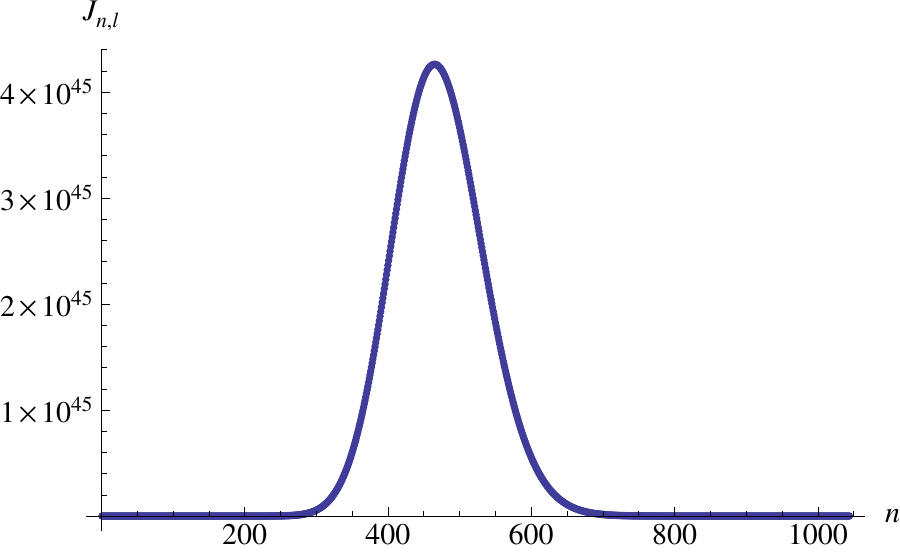}
  \includegraphics[width=8cm]{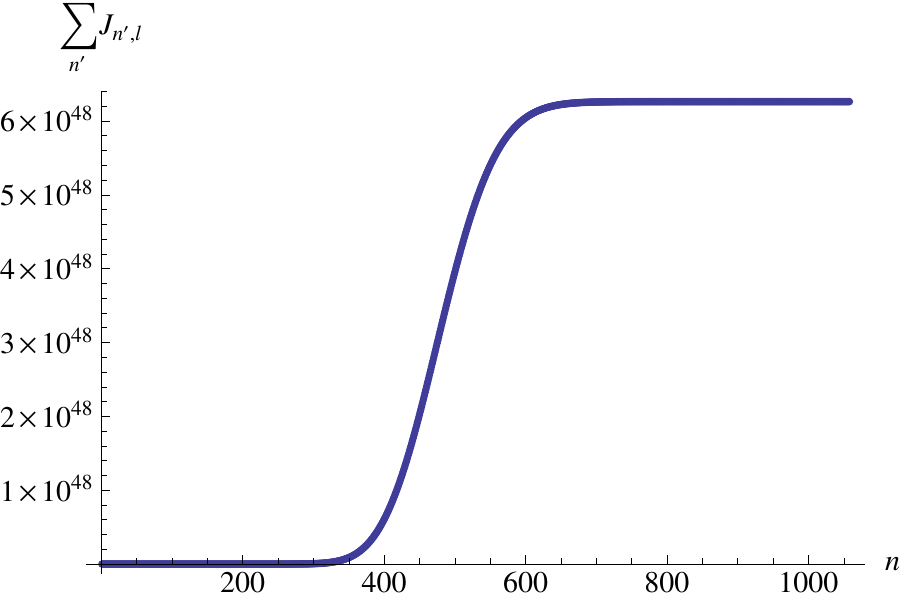}
   \includegraphics[width=8cm]{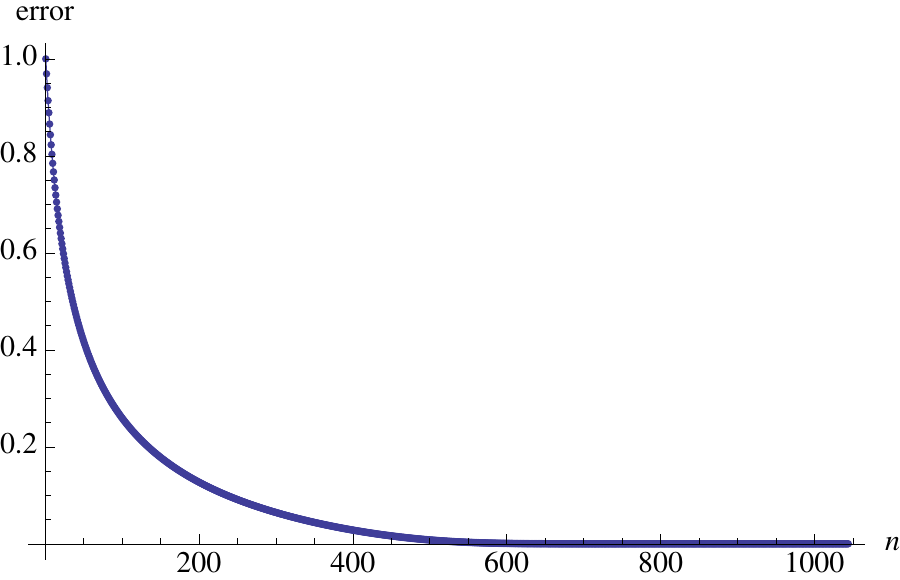}
      \includegraphics[width=8cm]{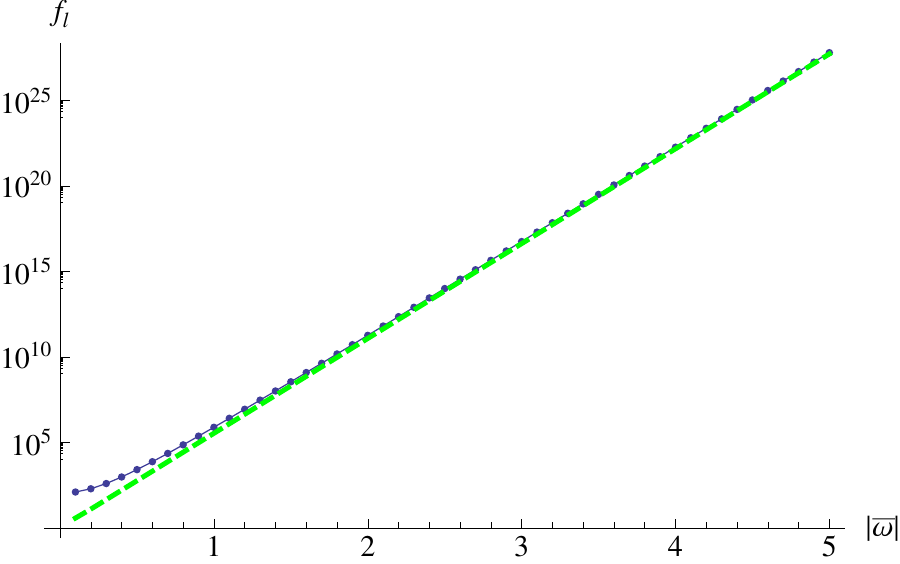}
\caption{
Construction of $\f(r,\omega)$ on the PIA using Eq.(\ref{eq:Jaffe}) for $s=2$, $\ell=2$, $r=10M$.
Plots (a)--(c) are with $\an{0}=1$, for the value $\bar\omega=9.8002i$ and with $n$ on the x-axis.
(a) Partial term $\Jn$.
(b) Partial sum $\sum_{n'=0}^{n}J_{n',\ell}$.
(c) Partial error $\left|\Jn/\sum_{n'=0}^{n}J_{n',\ell}\right|$.
(d) Log-plot of
$\f(r,\omega)$ using Eq.(\ref{eq:Jaffe})
(blue dots interpolated by the continuous blue curve) as a function of $|\ob|$
and log-plot of $\left(e^{i\omega r_*}+e^{-i\omega r_*}\right)$
(dashed green curve),
which corresponds to the asymptotics
of Eq.(\ref{eq:f,large-r}) ignoring the incidence and reflection coefficients.
}
\label{fig:ln(f),l=2,s=0}
\end{figure}

\begin{figure}[h!]
   \includegraphics[width=8cm]{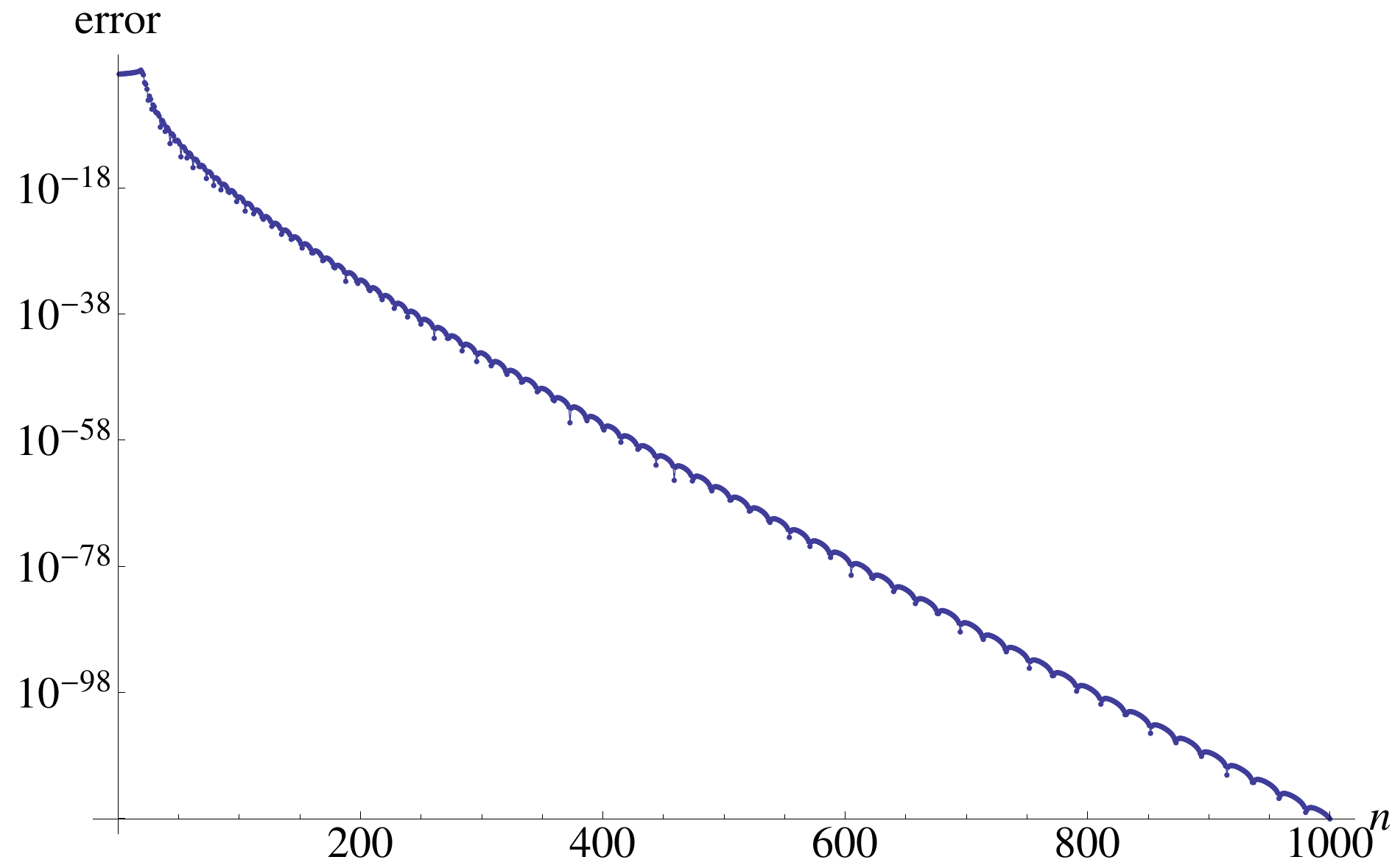}
      \includegraphics[width=8cm]{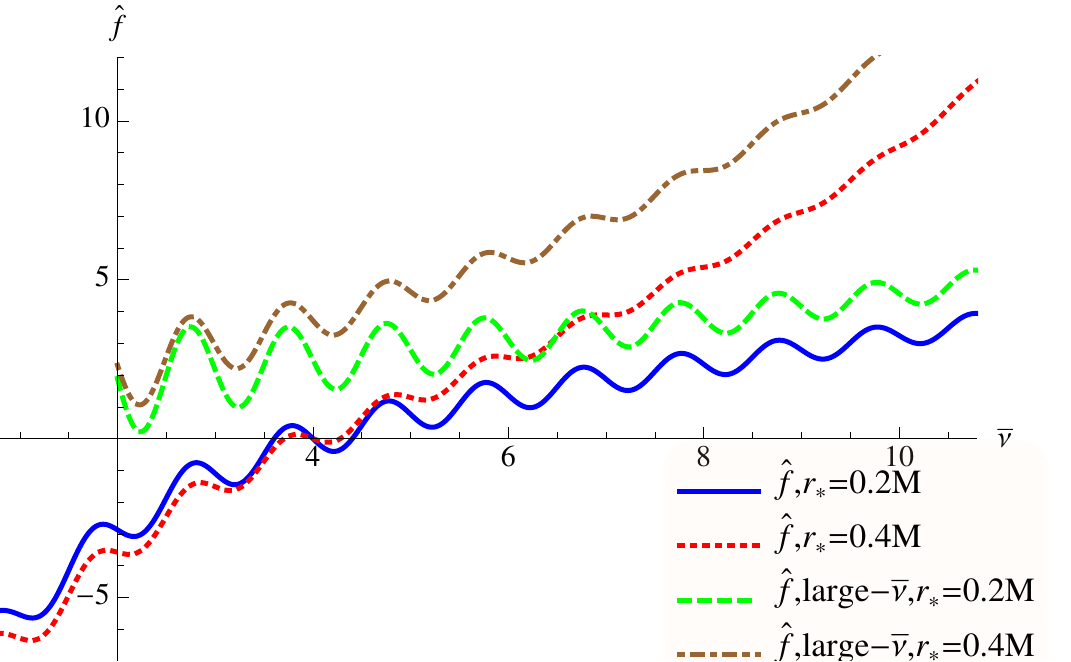}
\caption{
Construction of $\fb$ on the NIA using Eq.(\ref{eq:Jaffe}) for $s=2$, $\ell=2$.
(a) Partial error $\left| \Jnb{n}/\sum_{n'=0}^{n}\Jnb{n'}\right|$ as a function of $n$ with $\an{0}=1$ for $\nb=9.8002$ and $r=10M$.
(b) $\fb$ from Eq.(\ref{eq:Jaffe}) as a function of $\nb$.
The continuous blue and dotted red curves are calculated using the Jaff\'e series Eq.(\ref{eq:Jaffe}) for $r_*=0. 2M$ and $r_*=0.4M$ respectively.
Note that they are both zero at $\nb=\nb_{AS}$ because of the fact that, for $s=2$,  $\nb_{AS}$ is not a pole of $\fb$.
The dashed green and dot-dashed brown curves are the large-$\nb$ asymptotics in~\cite{Casals:2011aa} for $r_*=0. 2M$ and $r_*=0.4M$ respectively.
\fixme{Why do the curves start at different values of $\nb$?}
}
\label{fig:ln(f),l=2,s=0 NIA}
\end{figure} 

\begin{figure}[h!]
 \includegraphics[width=8cm]{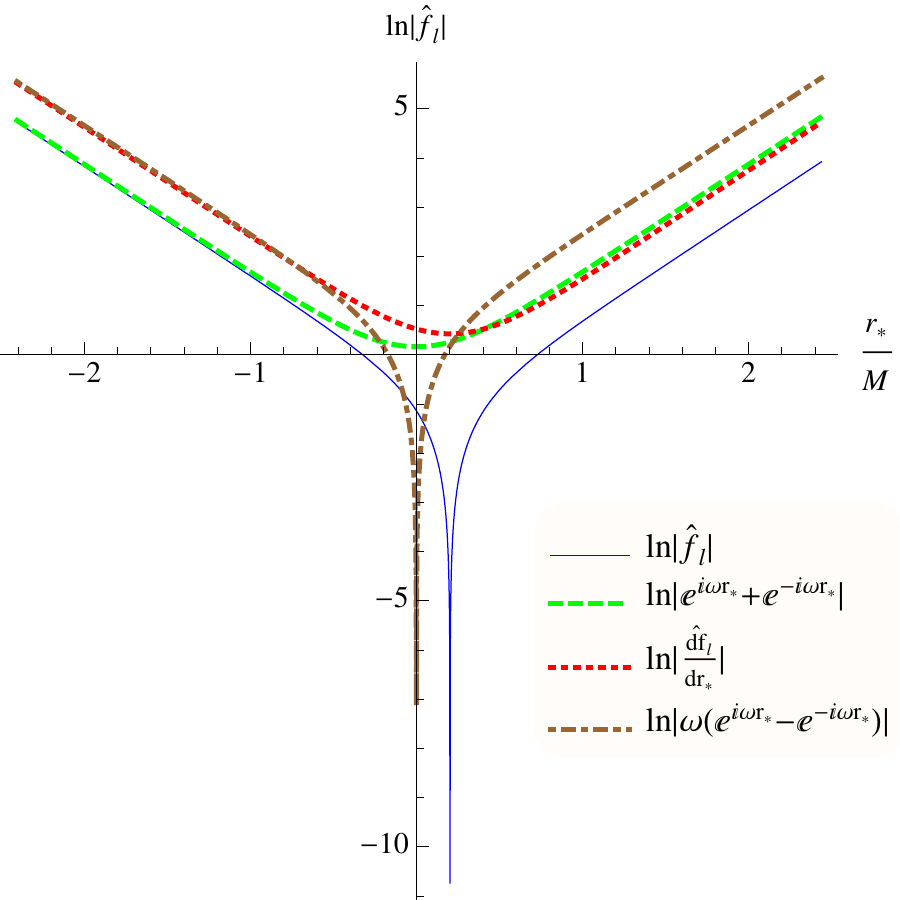}            
\includegraphics[width=8cm]{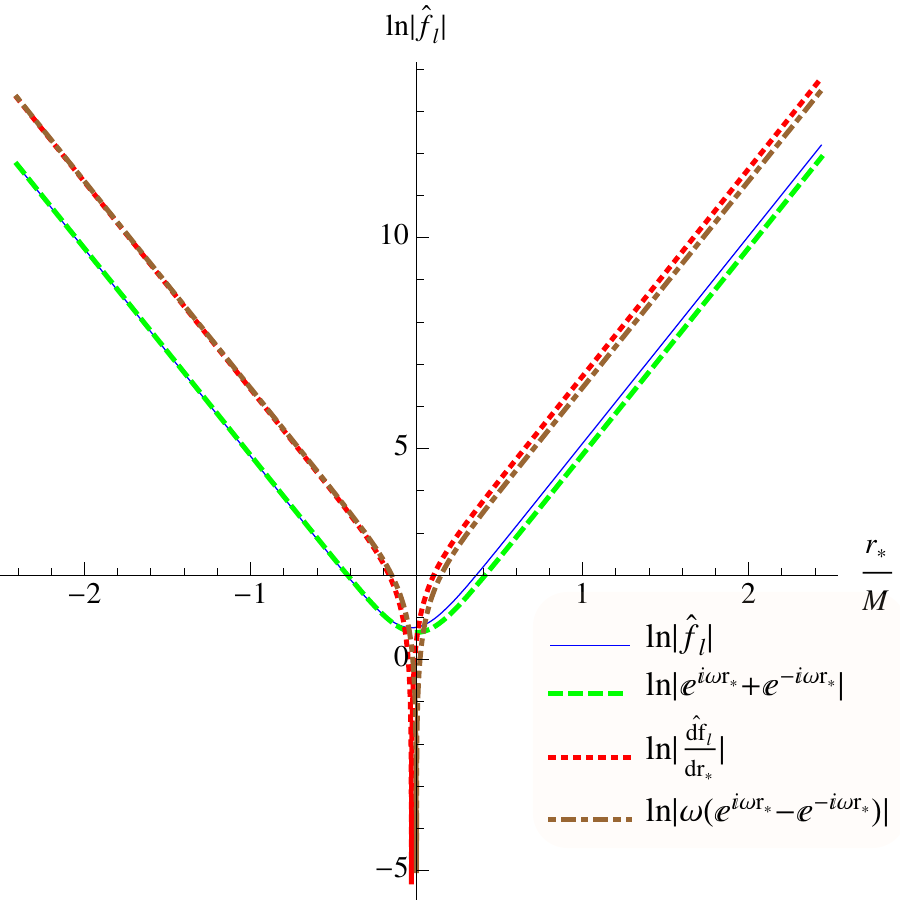}
\caption{
Plots of $\ln|\fb|$ as a function of $r_*/M$ for $s=2$, $\ell=2$ and (a) $\nb=4.4002$, (b) $\nb=9.8002$.
Continuous blue curve: $\ln|\fb|$ obtained using Eq.(\ref{eq:Jaffe}).
Dashed green curve: $\ln\left|e^{i\omega r_*}+e^{-i\omega r_*}\right|$.
Dotted red curve: $\ln|d\fb/dr_*|$ obtained using Eq.(\ref{eq:df/dr*}).
Dot-dashed brown curve:  $\ln\left|\omega (e^{i\omega r_*}+e^{-i\omega r_*})\right|$.
The curves for $\ln\left|e^{i\omega r_*}+e^{-i\omega r_*}\right|$ and its radial derivative only serve as very crude approximations
to the large-$r$ asymptotics of $\fb$ in Eq.(\ref{eq:f,near hor}).
}
\label{fig:ln(f),l=2,s=0 wrt r}
\end{figure}




\section{Calculation of $\gt$ and  the BC `strength'}


\subsection{Recurrence relation}

The terms in the series Eq.(\ref{eq:Leaver-Liu}) for  $\g$,  Eq.(\ref{eq:Leaver-Liu series for Deltagt}) for $\Dgt$ 
and Eq.(\ref{eq:Delta h}) for $\Dg$ all consist on the coefficient $a_n$ times a factor, which we denote by $\Tnm{n}$, $\Tno{n}$
and $\Tnp{n}$, respectively (times a factor independent of $n$).
All three factors, $\Tnm{n}$, $\Tno{n}$ and $\Tnp{n}$, satisfy
 the same, following recurrence relation~\cite{Temme:1983,bk:onlineAS,Liu-1997}:
 \begin{equation} \label{eq:recurr rln Tn}
(n+1-2\nb)(n-2\nb)
T_{n-1}
-(n+1-2\nb)(2n+1-4\nb-2\nu r)
T_n
+(n+1-2\nb+s)(n+1-2\nb-s)
T_{n+1}
=0
\end{equation}
where $T_n$ denotes any of $\Tnm{n}$, $\Tno{n}$ and $\Tnp{n}$.

In the subsections below we will show the following.
On the NIA, all three factors, $\Tnm{n}$, $\Tno{n}$ and $\Tnp{n}$, have the same  leading-order behaviour $O(n^{-1/4}e^{-2i\sqrt{2\nu r n}})$ 
for large-$n$
--  see, respectively, Eqs.(\ref{eq:Tnm large-n}), (\ref{eq:Tno large-n}) and (\ref{eq:large-n M}) below.
\fixme{The following assumes that the large-$n$ asymptotics for $\Tno{n}$ and $\Tnp{n}$ off the NIA are of the same order as those for $\Tnm{n}$ 
in Eq.(\ref{eq:Tnm large-n}) - to be checked}
The solution $\Tnm{n}$ and either the solution $\Tno{n}$ or $\Tnp{n}$ of the recurrence relation Eq.(\ref{eq:recurr rln Tn}) are linearly independent \fixme{check}.
We show that, when solving the recurrence relation (\ref{eq:recurr rln Tn}) for $\omega$ anywhere except on the NIA,
$\Tnm{n}$ is a subdominant solution and $\Tno{n}$ and $\Tnp{n}$ are dominant solutions. 
In this case of $\omega\notin \text{NIA}$, if one wishes to find $\Tnm{n}$, 
one expects that forward recurrence will be unstable in that the dominant solution will `creep in' as $n$ is increased.
One should then use instead Miller's algorithm of backward recursion~\cite{Liu-1997}.
On the other hand, when solving the recurrence relation (\ref{eq:recurr rln Tn}) {\it on} the NIA
there are no dominant nor subdominant solutions, 
all solutions asymptoting like $O(n^{-1/4})$. 
In this case of $\omega\in \text{NIA}$, there is no danger
of a dominant solution `creeping in' and so there is no need for using Miller's algorithm of backward recursion for finding 
any of the three solutions.


\subsection{Calculation of $\gt$}

We need a method for calculating the radial solution $\gt$ on the 
PIA,
as required by $q(\nu)$ in Eq.(\ref{eq:dg=qg}), as well as
on the NIA, as required by the Wronskian Eq.(\ref{eq:Wronskian}).
We will calculate $\gt$ directly \textit{on} the NIA, as well as on the PIA, using the Leaver-$U$ series Eq.(\ref{eq:Leaver-Liu}).
As mentioned in Sec.~\ref{sec:series for g}, the Leaver-$U$ series allows us to calculate $\g$ directly on the NIA, specifically as the limit
from the third quadrant, i.e.,  $\gpt$.
The advantage of evaluating  $\gt$ {\it on} the NIA is twofold. First, no extrapolating procedure $\epsilon\to 0^+$ onto the NIA is then needed.
Secondly, on the NIA there are no dominant/subdominant solutions to the recurrence relation (\ref{eq:recurr rln Tn}) and so there is no need for 
Miller's algorithm of backward recursion.

The confluent hypergeometric $U$-functions, however, are notoriously difficult to evaluate.
We have three options in order to calculate the factors $\Tnm{n}$ in the Leaver-$U$ series:
(1) from their definition (\ref{eq:Leaver-Liu}) and using the in-built $U$-function in Mathematica,
(2) from their definition (\ref{eq:Leaver-Liu}) and using the integral representation  Eq.(\ref{eq:U num int}) for the  $U$-function, and
(3) from the above recurrence relation Eq.(\ref{eq:recurr rln Tn}).
We note that method (1) is highly unstable, whereas it is a lot more stable to calculate $\Tnm{n}$ using method (2)
(see Eq.4.16~\cite{Liu-1997}, App.B~\cite{Leung:1999iq},~\cite{Temme:1983}).



\subsubsection{Large-$n$ asymptotics} \label{sec:g large-n}

We obtain the large-$n$ behaviour of the terms in the Leaver-$U$ series Eq.(\ref{eq:Leaver-Liu}) in order to investigage its convergence properties.
From Eq.(\ref{eq:Leaver-Liu}) and (\ref{eq:large-a U}), we have
\begin{align}\label{eq:Tnm large-n}
&
\Tnm{n}\sim
\frac{e^{-\nu r}
}{\Gamma(1-2\nb)
(2|\nu| r)^s
}
\sqrt{\frac{\pi}{\sqrt{2|\nu| r}}}e^{-i\chi (s+1/4)}n^{-1/4}e^{-2\sqrt{2|\nu| r n}e^{i\chi /2}}, 
\quad n\to \infty, \quad 
s\le 0,
\ \chi\equiv \arg(-\nu)\in (-\pi,+\pi],\ r>0
\end{align}
\fixme{check Eq.(\ref{eq:Tnm large-n}) is valid $\forall \chi$ and for $s>0$}
agreeing with Eq.4.19~\cite{Liu-1997}.
It follows from Eq.(\ref{eq:Tnm large-n}) that  the series for $\hnt$ is absolutely convergent everywhere except, maybe, on the NIA.
\fixme{Removed the following statement because maybe for such a statement we need to show `uniform convergence'
instead of `absolute convergence':
``Taking also into account that the only singularities of the coefficients $\an{n}$ consist on poles on the NIA, it follows that  $\g$ is analytic everywhere
on the complex-frequency plane except on the NIA."}

The convergence properties of the series
$\h=\sum_n\hn$ 
on the NIA (which would yield $\gp$)  is the same as that for $\Dht=\sum_n \Dhnt$ in Sec.\ref{sec:large-n Dhnt} 
and as that for $\Dh=\sum_n\Dhn$ in Sec.\ref{sec:large-n Dhn}, and so we refer the reader to this latter section.
\fixme{Say convergence properties of the series for $\hnt$ on the PIA, as that is not said in Sec.\ref{sec:large-n Dhn}}


\subsubsection{Results} \label{sec:results gt}

In Fig.\ref{fig:tilde g,r=5} we plot $\gpt$ as a function of the frequency on the NIA. 
Since $\gt\to e^{\nu r}\in\mathbb{R}$ for large $r$ and the asymptotic series for $\gt$ for large $r$ contains only real coefficients (e.g., Sec.B.1~\cite{Liu-1997}),
one would expect that $|\text{Im}(\gpt)|\ll|\text{Re}(\gpt)|$, especially as $\nb$ increases - this is indeed what happens in Fig.\ref{fig:tilde g,r=5}.
Fig.\ref{fig:tilde g,r=5} shows that $\text{Im}(\gpt)$ becomes round-off error at some stage \fixme{I don't understand why it's round-off error about a value
that increases -exponentially- with $\nb$?}.
In Fig.2~\cite{Casals:2011aa} we show similar plots for $s=0$.

When calculating the terms in the series Eq.(\ref{eq:Leaver-Liu}) for $\gpt(r,-i\nu)$ in practise using Mathematica, the first two terms $\Tnm{n=0}$ and $\Tnm{n=1}$
on the NIA we calculate using 
Eq.(\ref{eq:U num int}) and a particular splitting of some expression \fixme{which one?} for the $U$-function instead of obtaining it using 
Mathematica's in-built {\it HypergeometricU} function.  \fixme{why? more accurate for large-$\nu$ or dominant sln. creeping in?}

In Fig.\ref{fig:tilde g,nu=4.9001} we plot both the radial function $\gpt$ and its radial derivative as functions of the radius.
The radial derivative as a function of the frequency has a very similar behaviour to that of $\gpt$ in Fig.\ref{fig:tilde g,r=5}.

On the PIA, as noted in Sec.\ref{sec:g large-n}, solving the recurrence relation Eq.(\ref{eq:recurr rln Tn}) to obtain $\gt$ is unstable since it corresponds to
the subdominant solution and the dominant solution would be `creeping in'. One option is to obtain $\gt (r,+i\nu)$ using 
Eqs.(\ref{eq:Leaver-Liu}) and (\ref{eq:large-a U}).
In Fig.\ref{fig:tilde g,PIA} we show that using the recurrence relation is unstable whereas the latter option does well.
The following is the method we follow:
when calculating the terms in the series Eq.(\ref{eq:Leaver-Liu}) for $\gt(r,+i\nu)$ in practise in Mathematica, we use a numerical
evaluation of the integral representation Eq.(\ref{eq:U num int}) in order to calculate
$\Tnm{n}$, $\forall n$, on the PIA instead of obtaining it by solving the recurrence relation that it satisfies.
 \fixme{why? more accurate for large-$\nu$ or dominant sln. creeping in?}


\begin{figure}[h!]
\begin{center}
  \includegraphics[width=8cm]{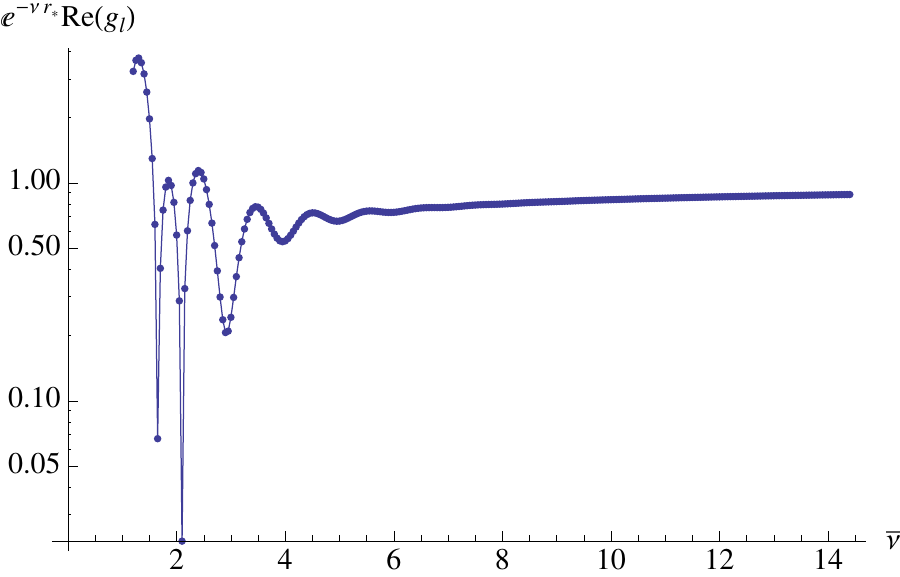}
   \includegraphics[width=8cm]{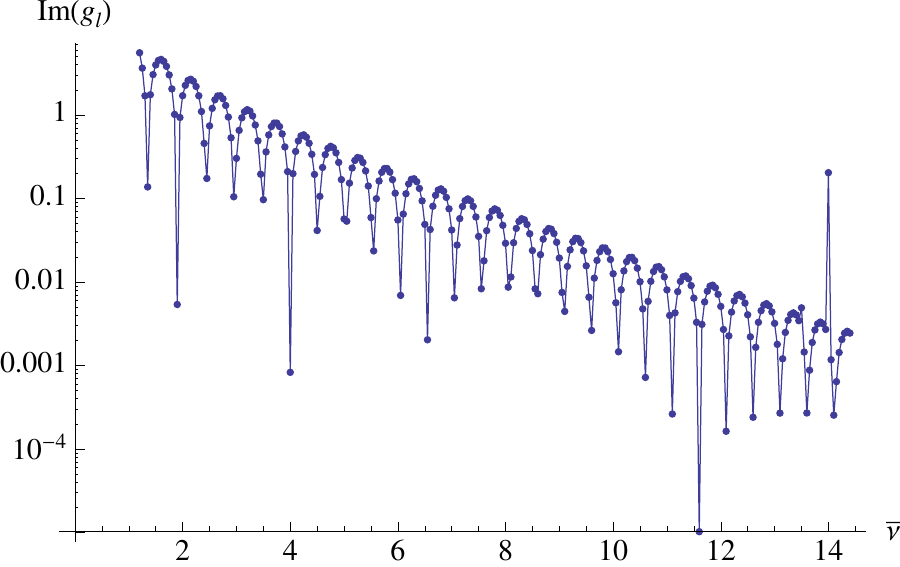}
  \includegraphics[width=8cm]{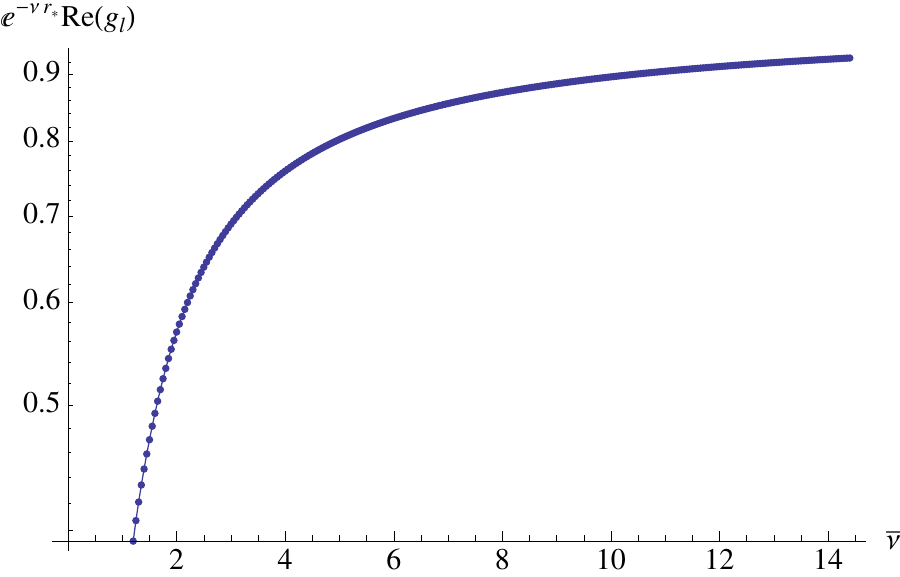}
   \includegraphics[width=8cm]{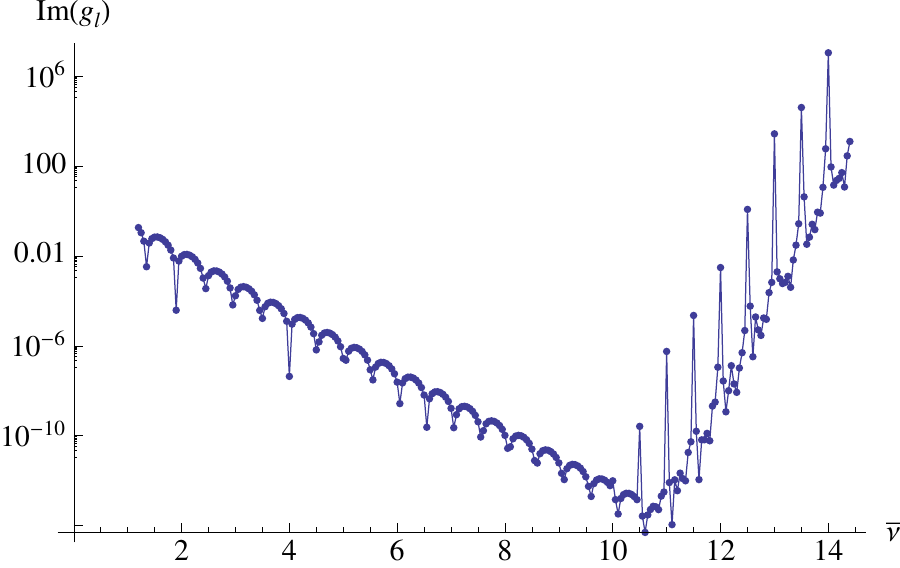}
 \end{center}
\caption{
Log-plots of the radial solution $\gpt$ as a function of $\nb$ for $s=2$, $\ell=2$.
Top plots (a) and (b) are for $r=2.8M$ and bottom plots (c) and (d) are for $r=5M$.
Left plots (a) and (c) are $e^{-\nu r_*}\text{Re}(\gpt)$ 
and right plots (b) and (d) are $\text{Im}(\gpt)$, all obtained using Eq.(\ref{eq:Leaver-Liu})
and finding $\Tnm{n}$ by solving the recurrence relation (\ref{eq:recurr rln Tn}) with $\Tnm{0}$ and $\Tnm{1}$ calculated using Mathematica's
in-built {\it HypergeometricU} function.
\fixme{Round-off error for the Im part from a certain value of $\nu$?}
Cf. Fig.2~\cite{Casals:2011aa}.
A similar behaviour is exhibited by the radial derivative $d\gpt/dr_*$.
}
\label{fig:tilde g,r=5}
\end{figure} 




\begin{figure}[h!]
\begin{center}
   \includegraphics[width=8cm]{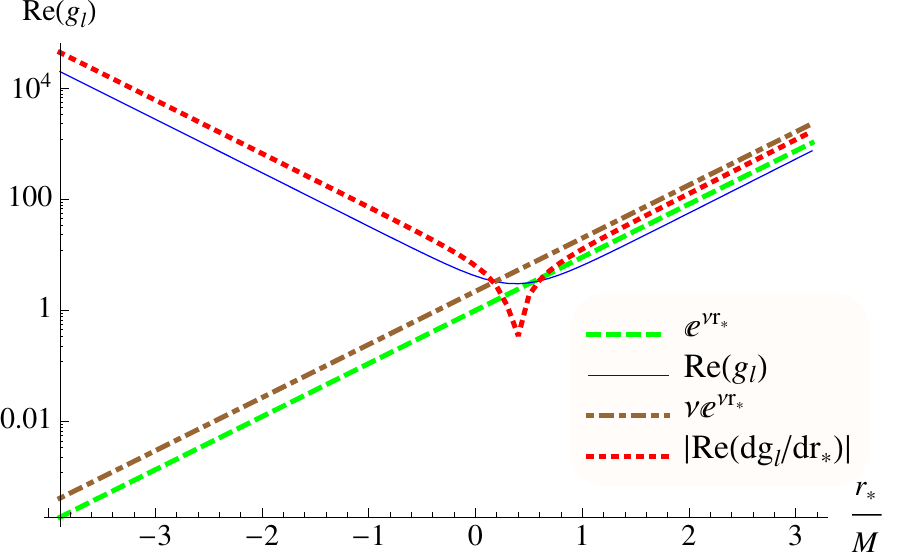}
      \includegraphics[width=8cm]{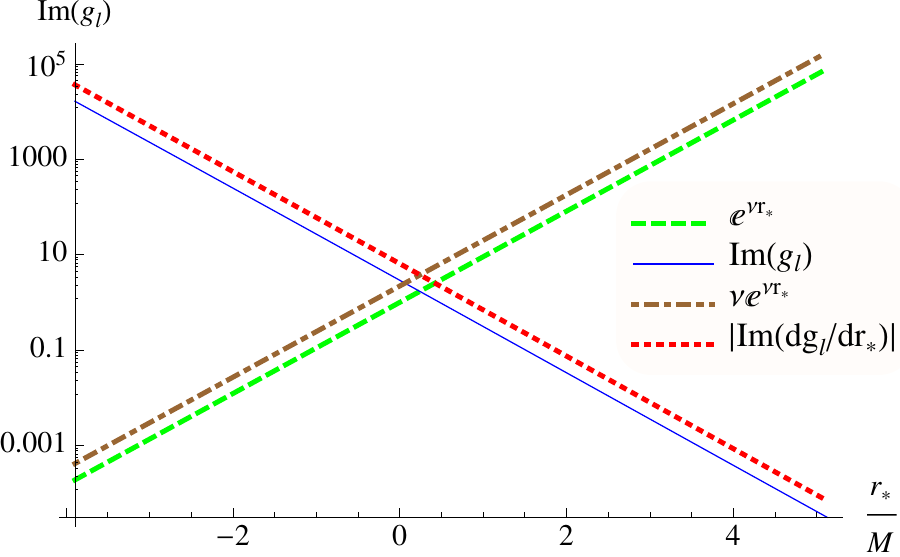}
    \end{center}
\caption{
Log-plots of $\gpt$ and $d\gpt/dr_*$ as functions of $r_*/M$ for $s=2$, $\ell=2$, $\nb=4.4002$.
These radial functions are obtained using Eqs.(\ref{eq:Leaver-Liu}) and (\ref{eq:d tilde g/dr*})
and finding $\Tnm{n}$ by solving the recurrence relation (\ref{eq:recurr rln Tn}) with $\Tnm{0}$ and $\Tnm{1}$ calculated using Mathematica's
in-built {\it HypergeometricU} function.
(a) Continuous blue curve: $\text{Re}(\gt)$;
dashed green curve: large-$r$ asymptotics $\gt\sim e^{\nu r_*}$;
dotted red curve: $\left|\text{Re}(d\gpt/dr_*)\right|$;
dot-dashed brown curve: large-$r$ asymptotics $d\gpt/dr_*\sim \nu e^{\nu r_*}$.
(b) Similar as (a) but the continuous blue curve and the dotted red curve here correspond to the imaginary -- instead of real -- part of $\gpt$ and $d\gpt/dr_*$ respectively.
We note that for $\nb=9.8002$ the calculation of the imaginary part using the method as in these plots becomes unstable for $r_*\gtrapprox 8M$.
}
\label{fig:tilde g,nu=4.9001}
\end{figure}


\begin{figure}[h!]
\begin{center}
      \includegraphics[width=8cm]{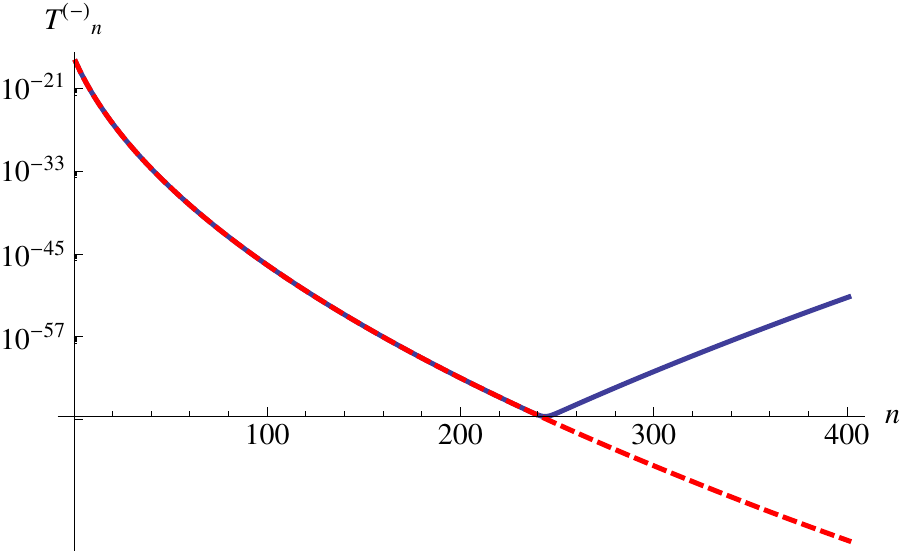}  
   \includegraphics[width=8cm]{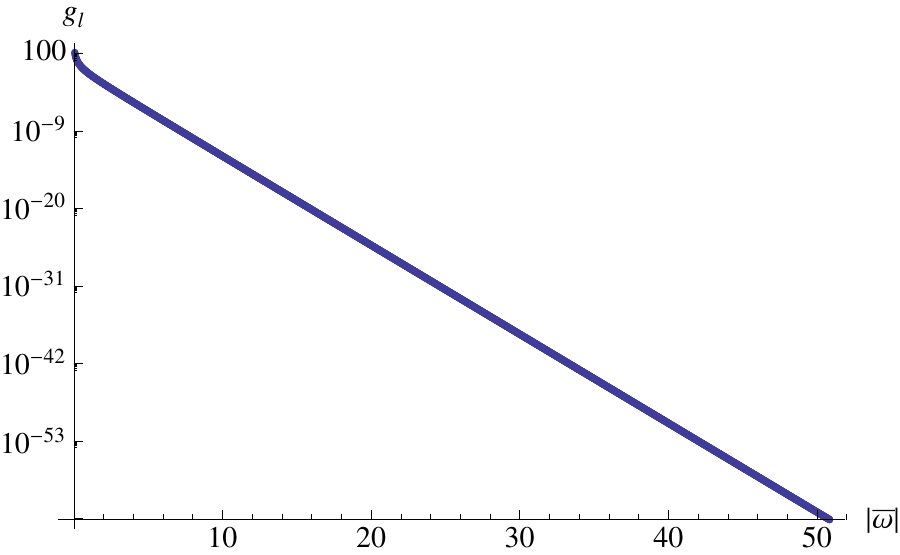}
\end{center}
\caption{
Radial solution $\g$ on the PIA for $s=2$, $\ell=2$, $r=5M$.
(a) Continuous blue curve: $\Tnm{n}$ in Eq.(\ref{eq:Leaver-Liu}) as a function of $n$ for $\ob=4.4002i$ obtained using Mathematica's {\it HypergeometricU} function
(a similar curve is obtained by solving the recurrence relation Eq.(\ref{eq:recurr rln Tn}));
this curve shows how the dominant solution `creeps in' in the solution of the recurrence relation.
Dashed red curve:  similar to the continuous blue curve but $\Tnm{n}$ is obtained
using Eq.(\ref{eq:U num int}).
(b) $\gt(r,+i\nu)$ as a function of $|\ob|$ on the PIA obtained by numerically evaluation the numerical representation Eq.(\ref{eq:U num int}): convergence is achieved up to large 
values
of $|\ob|$ whereas using the recurrence relation Eq.(\ref{eq:recurr rln Tn}) the series coincides with the curve plotted up to $|\ob|\approx14$ but
ceases to converge after that value.
}
\label{fig:tilde g,PIA}
\end{figure}

 
\subsection{Calculation of $\Dgt$} \label{sec:calc Dgt}

From Eq.13.4.15~\cite{bk:AS} and the property $\Gamma(z+1)=z\Gamma(z)$ we easily 
see that $\Tno{n}$ satisfy the recurrence relation Eq.(\ref{eq:recurr rln Tn}).
In Fig.~\ref{fig:Delta Gw,r=10} we plot
$e^{13M\nu}\Dgt(r,\nu)$ 
as a function of $\nb$ obtained with Eq.(\ref{eq:Leaver-Liu series for Deltagt}).
Note that, when carried out in practise in Mathematica, it is better to numerically evaluate the integration representation
Eq.(\ref{eq:U num int})
 instead of using Mathematica's in-built {\it HypergeometricU} 
function in order to calculate the two initial values $\Tno{n=0}$
and $\Tno{n=1}$. \fixme{why? more accurate for large-$\nu$ or dominant sln. creeping in?}


\begin{figure}[h!]
\begin{center}
 \includegraphics[width=10cm]{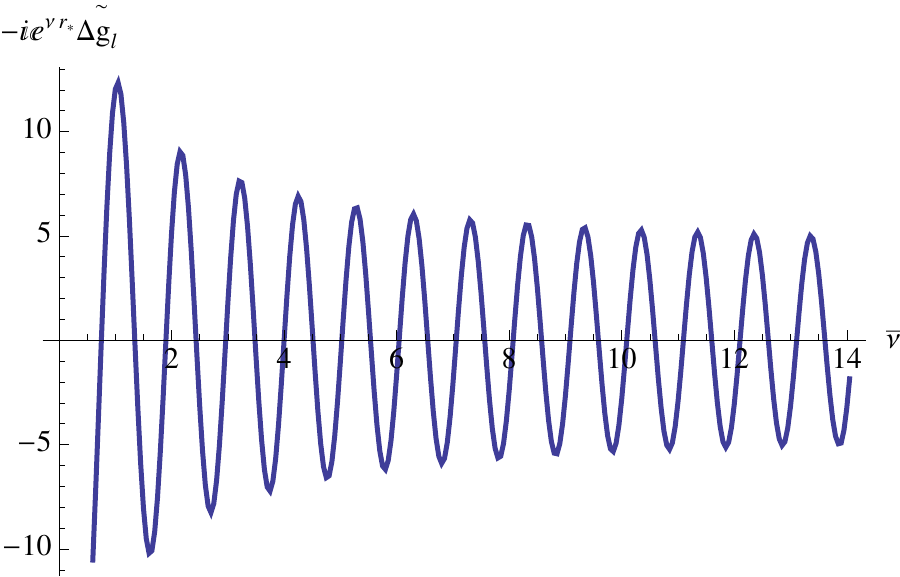}
 \end{center}
\caption{
Plot of `$-ie^{\nu r_*}\Dgt$' as a function of $\nb$ for  $s=2$, $\ell=2$, $r=10M$.
We have obtained $\Dgt$ from Eq.(\ref{eq:Leaver-Liu series for Deltagt}) by
solving the recurrence relation Eq.(\ref{eq:recurr rln Tn}) with $\Tno{0}$ and $\Tno{1}$ calculated using Mathematica's
in-built {\it HypergeometricU} function.
We have removed from $\Dgt$ the large-$r$ behaviour of Eq.(\ref{eq:dg=qg}): $\gt(r,+i\nu)\sim e^{-\nu r_*}$.
The zeros of the curve occur  at $\nb$-steps of approximately $1/2$, agreeing 
with the large-$\nb$ asymptotics of $q(\nu)$ in Eq.38~\cite{Casals:2011aa}.
}
\label{fig:Delta Gw,r=10}
\end{figure}


\subsubsection{Zeros and singularities of $\Dhnt$} \label{sec:zeros of Dhnt}

Any possible zeros and singularities of the terms $\Dhnt$ in Eq.(\ref{eq:Leaver-Liu series for Deltagt}) may only come from the $\ant{n}$, the $U$-function
and the four $\Gamma$-functions. 
The $U$-function has no singularities other than its branch point, and the $\Gamma$-function has no zeros.
We do not know analytically the possible zeros of $\ant{n}$ nor of the $U$-function, so 
we can only determine some of the zeros of $\Dhnt$, and we cannot be sure that any of the poles we might find are not actually cancelled out by zeros
of $\ant{n}$ and/or the $U$-function.

From Sec.\ref{sec:sings of an} we know that $\an{n}$ have simple poles $\forall n\ge k$ if $\bar\nu=k/2$ for some $k\in \mathbb{N}$
(with the exception of $\nu=\nu_{AS}$ when $s=2$). The function $\Gamma(z)$ has simple poles at $z\in\mathbb{Z}^-$.
Let us distinguish two cases:

\begin{itemize}
\item Case $1-2\nb\notin\mathbb{Z}^-\cup{0}$

Neither $\an{n}$ nor the $\Gamma$-functions have any pole. So $\Dhnt$ has no zeros (other than any
coming from $\an{n}$ or the $U$-function) and it has no poles.



\item $1-2\nb\equiv -k\in\mathbb{Z}^-\cup{0}$

In the numerator, $\an{n}$ has simple poles (except if $\nu=\nu_{AS}$ when $s=2$) at $n=k,k+1,k+2,\dots$  and $\Gamma(1+n-2\bar\nu)$ at $n=0,1,2,\dots ,k$.
In the denominator, $\Gamma(1+s+n-2\bar\nu)\Gamma(1-s+n-2\bar\nu)$ has double poles at $n=0,1,2,\dots,k-|s|$ and simple poles at
$n=k-|s|+1,k-|s|+2,\dots ,k+|s|$. 
Also in the denominator, $\Gamma(1-2\bar \nu)$ has a simple pole $\forall n\in\mathbb{N}$.
Therefore, $\Dhnt$ has no poles.
Regarding the zeros (other than any coming from $\an{n}$ or the $U$-function), if $n\neq k$, $\Dhnt$ has double zeros at $n=0,1,2,\dots ,k-|s|$ and
simple zeros at $n=k-|s|+1,k-|s|+2,\dots,k+|s|$; the term $n=k$ is not a zero if $s\neq 0$ and it is just a simple zero if $s=0$.

In the particular case $\nu=\nu_{AS}$ for $s=2$, $\an{n}$ does not have a pole for any $n\in\mathbb{N}$. In this case, a similar analysis to the one in the above paragraph
indicates that $\Dhnt$ has a zero there. \fixme{How to show that $\Dgt$ has a zero at $\nu=\nu_{AS}$ as observed in Fig.\ref{fig:Delta Gw,r=10}?}. 

\fixme{Eq.13.5.2~\cite{bk:AS} seems to indicate that $U(a,b,z)$ does not have a cut when $a\in \mathbb{Z}$ and that $z^a U(a,b,z)$ never has a cut,
both seemingly in contradiction with Eq.13.1.6~\cite{bk:AS} where the cut does not seem to depend on the value of $a$ (if $b=n+1$)?
In our case, we effectively have $z^a U(a,b,z)$ for $\gt$ via Eq.(\ref{eq:Leaver-Liu}) but we know that $\gt$ has a cut - this could be explained by the fact
that the remnant in Eq.13.5.2~\cite{bk:AS} might have a cut. However, giving values to $U(a,b,z)$ when $a\in \mathbb{Z}$  in Mathematica does not seem
to give a cut - this would imply that whenever $\nb$ is a half-integer, $\Dhnt$ and $\Delta G_{\ell}$ have zeros, which we know is not the case from our Figs.??}

\end{itemize}


\subsubsection{Large-$n$ asymptotics} \label{sec:large-n Dhnt}

Using Eq.(\ref{eq:large-n U}) and Eq.5.11.3~\cite{bk:onlineAS} we obtain 
\begin{equation}\label{eq:Tno large-n}
\Tno{n}\sim
\frac{e^{\nu r}(2\nu r)^{-s-1/4}}{\sqrt \pi}
(-1)^n n^{-1/4}
\cos\left(\sqrt{4\nu r\left(2n-4\bar\nu+1\right)+\pi \left(2\bar\nu -n-\frac{1}{4}\right)}\right),
\quad n\to \infty, \quad \text{if } s\ge 0\ \text{and}\ \nu r>0 
\end{equation}

Finally, together with the large-$n$ asymptotics (\ref{eq:a_n large-n Wimp}) for $a_n$ we can obtain 
the large-$n$ asymptotics of the terms in the series for $\Dht$:

\begin{align} \label{eq:large-n Delta tilde h}
&
\Dhnt\sim
\tilde C(\nu,r)
(-1)^n n^{-\bar\nu-1}e^{\pm2\sqrt{2n\bar\nu}i}
\cos\left(\sqrt{4\nu r\left(2n-4\bar\nu+1\right)+\pi \left(2\bar\nu -n-\frac{1}{4}\right)}\right),
\quad n\to \infty, \quad \text{if } s\ge 0\ \text{and}\ \nu r>0
\\&
\tilde C(\nu,r)\equiv \frac{2\sqrt\pi ie^{-\nu r}e^{\pi i (s+1-2\bar\nu)}(-2\bar\nu)^{s+1-2\bar\nu}(2\nu r)^{-s-1/4}}{\Gamma(1-2\bar\nu)}
\nonumber
\end{align}
\fixme{for Eqs.(\ref{eq:Tno large-n}) and (\ref{eq:large-n Delta tilde h}):
or is it just $\nu r\in\mathbb{R}$? and does $a\to -\infty$ also require $\nb\in \mathbb{R}$?}
for fixed $\nu$. \fixme{I don't think the actual value of $\tilde C(\nu,r)$ can be determined, as the value of $\ant{0}$ does not play any part in the asymptotics
of (\ref{eq:a_n large-n Wimp})?}
The modulus of the large-$n$ asymptotics of $\Dhnt$ in Eq.(\ref{eq:large-n Delta tilde h}) is basically the same as that for $\Dhn$ in Eq.(\ref{eq:large-n Delta h}) below.
Therefore the conclusions below Eq.(\ref{eq:large-n Delta h}) regarding the convergence of 
the series $\Dh=\sum_n\Dhn$ apply equally to the series $\Dht =\sum_n\Dhnt$.


\subsection{Calculation of $\Dg$}

From Eq.13.4.1~\cite{bk:AS} and the property $\Gamma(z+1)=z\Gamma(z)$ we easily find that the $\Tnp{n}$ 
satisfy the same recurrence relation Eq.(\ref{eq:recurr rln Tn}) as the $\Tnm{n}$ in the series for $\gt$ and the $\Tno{n}$ in the series for $\Dgt$ --
see~\cite{Liu-1997,Temme:1983}.
To investigate the convergence properties of this series, we require the large-$n$ asymptotics of its terms.


\subsubsection{Large-$n$ asymptotics} \label{sec:large-n Dhn} 

From
Eqs.13.1.27 and 13.5.14~\cite{bk:AS}
we obtain
\begin{align}
\label{eq:large-n M}
&
\Tnp{n}\sim
\pi^{-1/2}e^{-\nu r}\Gamma(2s+1)
\left(2\nu r\right)^{-1/4-s}n^{-1/4}
\cos\left(\sqrt{\left(1-4\bar\nu+2n\right)4\nu r}-\frac{\pi (4s+1)}{4}\right),
\quad n\to \infty,
\end{align}
which is valid for $(n-2\bar{\nu}-s)\to +\infty$, $s$ bounded and $\nu r\in\mathbb{R}$.
This agrees with Eq.4.20~\cite{Liu-1997} (except for a typo they have in the sign of $s$ inside the $\Gamma$-function \fixme{and our extra phase $-\pi/4$?}).
[Note: Eq.(\ref{eq:large-n M}) differs from Eq.12 in Sec.6.13.2 of~\cite{Erdelyi:1953} in having an extra factor $1/2$ and also a power `$-1/4-s$' instead
of a `$+1/4-s$'. We have, however, checked with Mathematica for certain values of the parameters that Eq.(\ref{eq:large-n M}) is the correct
expression.]

From Eq.(\ref{eq:large-n M}) and the large-$n$ asymptotics (\ref{eq:a_n large-n Wimp}) for $a_n$ we can now obtain the large-$n$ asymptotics of the terms in the
 series for $\Dg$:
\begin{align} \label{eq:large-n Delta h}
& 
\Dhn\sim
C(\nu,r)
\left(n-2\bar\nu+\frac{1}{2}\right)^{-\bar\nu-1}e^{\pm2\sqrt{-2n\bar\nu}}
\cos\left(\frac{\pi (4s+1)}{4}-2\sqrt{\left(n-2\bar\nu+\frac{1}{2}\right)2\nu r}\right),
\quad n-2\bar\nu\gg 1, 
\\& 
C(\nu,r)\equiv 
\frac{(-1)^{2s}2\sqrt{\pi} i\left(2\nu r\right)^{-1/4-s}e^{-\nu r}}{\Gamma(1-2\bar\nu)} 
\nonumber
\end{align}
The ratio test yields $\left|\Delta h_{n+1,\ell}/\Dhn\right|\to 1$ as $n\to\infty$, and so it is inconclusive.
However, we may apply the integral test as follows.
For $ \nb>0$, the function $k(n)\equiv \left| C(\nu,r)\right| n^{-\nb-1}=|k(n)|$
\fixme{Isn't $C(\nu,r)$ meaningless because the value $\ant{0}$ does not play any part in the asymptotics?}
 is positive and monotone decreasing 
with $n$ and it satisfies
\begin{equation}
\int_1^{\infty}dn |k(n)|=
\left| C(\nu,r)\right| \int_1^{\infty}dn\ n^{-\nb-1}=
\frac{\left| C(\nu,r)\right| }{\bar\nu}\left[1-\lim_{n\to \infty}n^{-\bar\nu}\right]< \infty,\qquad \text{if}\quad \nb>0.
\end{equation}
Therefore $|k(n)|$ satisfies the
integral test and so the series $\sum_n k(n)$ is absolutely convergent for $ \nb>0$.
Since $\left|\Dhn\right|<|k(n)|$ for sufficiently large $n$, from the comparison test we have that  $\sum_n  \Dhn$ is also absolutely convergent for $ \nb>0$.
Indeed, in our calculations, the series has converged for arbitrary values of $\nb$. However, while the series is fast convergent for large $\nb$ the speed of convergence
becomes slower for smaller values $\nb$.

\subsection{Calculation of $q(\nu)$}

We calculate the BC `strength' $q(\nu)$ using Eq.(\ref{eq:dg=qg}), where 
we calculate $\Dgt$ using the method described in Sec.\ref{sec:calc Dgt} and
 $\gt(r,+i\nu)$ on the PIA using Eq.(\ref{eq:Leaver-Liu}) with $\nu\to -\nu$ everywhere. We note that an alternative method, which we have not explored, for calculating 
 $\gt(r,+i\nu)$ might be to use Eq.74~\cite{Leaver:1986a}.
In Fig.~\ref{fig:q,l=2=s} we plot $q(\nu)$ as a function of $\nb$.
This figure is to be compared with Fig.2~\cite{Leung:2003ix} (also Fig.2.~\cite{Leung:2003eq}).

\begin{figure}[h!]
\begin{center}
 \includegraphics[width=10cm]{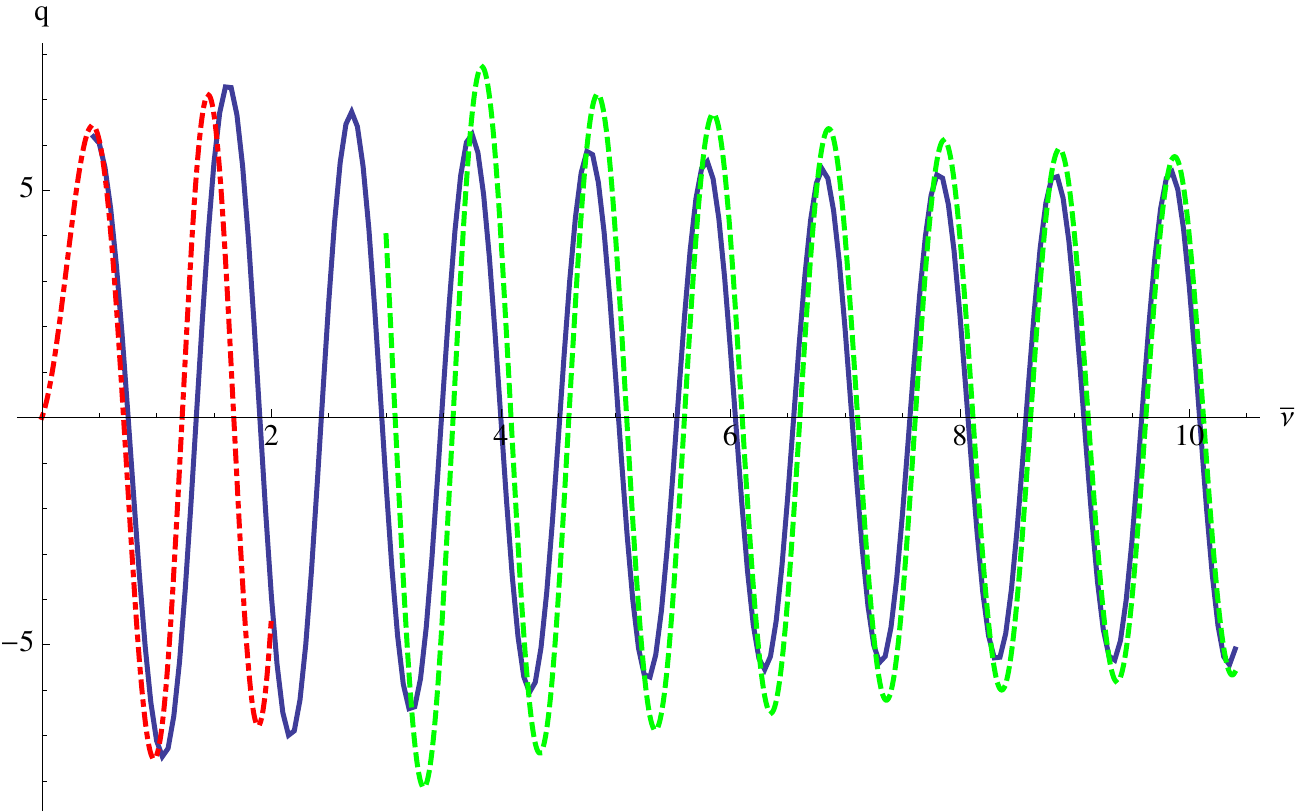}
 \end{center}
\caption{
BC `strength' $q(\nu)$ of Eq.(\ref{eq:dg=qg}) as a function of $\nb$ for $s=2$, $\ell=2$.
Continuous blue curve: $q(\nu)$ obtained using Eq.(\ref{eq:dg=qg}) with the value $r=5M$ and $\gt (r,+i\nu)$ obtained using Eq.(\ref{eq:U num int}).
Dashed green curve:  large-$\nb$ asymptotics of Eq.38~\cite{Casals:2011aa} (see Fig.5~\cite{Casals:2011aa} for a better agreement for larger values of $\nb$).
Dot-dashed red curve: small-$\nb$ asymptotics of Eqs.2.2 and 2.5~\cite{Leung:2003ix} 
(we note that the small-$\nb$ asymptotics in~\cite{Leung:2003ix}  do not work well for other spins; 
see~\cite{Casals:Ottewill:2011smallBC} for better agreement for small-$\nb$ for any spin).
Cf. Fig.2~\cite{Leung:2003ix} (also Fig.2.~\cite{Leung:2003eq}).
We note that when $\gt (r,+i\nu)$ is obtained using the recurrence relation Eq.(\ref{eq:recurr rln Tn}) instead of Eq.(\ref{eq:U num int}), 
the result for $q(\nu)$ is incorrect from
$\nb\approx 11$ - see Fig.\ref{fig:tilde g,PIA}(a).
}
\label{fig:q,l=2=s}
\end{figure}

\section{Calculation of  the Wronskian}

We calculate, on the NIA, the Wronskian $\Wh_+$ of the radial solutions 
$\fb\equiv -\sin(2\pi i\ob) \f$ 
and $\gpt$  using the methods described in the previous sections: 
the Jaff\'e series Eq.(\ref{eq:Jaffe}) for $\f$, Eq.(\ref{eq:df/dr*}) for $d\f/dr_*$, 
the Leaver-$U$ series Eq.(\ref{eq:Leaver-Liu}) for $\gpt$ and Eq.(\ref{eq:d tilde g/dr*}) for $d\gpt/dr_*$.

In Figs.\ref{fig:ln(f),l=2,s=0 NIA}--\ref{fig:tilde g,nu=4.9001} we plot, on the NIA, the radial solutions  $\fb(r,-i\nu)$ and $\gpt(r,-i\nu)$ and their radial derivatives,
 which are required for the Wronskian.

Let us define $\Wh_1\equiv \gpt\fb'$ and $W_2\equiv \fb\gpt'$, so that $\Wh_+=\Wh_1-\Wh_2$.
Figs.\ref{fig:wronskian,nu=2.2001}(a) and (b) show that the magnitudes of the two contributions $\Wh_1$ and $\Wh_2$ are very close for all $r_*$ except
near $r_*=0$.
Therefore, the computation of $\Wh=\Wh_1-\Wh_2$ would require the knowledge of these two contributions to very high accuracy away from this `window' near $r_*=0$.
We note that for the imaginary part, for $r_*\ge 0$ the two contributions  $\Wh_1$ and $\Wh_2$ actually add up and so there is no computational difficulty there either.
Figs.\ref{fig:wronskian,nu=2.2001}(c) shows that, indeed, there is a `window' near $r_*=0$ where the calculation of the absolute value of the Wronskian is reliable.
We note that in this `window' it is 
$\text{Im}(\Wh) \gg \text{Re}(\Wh)$,
so the imaginary part dominates
but, for accuracy, the real part cannot be neglected. 
We have checked that a similar `window' near $r_*=0$ occurs for different values of the spin, the multipole number $\ell$ 
and the frequency on the NIA.


In Figs.\ref{fig:W1,W2,r=2.8} we plot $\Wh_1$, $\Wh_2$ and $\Wh$ at $r=2.8M$ as functions of $\nb$.
Fig.\ref{fig:W1,W2,r=2.8}(c), together with Figs.7--9~\cite{Casals:2011aa} where these `mid'-frequency results are compared to large-$\nb$ asymptotics, show
 that the calculation of the Wronskian at this value of the radius yields a reliable result. 
 In~\cite{Casals:Ottewill:2011smallBC} we show that the `mid'-frequency results for the Wronskian agree well with small-$\nb$ asymptotics.
Fig.\ref{fig:W1,W2,r=2.8}(c) also shows -- for the particular value $\ell=2$ -- that for $s=2$ the
Wronskian $\hat W_+\equiv W[\gpt,\fb]$ has a  zero of order one \fixme{order 2 for just the imaginary part?} at $\nu=\nAS$. This is as expected because of the definition 
$\fb\equiv -\sin(2\pi i\ob) \f$ 
and the fact that $\nb=\nb_{AS}$ is not a pole of $\f$ for $s=2$ and
it agrees with ~\cite{Leung:2003ix}.

\begin{figure}[h!]
\begin{center}
   \includegraphics[width=8cm]{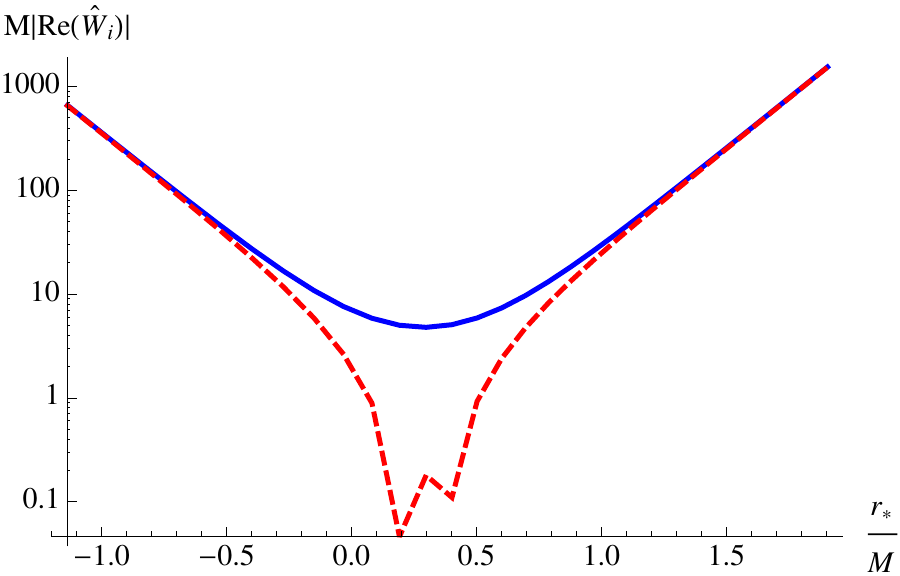}
      \includegraphics[width=8cm]{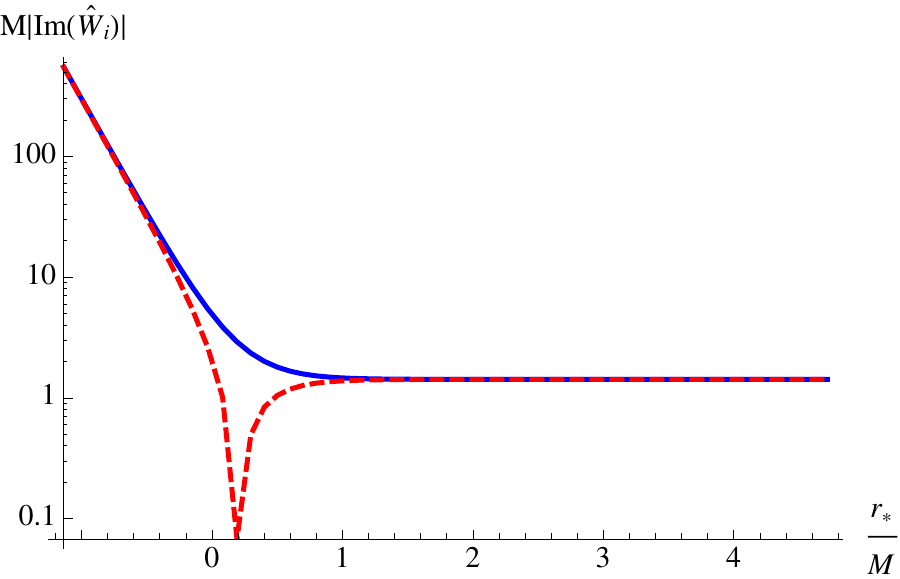}
                \includegraphics[width=8cm]{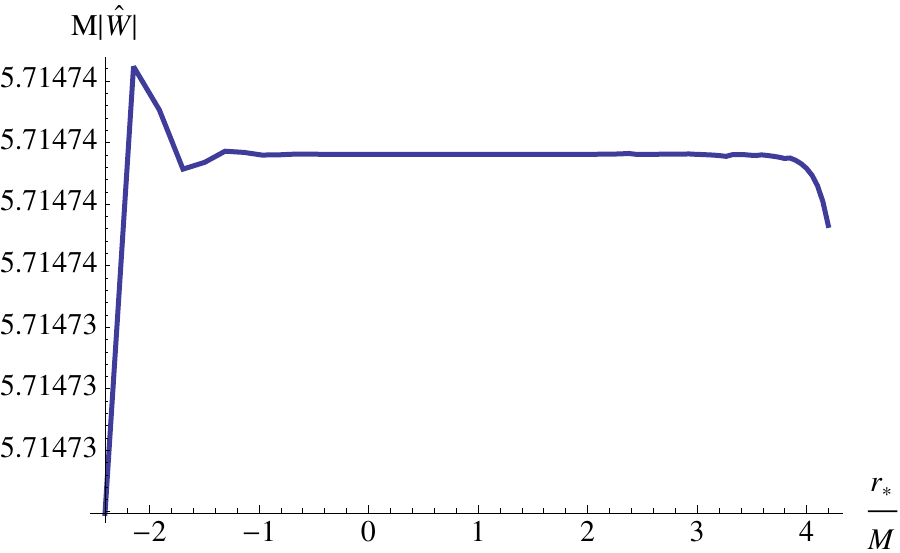}
            \end{center}
\caption{
Wronskian $\hat W_+\equiv W[\gpt,\fb;\omega]=\hat W_1-\hat W_2$ as a function of $r_*/M$  for $s=2$, $\ell=2$ and  $\nb=4.4002$.
Figs.(a) and (b): log-plots of the absolute values of, respectively, the real and imaginary parts of
$M\cdot \hat W_1\equiv M \gpt\fb'$ (continuous blue curve) and $M\cdot \hat W_2\equiv M\fb\gpt'$ (dashed red curve).
Fig.(c): Log-plot of $M|\hat W|$.
}
\label{fig:wronskian,nu=2.2001}
\end{figure} 

\begin{figure}[h!]
\begin{center}
   \includegraphics[width=8cm]{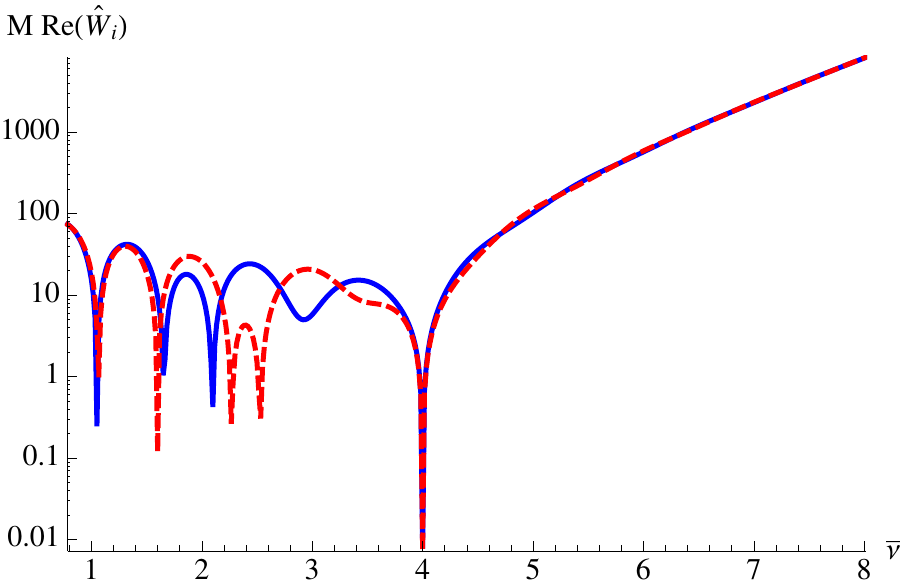}
      \includegraphics[width=8cm]{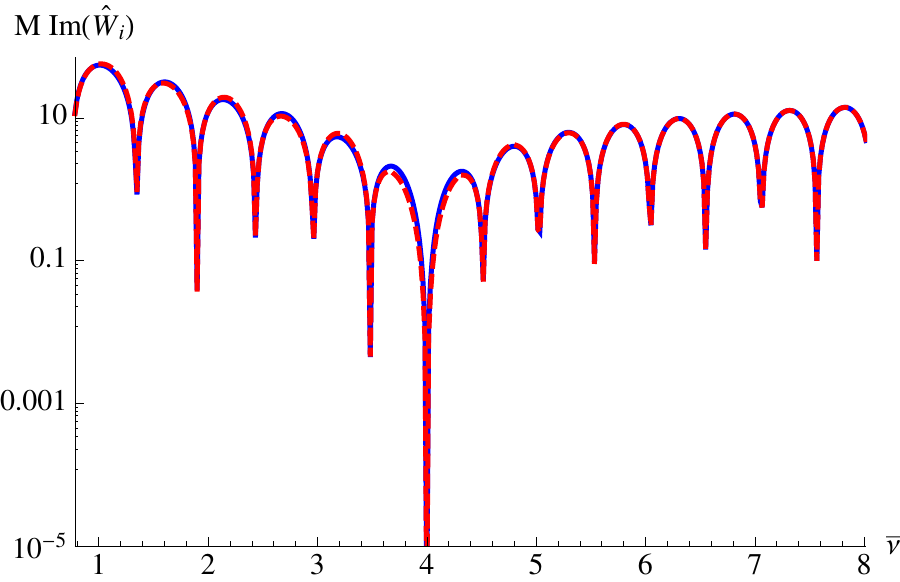}
            \includegraphics[width=10cm]{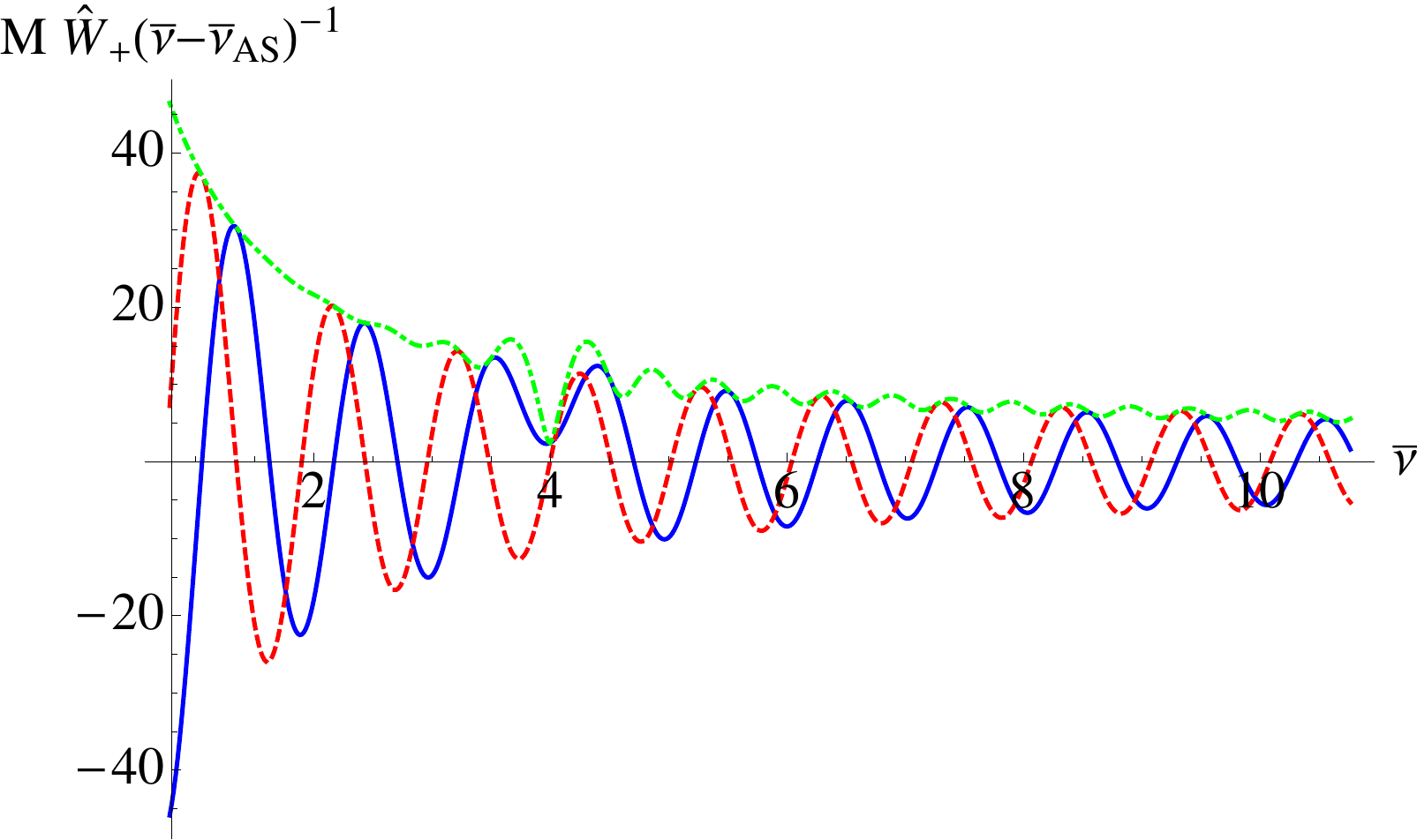}
            \end{center}
\newpage
\caption{
Wronskian $\hat W\equiv W[\gt,\fb;\omega]=\hat W_1-\hat W_2$ as a function of
$\nb$ for $s=2$, $\ell=2$, $r=2.8M$ ($r_*\approx 0.97M$).
(a) and (b): log-plots of, respectively,  the real and imaginary parts of both $M\hat W_1$ (continuous blue curve) and $M\hat W_2$ (dashed red curve).
(c) Plot of the real part (continuous blue curve), imaginary part (dashed red curve) and absolute value (dot-dashed green curve) of $M\hat W_+ /(\nb-\nb_{AS})$.
We note the zero of order one of $|\hat W_+|$ at the algebraically special frequency $\nb=\nb_{AS}$.
}
\label{fig:W1,W2,r=2.8}
\end{figure}

\section{Calculation of  BC modes} \label{sec:BCMs}

\begin{figure}[h!]
\begin{center}
   \includegraphics[width=10cm]{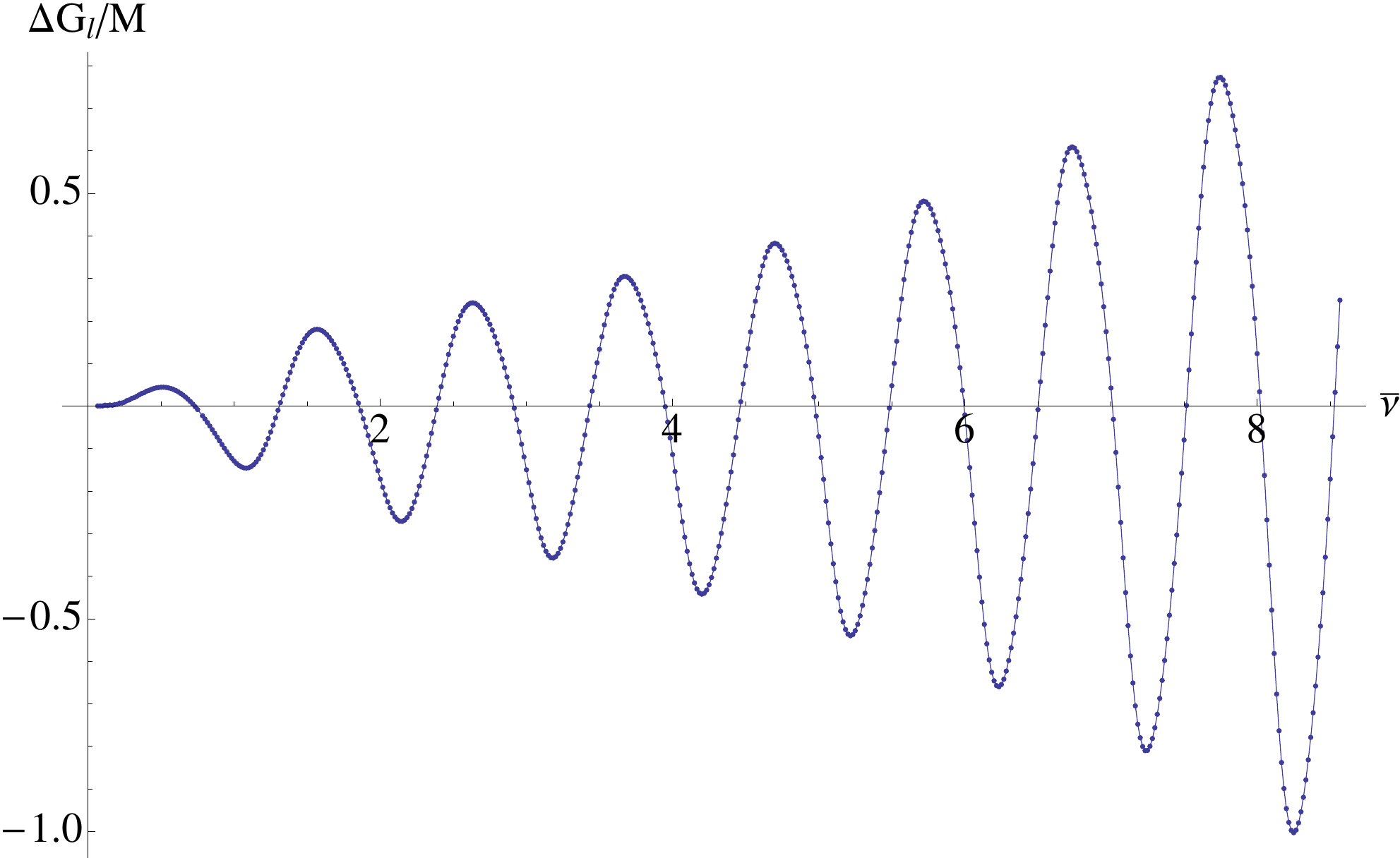}
     \includegraphics[width=10cm]{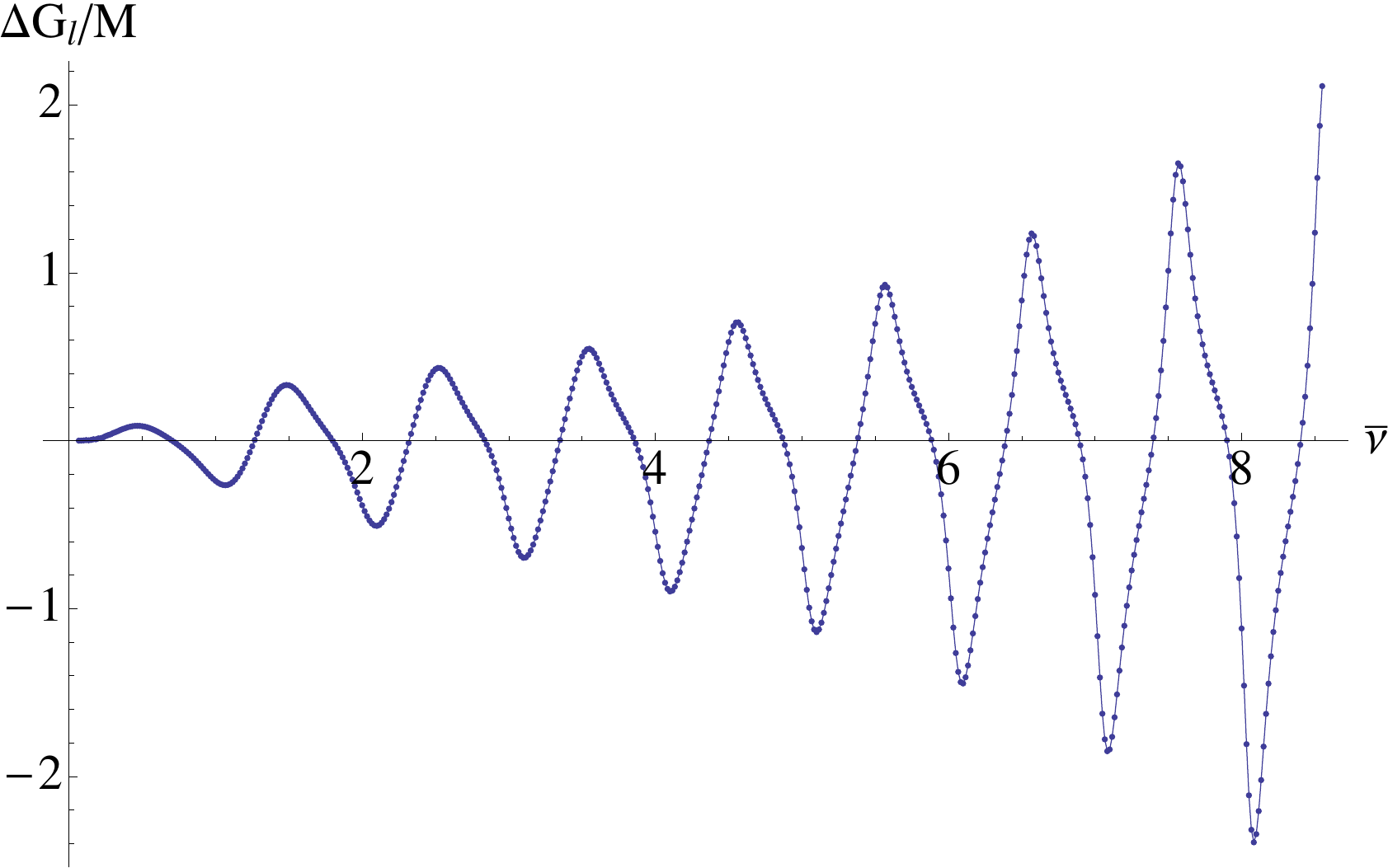}
              \includegraphics[width=10cm]{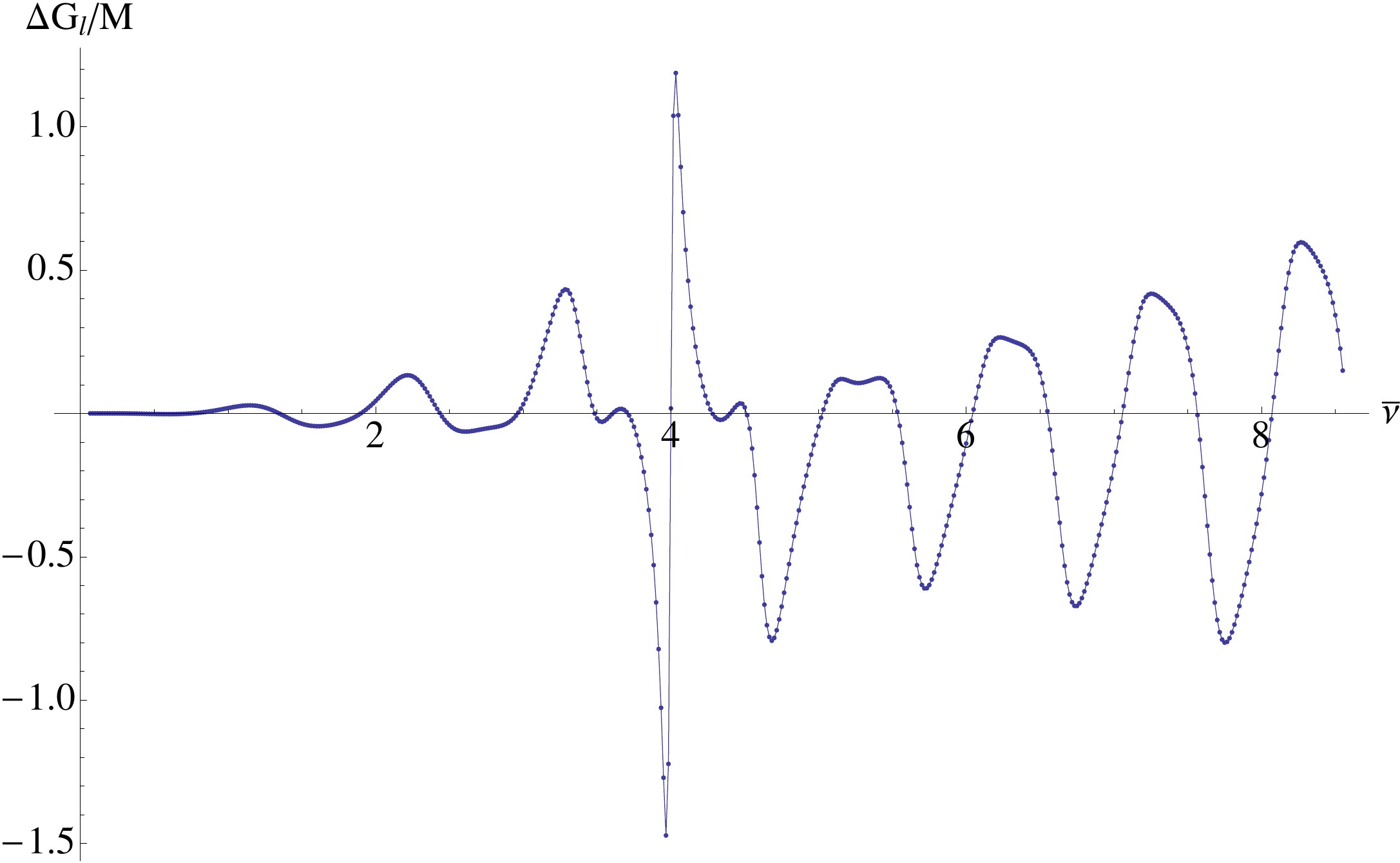}
\end{center}
\caption{
Branch cut mode $\DGl/M$ of Eq.(\ref{eq:DeltaG in terms of Deltag}) as a function of $\nb$ for $r_*=0.2M$
and $r'_*=0.4M$.
The Wronskian has been calculated at $r=2.8M$ while $q(\nu)$ has been calculated with functions at $r=5M$.
(a) $s=0$, $\ell=1$.
(b) $s=1$, $\ell=1$.
(c) $s=2$, $\ell=2$ -- cf. Fig.3~\cite{Leung:2003ix} (also Fig.3~\cite{Leung:2003eq}).
}
\label{fig:DeltaG Fig.3LMMY'03a}
\end{figure}

We obtain the branch cut modes 
$\DGw{r}{r'}{\nu}$ 
by calculating the different quantities  in Eq.(\ref{eq:DeltaG in terms of Deltag}) using the methods described in the previous sections.
In particular, for the calculation of the Wronskian we have evaluated the radial functions $\fb$ and $\g$ at $r=2.8M$ while
for the branch cut `strength' $q(\nu)$ we have evaluated the radial functions at $r=5M$.
In Fig.\ref{fig:DeltaG Fig.3LMMY'03a} we plot $\DGl$ as a function of $\nb$ for different spins.
In the spin-2 case the plot is to be compared with Fig.3~\cite{Leung:2003ix} (also Fig.3~\cite{Leung:2003eq}).
From  Figs.\ref{fig:ln(f),l=2,s=0 NIA}, \ref{fig:q,l=2=s}, \ref{fig:W1,W2,r=2.8}(c) respectively, 
the radial solution $\fb$, the BC `strength' $q$ and the absolute value of the Wronskian 
$|\Wh|$
all have a simple zero at $\nb=\nb_{AS}$
in the case $\nu=\nu_{AS}$ and $s=\ell=2$.
From Eq.(\ref{eq:DeltaG in terms of Deltag}) it then follows that 
$\DGl$
 has a simple zero at that frequency, as Fig.\ref{fig:DeltaG Fig.3LMMY'03a}(c)
reflects.
In Fig.\ref{fig:DeltaG s=l=2 r=5,10} we plot again
$\DGw{r}{r'}{\nu}$
 but in this case for larger values of the radius $r'$: the magnitude
of $\DGl$ increases rapidly with the radius, as expected from Fig.\ref{fig:ln(f),l=2,s=0 NIA}(b).

The spin-$2$ case is quite distinct due to the algebraically special frequency $\nb_{AS}$ ($=4$ when $\ell=2$): 
while the branch cut mode $\DGl$ is zero at $\nb=\nb_{AS}$, \fixme{is that true for any $\ell$?} 
$\DGl$ is particularly large for frequencies near $\nu_{AS}$. This behaviour is explained, in the case $\ell=2$
and $r'_*\to -\infty$ and $r_*\to \infty$, as arising from nearby `unconventional damped modes', that is a pair of poles 
in the unphysical Riemann sheet.
The imaginary part of the QNM frequencies is negative and increases in magnitude as the overtone number $n$ increases, so that $n$ is an index for 
the speed of damping of the mode 
 with time.
For spin-2, QNM frequencies approach the algebraically special frequency as 
 $n$ is increased from the lowest damped mode, $n=0$,
until a certain value of $n$, say $n_M$, whose QNM frequency is very close to $\ob_{AS}$;
 for $\ell=2$ it is $n_M=9$.
 As $n$ is further increased from $n_M$, the real part of the spin-2 QNM frequencies in the 3rd quadrant increases monotonically.
 Therefore, in a certain sense, the  algebraically special frequency marks the start of the highly-damped asymptotic regime for QNMs.
 
In Fig.\ref{fig:q/W^2} we plot the radius-independent quantity  
`$|2\nu q /(M\hat W^2)|$', 
which is the branch cut mode 
$\DGl$ of Eq.(\ref{eq:DeltaG in terms of Deltag}) but without the $\f$ factors.
The zeros of this radius-independent quantity correspond to the zeros of $q(\nu)$.
Fig.\ref{fig:q/W^2} shows that, for the cases with $s=1$ and $2$,
 these zeros occur with a period in $\nb$ close to that of the increment in the imaginary part of the QNM frequencies $\ob_{QNM}$ at consecutive overtone numbers.
 For $s=2$,  (minus) the imaginary part of the QNM frequencies lie close to the zeros of $q(\nu)$.
For $s=1$,  for which the QNM frequencies approach the NIA particularly fast~\cite{Casals:2011aa}, 
(minus) the imaginary part of the QNM frequencies lie  close to
the maxima points of
`$|2\nu q /(M\hat W^2)|$'
which are directly related to nearby zeros of $\hat W$, i.e., the QNM frequencies by definition.
For $s=0$, on the other hand, the periods of the zeros of $q(\nu)$ and of $\text{Im}(\ob_{QNM})$ differ slightly for mid-$\nb$
while, for large-$\nb$, `$-\text{Im}(\ob_{QNM})$' tend to lie somewhere in-between the zeros of $q(\nu)$ and the maxima points of
`$|2\nu q /(M\hat W^2)|$'.
For all spins in the large-$\nb$ asymptotic regime,
 the separation of the zeros of $q(\nu)$ approaches $1/2$,
  which coincides with
the separation in the imaginary part of  
highly-damped
QNM frequencies for consecutive overtone numbers (see, e.g.,~\cite{Casals:2011aa} for the asymptotic expressions).

\begin{figure}[h!]
\begin{center}
 \includegraphics[width=8cm]{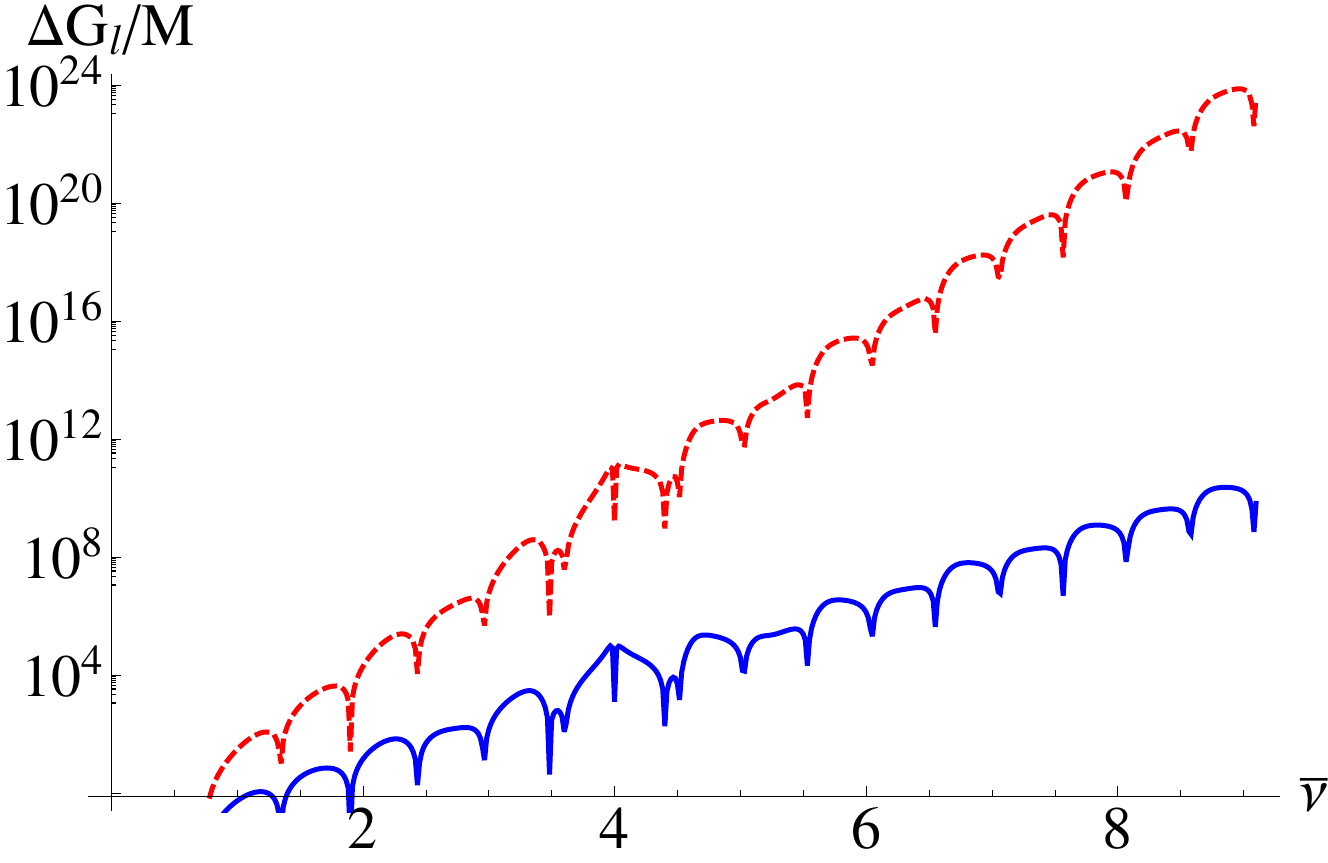} 
\end{center}
\caption{
Same as Fig.\ref{fig:DeltaG Fig.3LMMY'03a}(c), i.e., $\DGl/M$ as a function of $\nb$ for  $s=2$, $\ell=2$ and $r_*=0.2M$ but here it is
with $r'_*=r_*(r=5M)$ in the continuous blue curve and $r'_*=r_*(r=10M)$ in the dashed red curve.
\fixme{Worth including when we already have Figs.\ref{fig:q/W^2} and \ref{fig:ln(f),l=2,s=0}?}
}
\label{fig:DeltaG s=l=2 r=5,10}
\end{figure}

\begin{figure}[h!]
\begin{center}
\includegraphics[width=10cm]{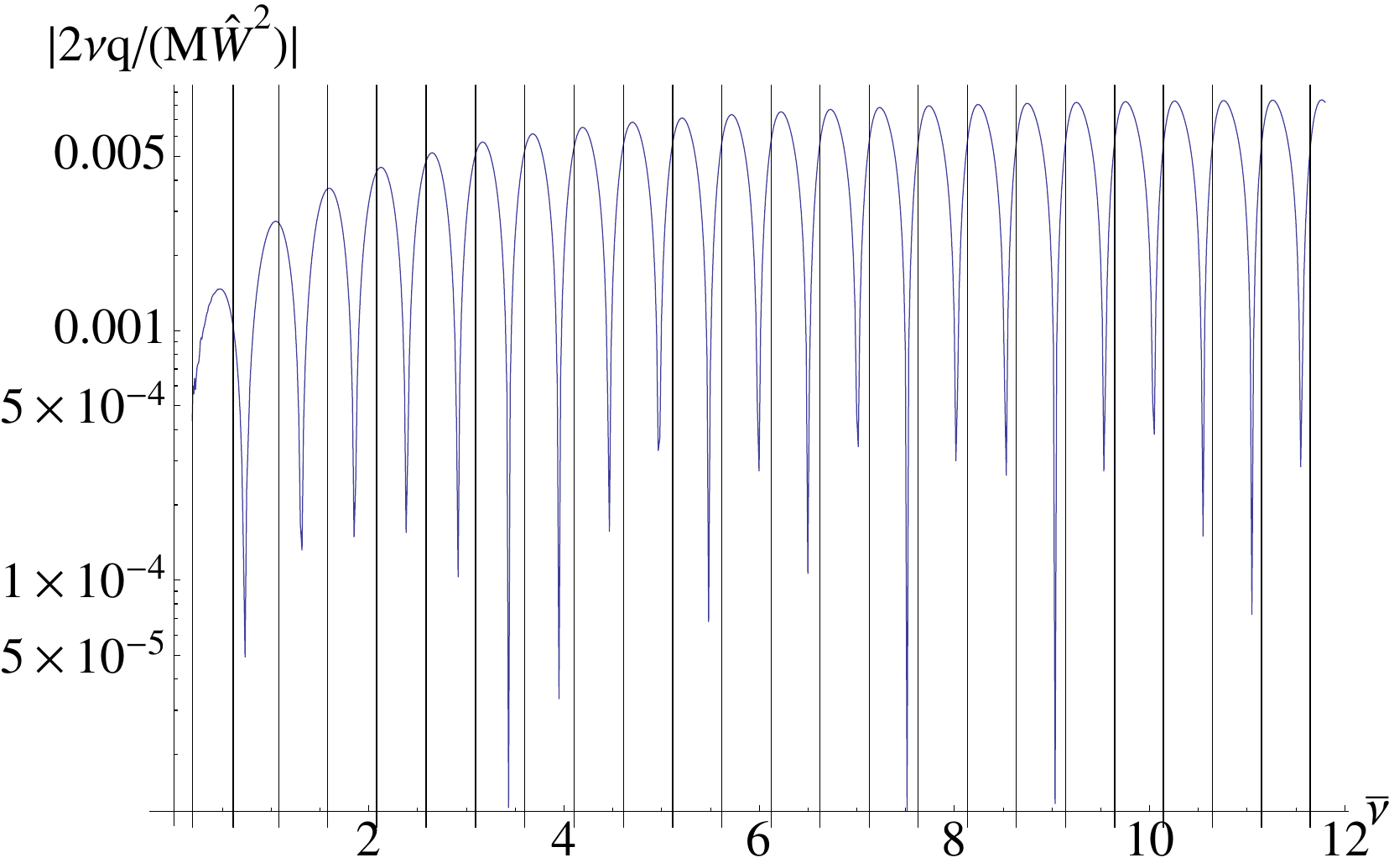}
              \includegraphics[width=10cm]{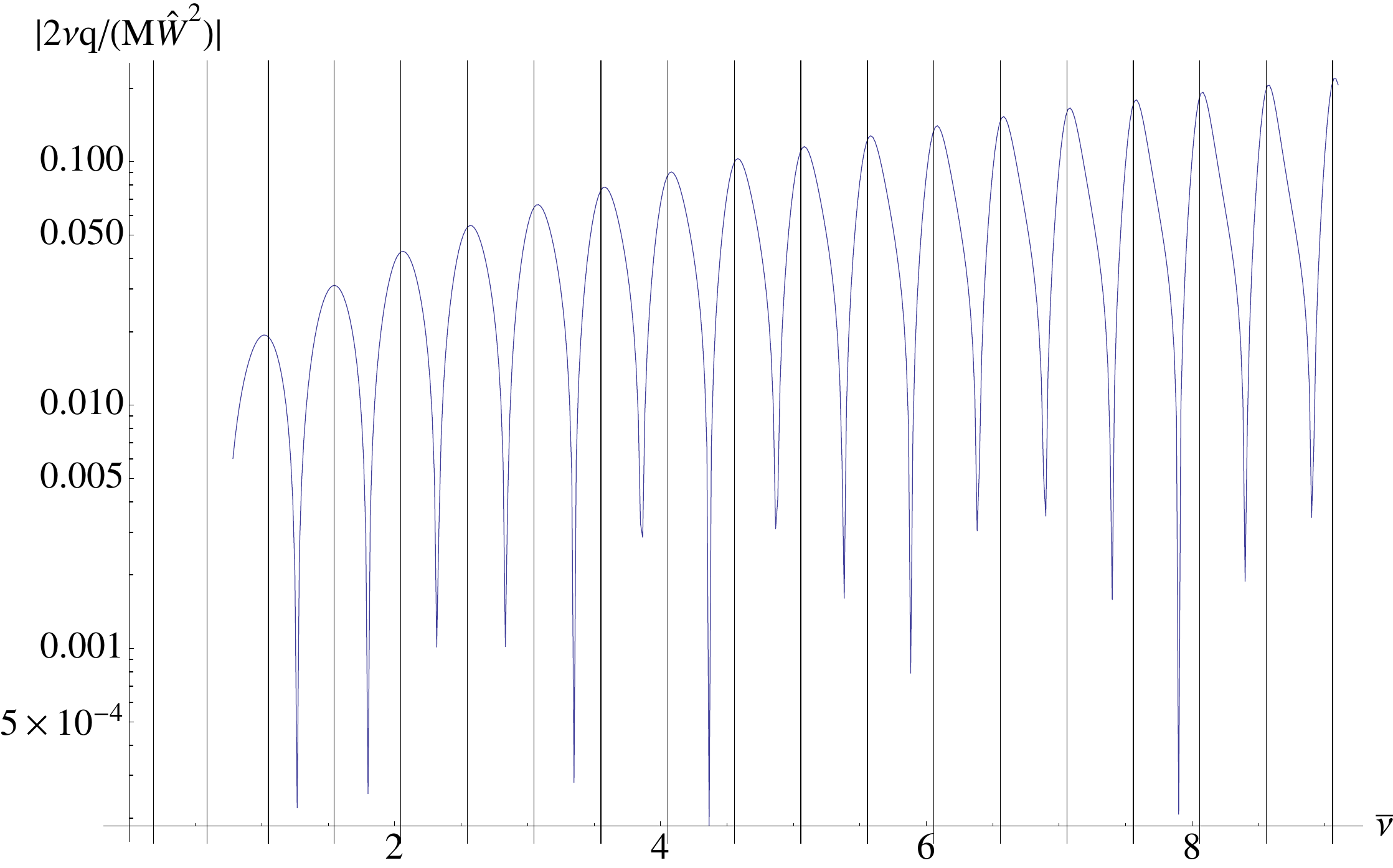}
              \includegraphics[width=10cm]{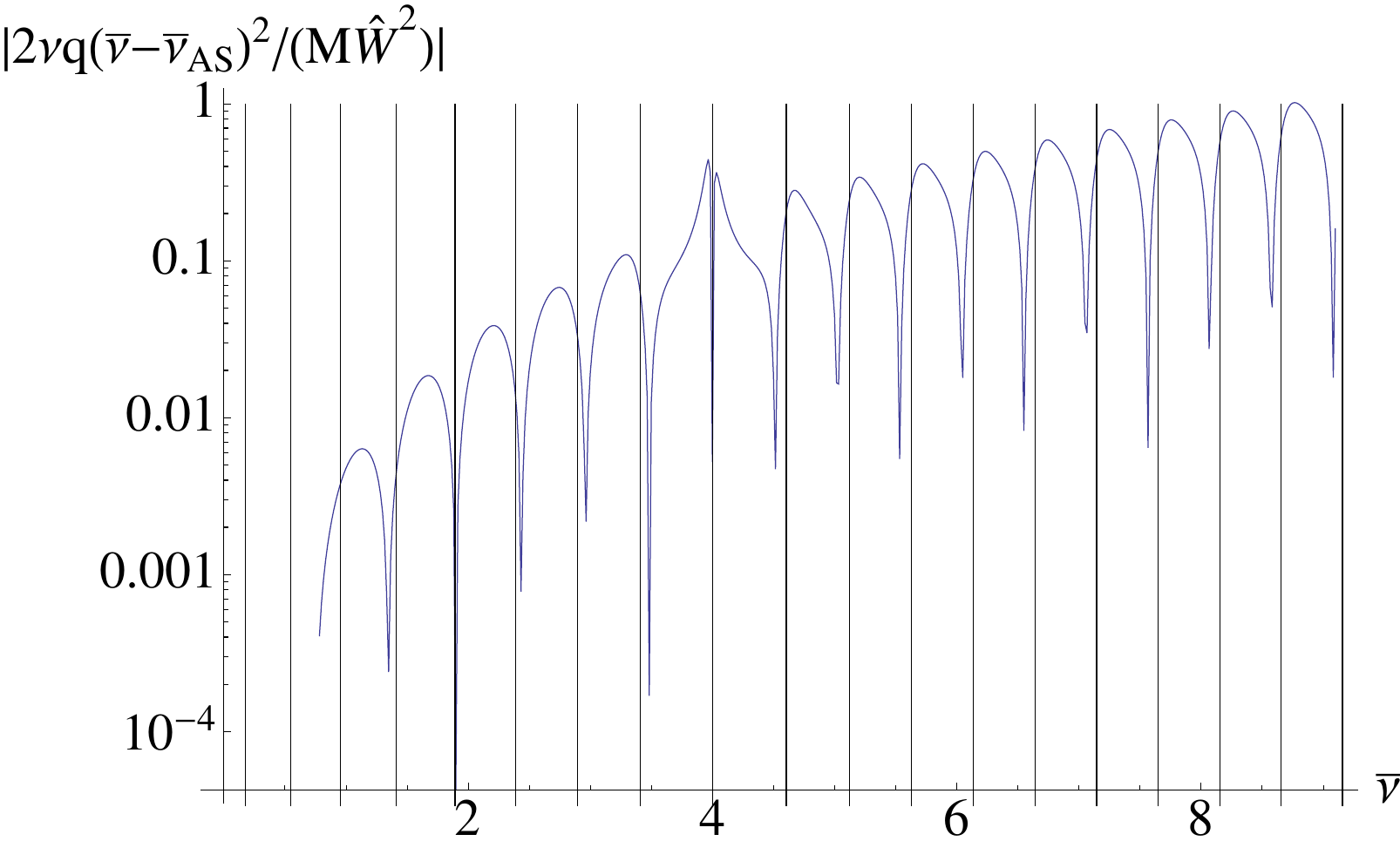}
\end{center}
\caption{
Log-plot of 
 `$|2\nu q/(M\hat W^2)|$'
(this is the branch cut mode $\DGl$ of Eq.(\ref{eq:DeltaG in terms of Deltag}) but without the $\fb$ factors) as a function of $\nb$. 
The vertical lines are located at the values of (minus) the imaginary part of the QNM frequencies -- obtained from~\cite{QNMBerti} for the cases $s=1$ and $s=2$
and using~\cite{DolanOttewill:QNMMST} for $s=0$.
The Wronskian $\hat W$
and the BC `strength' $q(\nu)$ have been calculated by evaluating the radial solutions at, respectively, $r=2.8M$ and $r=5M$.
(a) $s=0$, $\ell=1$.
(b) $s=1$, $\ell=1$.
(c) $s=2$, $\ell=2$: in this case we log-plot  `$|2\nu q (\nb-\nb_{AS})^2M/\hat W^2|$', i.e, we include an extra
factor $|\nb-\nb_{AS}|^2$ to account for the double zero of $|\hat W|^2$ at $\nb=\nb_{AS}$).
}
\label{fig:q/W^2}
\end{figure} 



\section{Self-force} \label{sec:SF}

The motion of a (non-test) point particle moving on a background space-time deviates from geodesic motion  
of that space-time due to a 
self-force (see, e.g.,~\cite{Poisson:2011nh} for a review).
The self-force may be calculated via an integration of the covariant derivative of the retarded Green function 
integrated over the whole past worldline 
of the particle.
In particular, for a scalar charge $q$ moving on Schwarzschild background space-time, the $\mu$-component of the self-force is given by
\begin{equation}
f_{\mu}(\tau)=q\int_{-\infty}^{\tau^-}d\tau'\ \nabla_{\mu}G_{ret}(z(\tau),z(\tau'))
\end{equation}
where $z(\tau)$ is the worldline of the particle and $\tau$ is its proper time.
In the rest of this section we will  deal with the case of a scalar charge ($s=0$) only, although
the self-force in the cases of an electromagnetic charge ($s=1$) and of a point mass ($s=2$) also involve the integration of the Green function in a similar way.
We will investigate the contribution to the scalar self-force from a single BC multipole mode $G_{\ell}^{BC}(r,r';t)$ in the case
of a particle on a worldline 
at constant radius.
\fixme{Check that this implies $dt/d\tau=const.$ for any motion of the particle, geodesic or not}.

\begin{figure}[h!]
\begin{center}
\includegraphics[width=10cm]{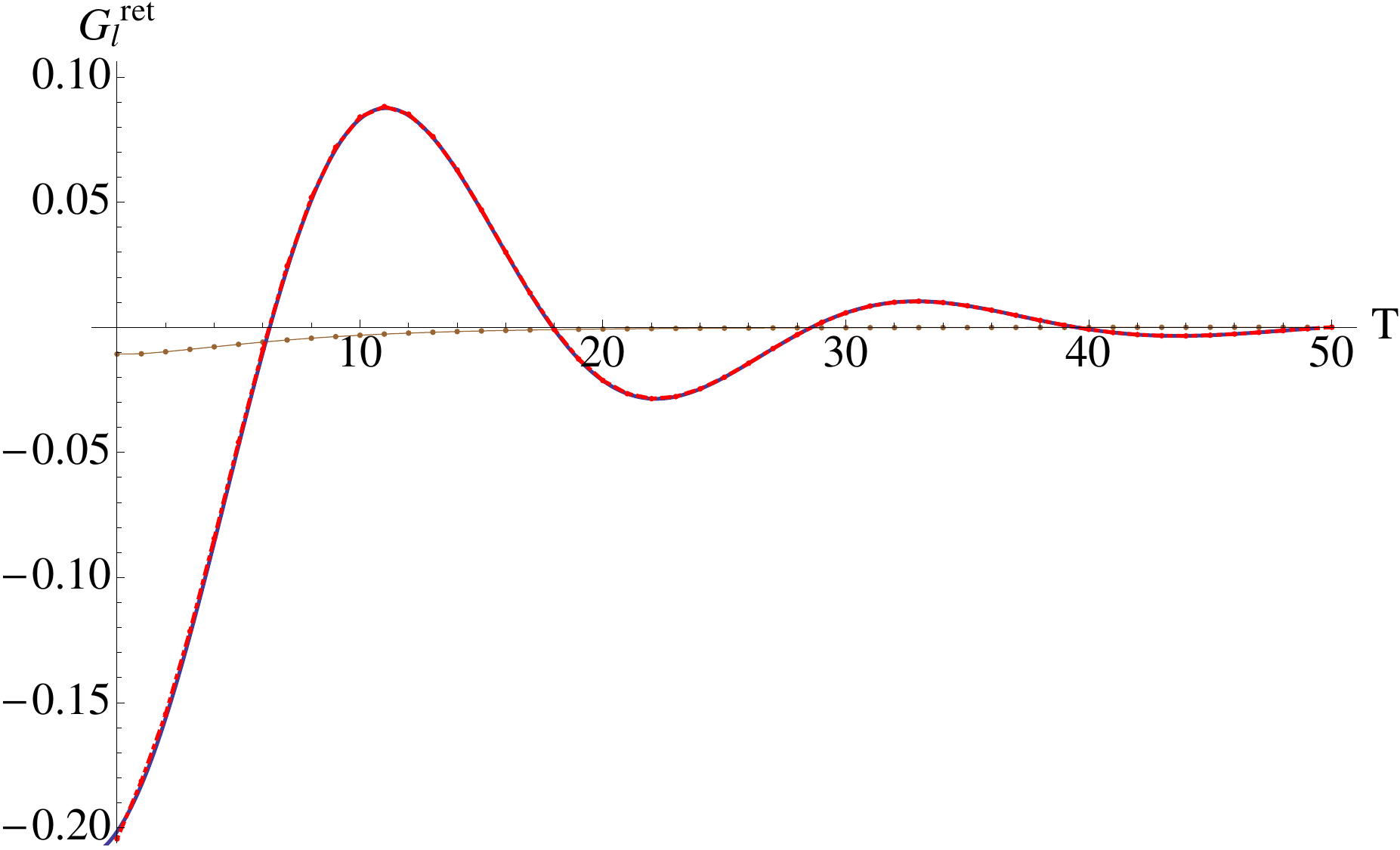}
\end{center}
\caption{
Mode $\ell=1$ of the retarded Green function, i.e., $G^{ret}_{\ell}(r,r';t)$, for $s=0$ and $r=r'=10M$ as a function of the `time' $T\equiv t-|r_*|-|r'_*|$.
Continuous blue curve: `exact'  numerical solution $G^{ret}_{\ell}$ obtained using~\cite{Wardell-scalarwave1d}.
Dashed black curve: $G_{\ell}^{BC}$ of Eq.(\ref{eq:GBC}).
Dot-dashed red curve (overlapping with the continuous blue curve):
 $G_{\ell}^{BC}$ plus the corresponding QNM series contribution (summing overtone numbers $n:0\to 23$) obtained using the method in~\cite{DolanOttewill:QNMMST}.
The matching between the numerical solution and the sum of $G_{\ell}^{BC}$ plus QNM series is excellent.
}
\label{fig:GlBC}
\end{figure}

In Fig.\ref{fig:GlBC} we construct the mode $\ell=1$ of the retarded Green function, $G^{ret}_{\ell}(r,r';t)$, in the scalar case $s=0$ at the radii $r=r'=10M$.
We plot: (1) the BC contribution to $G^{ret}_{\ell}$, i.e., $G_{\ell}^{BC}$, (2) the sum of  $G_{\ell}^{BC}$ and the QNM contribution to $G^{ret}_{\ell}$ (taking into account the 
QNMs for the first 24 overtones),
 and 
 (3)
 the full `exact' $G^{ret}_{\ell}$.
 We calculate $G_{\ell}^{BC}$ by integrating over the frequency the BC modes $\DGl$
obtained using two different methods in the `mid'-frequency and large-frequency regimes - in this case, the crossover frequency 
 is at $\nb=50$ \fixme{what about small-$\nb$?}.
In the `mid'-frequency regime, we interpolate the values of the BC modes $\DGl$ obtained  as described in the previous section.
In the large-frequency regime we use the asymptotics of~\cite{Casals:2011aa} for the BC modes.
 We calculate the QNM contribution to $G^{ret}_{\ell}$ using the method in~\cite{DolanOttewill:QNMMST}.
 We obtain the `exact' $G^{ret}_{\ell}$ by numerically integrating the (1+1)-dimensional partial differential equation
 $\left(-\partial^2_t+\partial^2_{r_*}-V\right)\phi_{\ell}(r,t)=0$
(with the potential $V$ given by  Eq.(\ref{eq:radial ODE})) 
 for the $\ell$-mode $\phi_{\ell}$ of the field
using the following initial data.
We choose zero data for the inital value of the $\ell$-mode of the field.
For the initial value of the time-derivative of the $\ell$-mode of the field, on the other hand, we choose a Gaussian distribution in $r_*$ 
 `peaked' at a certain value $r_{*0}$.
 From the Kirchhoff integral representation for the field (e.g.,~\cite{Leaver:1986}), the 
 solution  $\phi_{\ell}(r,t)$
 thus obtained should approximate the $\ell$-mode of the retarded Green function,  $G^{ret}_{\ell}(r,r_0;t)$, 
 where $r_0\equiv r(r_{*0})$ \fixme{is the order of the arguments correct?}.
 We used a Gaussian width of approximately $0.2M$ and we checked that the change in the numerical solution obtained by using
  smaller values of the width was negligible for our
 purposes. For the numerical integration of the (1+1)-dimensional partial differential equation
 we used Wardell's C-code available in~\cite{Wardell-scalarwave1d}.
A slightly different version of this numerical approach using a Gaussian distribution (though using it as the source, rather than as initial data)
 has recently been successfully applied in~\cite{Zenginoglu:2012xe} in the full (3+1)-dimensional case.
We observe from  Fig.\ref{fig:GlBC}  that the BC contribution becomes most significant for small values of the `time' $T\equiv t-|r_*|-|r'_*|$
but, in the regime plotted, the BC contribution  is always subdominant to the QNM contribution.
The matching between the numerical solution and the sum of $G_{\ell}^{BC}$ plus QNM series is excellent.
\fixme{See what it looks like for $T<0$}
For $T<0$ neither the QNM series nor the BC integral is expected to converge separately~\cite{Casals:2011aa}.

\begin{figure}[h!]
\begin{center}
\includegraphics[width=8cm]{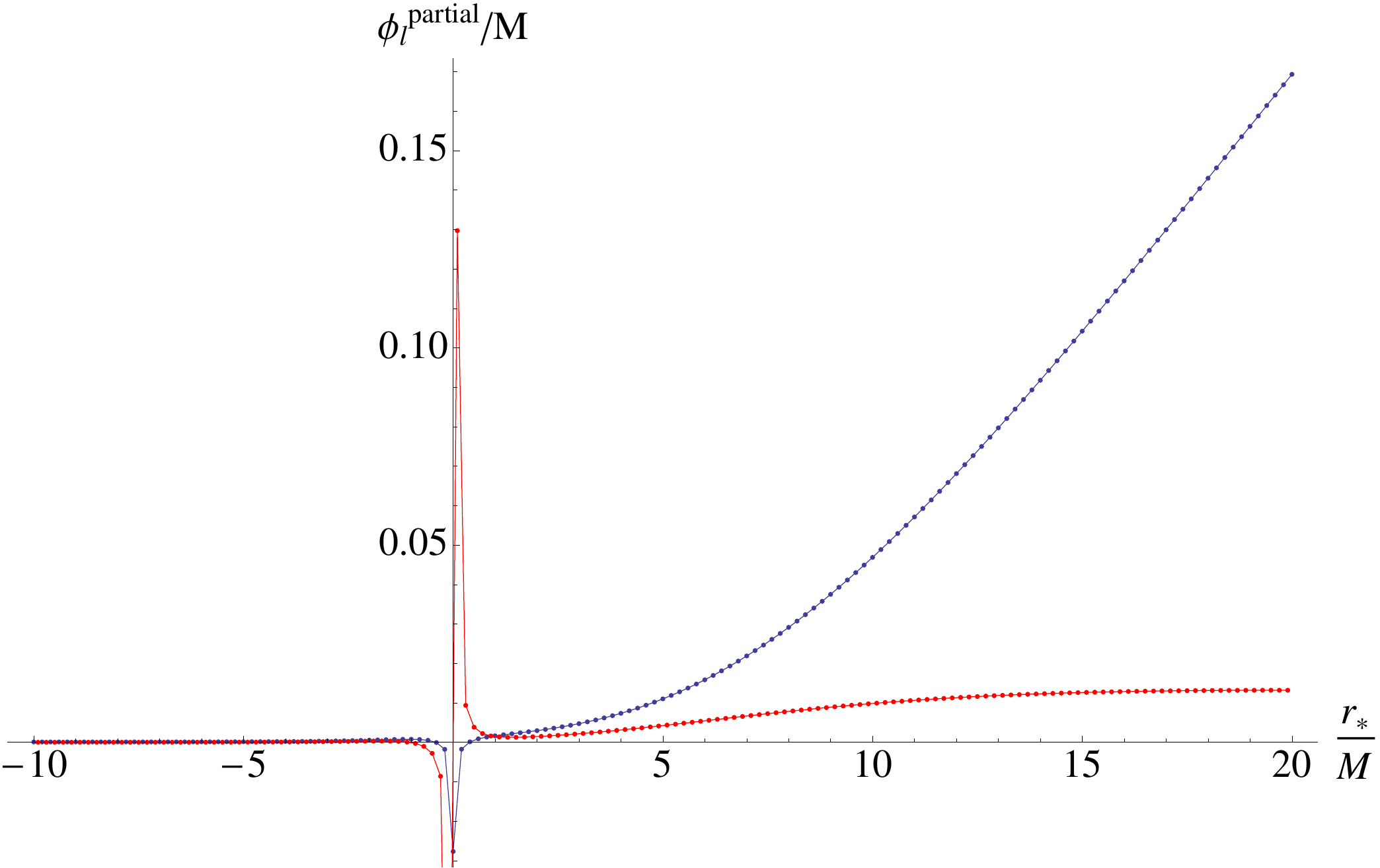}
\includegraphics[width=8cm]{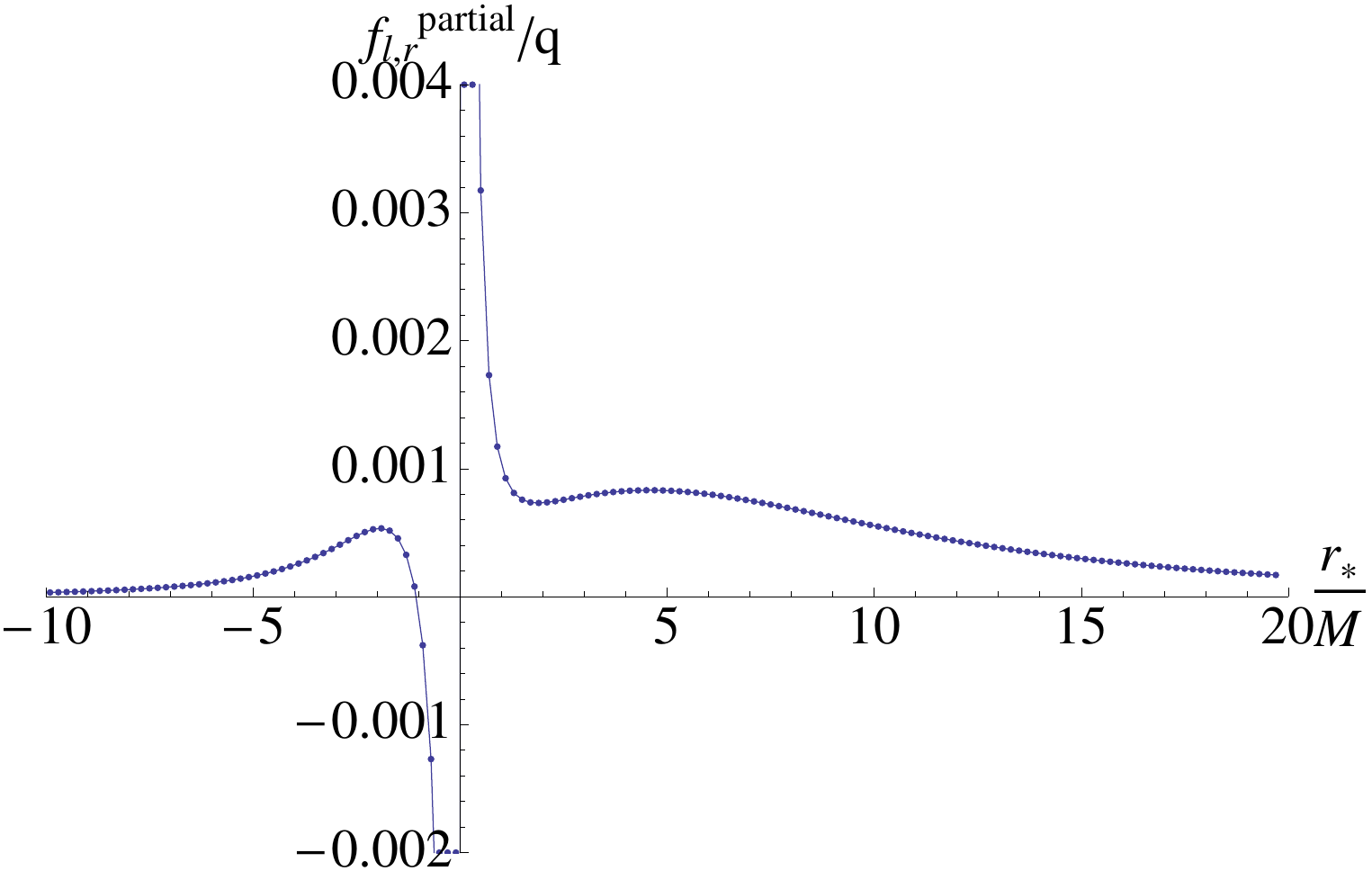}
\end{center}
\caption{
`Partial'  BC contribution to the $\ell$-mode for the radial component of the self-force
for $s=0$ and $\ell=1$ as a function of $r_*$. 
(a) Continuous blue curve: BC contribution to the `$\ell$-mode of the partial  field' $\phi_{\ell}^{partial}/M$ of Eq.(\ref{eq:partial field}).
Continuous red curve: its $r_*$-derivative.
(b)  
$f_{\ell, r}^{partial}(r)/q$ of Eq.(\ref{eq:partial r SF}) in the static case  (i.e., $\gamma=0$ and $d\tau/dt=(1-2M/r)^{1/2}$).
}
\label{fig:partial field}
\end{figure}

Let us now define the `$\ell$-mode of the partial field' as 
\begin{equation} \label{eq:partial field}
\phi_{\ell}^{partial}(r)\equiv \int_
{2|r_*|}
^{\infty}dt\ G^{ret}_{\ell}(r,r'=r;t)
\end{equation}
The contribution to the radial component of the self-force per unit charge in the
case of a particle at
constant radius
from the $\ell$-mode of the Green function from the segment of the 
worldline  lying between 
$t=2|r_*|$
and $t\to \infty$ is then obtained
as:
\begin{equation} \label{eq:partial r SF}
\frac{f_{\ell, r}^{partial}(r)}{q}\equiv \frac{(2\ell+1)}{r^2}P_{\ell}(\cos\gamma)\frac{d\tau}{dt} \left[\frac{d\phi_{\ell}^{partial}}{dr} -\frac{\phi_{\ell}^{partial}}{r}\right]
\end{equation}
 \fixme{and minus sign?}
\fixme{Units are wrong? }
This clearly only yields a partial contribution to the self-force from the $\ell$-mode since we are integrating from 
$t=2|r_*|$
instead of
from $t=0^+$, as required in order to obtain the self-force.
Because of the divergence of the BC and QNM contributions for $T<0$, in order to obtain the contribution from the worldline segment for 
$t: 0^+\to  2|r_*|$ 
we require
a different method for calculating the Green function, such as a quasi-local series (see, e.g.,~\cite{CDOWb}).
The BC contribution to $\phi_{\ell}^{partial}$ is obtained by inserting $G_{\ell}^{BC}$ in the place of
$G^{ret}_{\ell}$ in Eq.(\ref{eq:partial field}).
In Fig.\ref{fig:partial field}(a) we plot this contribution and its $r_*$-derivative
(evaluated using a central difference scheme)
 as functions of the radius in the case $s=0$ and $\ell=1$.
In Fig.\ref{fig:partial field}(b) we plot the corresponding BC contribution to $f_{\ell, r}^{partial}(r)/q$ in the static case $\gamma=0$ and 
$d\tau/dt=(1-2M/r)^{1/2}$.
\fixme{Comments about the plots in  Fig.\ref{fig:partial field}?}

\section{Discussion} \label{sec:discussion}

In this paper we have presented the first analytic method for calculating the branch cut modes in the non-asymptotic, `mid'-frequency regime
 in Schwarzschild space-time for fields of any integral spin. 
We have investigated their properties, in particular regarding their relation to quasinormal mode frequencies and around the algebraically
special frequency.
We have applied our calculation of the BC modes to investigate their partial (i.e., from $t>2|r_*|$)
 contribution for one $\ell$-mode to the self-force on a scalar charge moving on Schwarzschild
background at constant radius.
We have found that, for the particular case investigated, 
the BC contribution becomes larger as $t$ approaches $2|r_*|$ 
(where the high-frequency asymptotics of the BC modes become important)
but the QNM contribution dominates the self-force at most times.


In~\cite{CDOWa} the first successful application of the so-called method of matched expansions for the calculation of the self-force was achieved.
This method consists on calculating the Green function at `early times' using a quasilocal expansion and in the `distant past' using a Fourier mode and multipole
decomposition of the Green function as in Eq.(\ref{eq:Green}). 
In~\cite{CDOWa} the method was applied to the specific case of a black hole toy model space-time,
namely the Nariai space-time, where the Green function possesses
QNMs but not a BC and so the Green function in the `distant past' is fully determined by the QNM series.
In Schwarzschild space-time, on  the other hand, the QNM series must be complemented by a BC integral.
In~\cite{Dolan:2011fh} the QNM series was calculated for large-$\ell$ and it was shown that it  yields an interesting global singularity structure of the Green function.
In this paper we have presented a method for calculating the BC integral and we have applied it to one $\ell$-mode.
In~\cite{CDOW:MatchExpSchw} we plan to apply a calculation of the BC integral for all $\ell$-modes, add it to a similar calculation of the QNM series
and supplement it with a quasilocal expansion at `early times' in order to calculate the full self-force in Schwarzschild space-time 
using the method of matched expansions.


Another situation where it is important to investigate the contribution of the BC is that of the response of a black hole to an initial perturbation.
In this case, we expect that the BC contributes at `early' times (that is, for $t$ close to $|r_*|+|r_*|$) 
as well as at late times (where a logarithmic behaviour precedes the known power-tail decay~\cite{PhysRevLett.109.111101}).
We investigate the latter in depth in~\cite{Casals:Ottewill:2011smallBC}.

On the quantum side, 
 the asymptotically constant spacing in the imaginary part of the highly-damped QNM frequencies 
 led to suggestions of a link with
the quantization of the black hole area~\cite{Hod:1998vk,Maggiore:2007nq}.
In~\cite{Casals:2011aa} we showed that, in the large-$\nb$  regime,
 the spacing in the imaginary part of the QNM frequencies asymptotically equals
that of the zeros of the BC modes for all spins $s=0,1$ and $2$.
In this paper we have shown that, in the `mid'-frequency regime, these two spacings also remain very close 
 in the cases $s=1$ and $2$ studied, while they
differ more significantly in the case $s=0$.
 Intriguingly, the algebraically special frequency for gravitational perturbations plays
a special r\^ole 
in the connection between QNMs and the BC,
 not only harbouring in its neighbourhood an almost purely imaginary QNM frequency but also marking the onset
of the highly-damped regime for QNMs.
In a different work, highly-damped QNMs in Kerr
space-time have been interpreted as semiclassical bound
states along a specific contour in the complex-$r$ plane and have been linked to Hawking radiation~\cite{Keshet:2007be}.
The least-damped QNMs have also been linked to Hawking radiation~\cite{York:1983}.
Given that QNM frequencies, particularly in the spin-1 case, `approach' the branch cut~\cite{Casals:2011aa} in the high-damping limit
and that a connection with the branch cut also appears to exist in the `mid'-frequency regime, 
it would be interesting to investigate
whether branch cut modes may play any r\^ole in the quantum properties of black holes.

An impending generalization of our current results in Schwarzschild is that to a rotating, Kerr black hole space-time.
In principle, our method is readily generalizable to the rotating case, since Leaver's series representations for the radial
solutions are already valid in Kerr~\cite{Leaver:1986a}.
Immediately, however, some significant differences appear with respect to the non-rotating Schwarzschild case.
For example, the corresponding symmetry (\ref{eq:symms f,g}) in Kerr also involves a change in the sign of the azimuthal angular number,
on which the radial solutions depend in the rotating case.
As a consequence, the radial solution $\f$ is not necessarily real along the branch cut and
the discontinuity of $\g$ across the branch cut is generally not only in its imaginary part but also in its real part, and so
the BC `strength' $q(\nu)$ is not necessarily real-valued.
To further 
spice up the analysis in Kerr,
the angular eigenvalue has various branch points in the complex-frequency plane (see, e.g.,~\cite{BONGK:2004}).
This intricate and delicate structure of the modes in Kerr has various physical manifestations.
We plan to investigate these issues in a future publication.

\appendix

\section{Irregular confluent hypergeometric $U$-function} \label{sec:App}

In this appendix we give some properties of the irregular confluent hypergeometric $U$-function, which we have used in the main body of the paper.

A useful integral representation of the $U$-function is given in, e.g., Eq.13.4.4~\cite{bk:onlineAS}:
\begin{equation}\label{eq:U num int}
U(a,b,z)=\frac{1}{\Gamma (a)}\int_0^{\infty}dt\ e^{-zt}t^{a-1}(1+t)^{b-a-1},\quad \text{Re}(a)>0, |\text{ph}(z)|<\pi/2
\end{equation}

We require asymptotics for large values of the first argument of the $U$-function.
For large, negative values of the first argument we  use Eqs.13.8.10 and 10.17.3~\cite{bk:onlineAS} to obtain
\begin{equation} \label{eq:large-n U}
U(a,b,x)\sim
\Gamma\left(\frac{b}{2}-a+\frac{1}{2}\right)e^{x/2}x^{1/4-b/2}\left(\frac{\sqrt{2}}{\pi\sqrt{b-2a}}\right)^{1/2}
\cos\left(\sqrt{2x(b-2a)}+\pi\left(a-\frac{b}{2}+\frac{1}{4}\right)\right),\quad
a\to -\infty,b\ge 1,x>0
\end{equation}
\fixme{or is it just $x\in\mathbb{R}$? and does $a\to -\infty$ require $a<0$ or just $\text{Re}(a)<0$?}
\fixme{this differs from Eq.13.5.16~\cite{bk:AS} and p.68~\cite{Temme:1983} (in this case there's even an extra $b$ in the $\sqrt{}$)
in the factor $(\frac{\sqrt{2}}{\sqrt{b-2a}})^{1/2}$??}

For large, positive values of the first argument we  use  Eqs.13.8.8 and 10.25.3~\cite{bk:onlineAS} to obtain
\begin{align} \label{eq:large-a U}
&
U(a,b,x)\sim
\frac{e^{x/2}}{\Gamma(a)}\left(\sqrt{\frac{|x|}{a}}e^{i\psi/2}\right)^{1-b}\sqrt\frac{\pi}{\sqrt{|x|a}e^{i\psi/2}}e^{-2\sqrt{|x|a}e^{i\psi/2}},\quad
a\to +\infty,\ b\le 1,
\ \psi\equiv \arg(x)\in (-\pi,+\pi]
\end{align}
\fixme{does $a\to -\infty$ require $a<0$ or just $\text{Re}(a)<0$?}
\fixme{The asymptotics in~\cite{bk:onlineAS} are valid ``uniformly with respect to $x\in[0,\infty]$. However, in Eq.(\ref{eq:large-a U}) I have
applied them to $x\in\mathbb{C}$ - is this correct?}.

Finally, from Eqs.13.2.4 and 13.2.41~\cite{bk:onlineAS} we obtain the following expression for the discontinuity across the branch cut
of the $U$-function,
\begin{equation} \label{eq:Delta z^aU}
\left(ze^{-2\pi i}\right)^aU(a,b,ze^{-2\pi i})-z^aU(a,b,z)=\frac{2\pi ie^z(-z)^a}{\Gamma(a)\Gamma(1+a-b)}U(b-a,b,-ze^{-2\pi i})
\end{equation}




%
%


\begin{acknowledgments}
We are thankful to Sam Dolan and Barry Wardell for
many 
helpful discussions.
M.C. also thanks Yuk Tung Liu for useful discussions and, particularly, for providing us with Ref.~\cite{Liu-1997}.
M.C. gratefully acknowledges support by a IRCSET-Marie Curie International Mobility Fellowship in Science, Engineering and Technology.
A.O. acknowledges support from Science Foundation Ireland under grant no 10/RFP/PHY2847.
\end{acknowledgments}



\bibliographystyle{apsrev}


\end{document}